\documentclass[traditabstract]{aa}
\usepackage{txfonts}
\usepackage{graphicx}
\usepackage{longtable}
\usepackage{afterpage}
\usepackage{lscape}

\newcommand{\kms}   {km~s$^{-1}$}
\newcommand{\mo}    {$M_{\sun}$}
\newcommand{\lo}    {$L_{\sun}$}
\newcommand{\nh}    {NH$_3$}
\newcommand{\cmd}   {cm$^{-2}$}

\newcommand{\hho}   {H$_2$O}
\newcommand{\hh}    {H$_2$}
\newcommand{\tco}   {$^{13}$CO}
\newcommand{\dco}   {$^{12}$CO}
\newcommand{\cdo}   {C$^{18}$O}
\newcommand{\mm}    {$\mu$m}
\newcommand{\hco}   {HCO$^+$}   
\newcommand{\kmp}   {km~s$^{-1}$~pc$^{-1}$}
\newcommand{\hii}   {H~{\small II}}
\newcommand{\cts}   {C$^{34}$S}

\begin{document}

\title{Dense gas and the nature of the outflows \thanks{The Appendix is only available in electronic 
form at {\tt http://www.edpsciences.org}}\fnmsep \thanks{Haystack Observatory data as CLASS files are 
available in electronic form at the CDS via anonymous ftp to cdsarc.u-strasbg.fr (130.79.128.5)
or via {\tt http://cdsweb.u-strasbg.fr/cgi-bin/qcat?J/A+A/vol/page}}}

\author{I. Sep\'ulveda\inst{1} 
\and G. Anglada\inst{2}
\and R. Estalella\inst{1}
\and R. L\'opez\inst{1}
\and J.M. Girart\inst{3}
\and J. Yang\inst{4}}
\institute{
 Departament d'Astronomia i Meteorologia, Institut de Ci\`encies del Cosmos,
Universitat de
Barcelona, Mart\'{i} i Franqu\`es 1, 08028 Barcelona, Catalunya, Spain.
 \and Instituto de Astrof\'{\i}sica de Andaluc\'{\i}a, CSIC, Camino Bajo 
de Hu\'etor 50, 18008 Granada, Spain.
 \and Institut de Ci\`encies de l'Espai (CSIC-IEEC), Campus UAB, Facultat de
Ci\`encies, Torre C5-parell 2, 08193 Bellaterra, Catalunya, Spain.
 \and Purple Mountain Observatory, Chinese Academy of Sciences, Nanjing 210008,
PR China.}

%\offprints{I. Sep\'ulveda}

\date{Received 17 July 2009; Accepted 12 October 2010}

\abstract{We present the results of the observations of the $(J,K)=(1,1)$ and
the $(J,K)=(2,2)$
inversion transitions of the \nh\ molecule toward a large sample of 40 regions
with molecular or optical outflows, using the 37 m radio telescope of the
Haystack Observatory. We detected \nh\ emission in 27 of the
observed regions, which we mapped in 25 of them. Additionally, we
searched for the $6_{16}-5_{23}$ \hho\ maser line toward six regions, detecting
\hho\ maser emission in two of them, HH265 and AFGL 5173.
We estimate the physical parameters of the regions mapped in \nh\ and analyze for each
particular region the distribution of high density gas and its
relationship with the presence of young stellar objects. In particular, we
identify the deflecting high-density clump of the HH270/110 jet. We were
able to separate the \nh\ emission from the L1641-S3 region into two overlapping
clouds, one with signs of strong perturbation, probably associated with the
driving source of the CO outflow, and a second, unperturbed clump, which is probably
not associated with star formation.
We systematically found that the position of the best candidate for the exciting source
of the molecular outflow in each region is very close to an \nh\ emission
peak. 
From the global analysis of our data we find that in general the highest
values of the line width are obtained for the regions
with the highest values of mass and kinetic temperature. We also found a
correlation between the nonthermal line width and the bolometric luminosity of
the sources, and between the mass of the core and the bolometric luminosity.
 We confirm with a larger sample of regions the conclusion of Anglada et al.\
(1997) that  the \nh\ line emission is
more intense toward molecular outflow sources than toward sources with optical
outflow, suggesting a possible evolutionary scheme in which young stellar
objects associated with molecular outflows progressively lose their neighboring
high-density gas, weakening both the \nh\ emission and the molecular outflow in
the process, and making optical jets more easily detectable as the total amount
of gas decreases.} 

\keywords{ISM: jets and outflows -- ISM: molecules -- stars:formation}

\maketitle

\section{Introduction}

Over the last decades, a great effort has been made to study the
processes that take place in the earliest stages of stellar evolution.  
It is now widely accepted that low-mass stars begin their lives in
the densest cores of molecular clouds and that the earliest stages of
stellar evolution are associated with processes involving a strong mass
loss, traced by molecular outflows, Herbig-Haro
objects, and optical jets, which emanate from the deeply embedded young stellar
objects.
These mass-loss processes have been proposed as a way to eliminate the
excess of material and of angular momentum as well as to
regulate the IMF (Shu, Adams \& Lizano 1987). The molecular outflow phase
is known as one of the earliest observable phases of the stellar
evolution. Several studies indicate that most, if not all, of the Class 0 and
Class I sources drive molecular outflows (e.g. Davis et al.\ 1999) and that an
important fraction of these sources are associated with Herbig-Haro objects, as
well as with molecular
outflows (Eiroa et al.\ 1994;  Persi et al.\ 1994). These results suggest
that both phenomena start and coexist in the early stages of the star-formation process. 

These outflows emanating from protostars collide with
the remaining molecular cloud and disperse the surrounding material, and
determine the evolution of the dense core where the star is born (Arce \& Sargent 2006). 
In this sense, the study of these mass-loss processes and the molecular environment of the 
embedded objects from which they emanate has became an
important tool in order to better understand the earliest stages of
stellar evolution.

Ammonia observations have proved to be a powerful tool for studying the
dense cores where the stars are born. Since the first surveys of dense cores
(Torrelles et al.\ 1983, Benson \& Myers 1989, Anglada et al.\ 1989), a clear
link was established between dense cores, star formation and outflows (see e.g., the review of Andr\'e et al.\ 2000).
From these surveys, it was clearly established that the driving sources of outflows
are usually embedded in the high-density gas, which is traced by the \nh\ emission, and are
located very close to the emission peak (Torrelles et al.\ 1983; Anglada et
al.\ 1989). Following these results, Anglada et al.\ (1997) (hereafter 
Paper I) undertook a survey of dense cores to investigate the
relationship between the type of outflow and the dense gas associated with their
exciting sources. A statistical study of the sources observed in that survey
reveals that the ammonia emission is more intense toward molecular outflow
sources than toward sources with only optical outflows, indicating that
molecular outflows are associated with a larger amount of high-density gas. 
From this result, a possible evolutionary scheme was suggested in which young
objects associated with molecular outflows progressively lose their neighboring
high-density gas, while both the \nh\ emission and the molecular
outflow become weaker in
the process, and the optical jets become more easily detectable as the total amount
of gas and extinction decreases. In this sense, the observations of high-density tracers, 
such as the \nh\ molecule, confirm the decrease of high-density gas around the stars.

We present here new ammonia observations. Additional
regions allow us to obtain a sample of outflow regions 
observed in ammonia that doubles the number used in Paper I.  We selected a
sample 
of 40 star-forming regions, taking into account the presence of molecular
outflows, 
optical outflows, or both, and mapped
with the Haystack 37 m telescope the \nh\ emission around the suspected outflow
exciting sources. In Sect.~\ref{obs4} we describe the observational procedure, 
in Sect.~\ref{sources4} we present the observational results (the discussion of individual sources
is presented in Appendix A), in Sect.~\ref{results} we discuss the global results, in
Sect.~\ref{evol} we describe the relationship between the high-density gas
and the nature of the outflow based on the sample, and in
Sect.~\ref{conclus4} we give our conclusions.

\section{Observations}\label{obs4}

We observed the $(J,K)$=(1,1) and the $(J,K)$=(2,2) inversion transitions of
the ammonia molecule with the 37 m radio telescope at Haystack
Observatory\footnote{Radio Astronomy at Haystack Observatory of the Northeast
Radio Observatory Corporation was supported by the National Science Foundation}
in 1993 January, 1996 May and 1997 December. At the frequencies of these
transitions (23.6944960 GHz and 23.7226320 GHz, respectively), the half power
beam width of the telescope was $1\farcm4$ and the beam efficiency at an
elevation of $40\degr$ was $\sim 0.41$ for the observations made in 1993 and
$\sim 0.33$ for the observations made in 1996 and 1997.  In all the observing
sessions, we used a cooled K-band maser receiver and a 5\,000-channel
autocorrelation spectrometer with a full bandwidth of 17.8 MHz. The calibration
was made with the standard noise-tube method. All spectra were corrected for
elevation-dependent gain variations and for atmospheric attenuation. The rms
pointing error of the telescope was $\sim 10''$. Typical system temperatures
were $\sim100$ K, $\sim 140$ K and $\sim 90$ K for the observations made in
1993, 1996, and 1997, respectively. The observations were made in the position
switching mode in 1993 and 1997 and in frequency switching mode in 1996. During
the data reduction the observed spectra were smoothed to a velocity resolution
of $\sim 0.11$ \kms, achieving a $1\sigma$ sensitivity of $0.2$ K per spectral
channel.

\begin{table*}
  \caption[]{Regions observed in \nh}
\resizebox{\hsize}{!}{
\centering
\begin{tabular}{l c c c c c c c r c l}
\noalign{\medskip\hrule\medskip}
Region&
\multicolumn{2}{c}{Reference Position\tablefootmark{a}}&
Observation &
Outflow\tablefootmark{b} &
Ref. &
Molecular &
Ref. &
\multicolumn{1}{c}{$D$} &
Ref. &
Alternative \\ \cline{2-3}
 &$\alpha(1950)$ & $\delta(1950)$  &Epoch &Type & &Observations & &\multicolumn{1}{c}{(pc)} & & Name \\

\noalign{\medskip\hrule\medskip}

M120.1+3.0-N\tablefootmark{c}
& $00^h 21^m 22\fs0$ & $+65\degr 30\arcmin 24\arcsec$ &1993 Jan & CO & 1
&- & - &850 &1  & -\\

M120.1+3.0-S\tablefootmark{d}
& $00^h 25^m 59\fs8$ & $+65\degr 10\arcmin 11\arcsec$ &1993 Jan & CO & 1
&- &- &850 &1  &- \\

L1287 
& $00^h 33^m 53\fs3$ & $+63\degr 12\arcmin 32\arcsec$ &1993 Jan
& CO & 2 &\hco,HCN,CS,\nh &2,3,4,5 &850 &2 & RNO1B/1C\\

L1293
& $00^h 37^m 57\fs4$ & $+62\degr 48\arcmin 26\arcsec$ &1993 Jan & CO & 6
& HCN,\hco & 6 &$850$ &6 &-\\

NGC 281 A-W
& $00^h 49^m 27\fs8$ & $+56\degr 17\arcmin 28\arcsec$ &1993 Jan & CO & 7
&CS,\nh,HCN,\hco,\cdo,\cts &8,9,10,80 & $3\,500$ &9  &S184\\

HH 156
& $04^h 15^m 34\fs8$ & $+28\degr 12\arcmin 01\arcsec$ &1997 Dec &jet  &11
&- &- &140 &11 &CoKu Tau 1 \\

HH 159
& $04^h 23^m 58\fs4$ & $+25\degr 59\arcmin 00\arcsec$ &1996 May &jet, CO
&12,13 &- &-  &160 &12 &DG Tau B\\

HH 158
& $04^h 24^m 01\fs0$ & $+25\degr 59\arcmin 35\arcsec$ &1996 May &jet &14
&\cdo, CS &15,16 &160 &14 &DG Tau\\

HH 31
& $04^h 25^m 14\fs4$ & $+26\degr 11\arcmin 04\arcsec$ &1996 May & jet,
CO? &11,79 &CS,\cdo &17,18 &160 &23 &-\\  

HH 265
& $04^h 28^m 21\fs0$ & $+18\degr 05\arcmin 35\arcsec$ &1996 May & HH &19 &CS,\nh,\cdo & 81
&160 &20 &L1551MC \\

L1551 NE
& $04^h 28^m 50\fs5$ & $+18\degr 02\arcmin 10\arcsec$ &1996 May & CO, jet
&21,22 & - & - &160 &20 &L1551 \\

L1642
& $04^h 32^m 32\fs0$ & $-14\degr 19\arcmin 18\arcsec$ &1993 Jan & CO, HH
&23,24 &\hco,\cdo &25 &200 & 25 &HH 123 \\

L1634
& $05^h 17^m 13\fs8$ & $-05\degr 54\arcmin 45\arcsec$ &1996 May, 1997 Dec& 
CO,\hh,jet &26,27 & - & - &460 & 28 &HH 240/241,RNO 40\\

HH 59
& $05^h 29^m 52\fs0$ & $-06\degr 31\arcmin 09\arcsec$ &1996 May &HH & 29
& - & &460 &29 &-\\

IRAS 05358
& $05^h 35^m 48\fs8$ & $+35\degr 43\arcmin 41\arcsec$ &1993 Jan & CO & 7
& HCN,\hco,CS,\nh,\cts &30,31,81 &$1\,800$ &7 &-\\

L1641-S3
& $05^h 37^m 31\fs7$ & $-07\degr 31\arcmin 59\arcsec$ &1993 Jan, 1996 May&
CO &32 & \nh,CS &33,34 &480 &32 &-\\

HH 68
& $05^h 39^m 08\fs7$ & $-06\degr 27\arcmin 20\arcsec$ &1996 May &HH &29
& & &460 &29 & - \\

CB 34
& $05^h 44^m 03\fs0$ & $+20\degr 59\arcmin 07\arcsec$ &1993 Jan & CO, jet,
\hh &35,36 & \cdo,CS,\nh,HCN &37,38,39,40 &$1\,500$ &41 &HH 290\\

HH 270/110
& $05^h 48^m 57\fs4$ & $+02\degr 56\arcmin 03\arcsec$ &1997 Dec & jet &42
&\cdo &83 &460 &42 &L1617\\

IRAS 05490   
& $05^h 49^m 05\fs2$ & $+26\degr 58\arcmin 52\arcsec$ &1993 Jan & CO & 7
& CS &8 &$2\,100$ &43 &S242\\

HH 111
& $05^h 49^m 09\fs1$ & $+02\degr 47\arcmin 48\arcsec$ &1993 Jan & CO, jet,
\hh &44,45,46 &CS &47 &460 &44 &L1617\\

HH 113
& $05^h 50^m 58\fs1$ & $+02\degr 42\arcmin 49\arcsec$ &1996 May & jet &44
&- &- &460 &44 &L1617 \\

AFGL 5173
& $05^h 55^m 20\fs3$ & $+16\degr 31\arcmin 46\arcsec$ &1996 May & CO &7
&CS &8 &$2\,500$ &7 &-\\

CB 54
& $07^h 02^m 06\fs0$ & $-16\degr 18\arcmin 47\arcsec$ &1993 Jan & CO &35
&\cdo,CS,HCN &37,38,40 &$1\,500$ &41 & LBN 1042 \\

L1709
& $16^h 28^m 33\fs5$ & $-23\degr 56\arcmin 32\arcsec$ &1996 May& CO &48 &-
 &- &160 &49 &-\\

L379
& $18^h 26^m 32\fs9$ & $-15\degr 17\arcmin 51\arcsec$ &1993 Jan & CO &32 &
\cdo,\nh &50,51 &$2\,000$ &32 &-\\

L588
& $18^h 33^m 07\fs6$ & $-00\degr 35\arcmin 48\arcsec$ &1997 Dec & CO?, HH
&48,52 & - & - &310 &52 &HH 108/109 \\

CB 188
& $19^h 17^m 57\fs0$ & $+11\degr 30\arcmin 18\arcsec$ &1993 Jan & CO &35 
&HCN,CS,\cdo  &40,38,37 &300 &41 &- \\

L673\tablefootmark{e}
& $19^h 18^m 30\fs8$ & $+11\degr 09\arcmin 48\arcsec$ &1996 May &CO &53 
&\nh,CS,\cts,HCN &54,55,56 & 300 &57 &RNO 109\\

HH 221
& $19^h 26^m 37\fs5$ & $+09\degr 32\arcmin 24\arcsec$ &1996 May &jet &58
&- &- &1\,800 &58 &Parsamyan 21\\

L797
& $20^h 03^m 45\fs0$ & $+23\degr 18\arcmin 25\arcsec$ &1993 Jan & CO &35
&HCN,CS,\cdo &59,38,37 &700 &41 &CB 216 \\

IRAS 20050
& $20^h 05^m 02\fs5$ & $+27\degr 20\arcmin 09\arcsec$ &1996 May & CO &60
&CS,\hco,HCN,\cdo &60,61,78,84 &700 &62 &-\\

V1057 Cyg
& $20^h 57^m 06\fs2$ & $+44\degr 03\arcmin 47\arcsec$ &1996 May &CO &63
& - &- &700 & 49 &-\\

CB 232
& $21^h 35^m 14\fs0$ & $+43\degr 07\arcmin 05\arcsec$ &1993 Jan &CO &35
& \cdo,CS &37,38 &600 &41 &B158\\

IC 1396E
& $21^h 39^m 10\fs3$ & $+58\degr 02\arcmin 29\arcsec$ &1993 Jan &CO &32
& \cdo,CS,\nh,HCN,\hco &64,65,66,10 &750 &32 &GRS 14,IC1396N\\

L1165
& $22^h 05^m 09\fs6$ & $+58\degr 48\arcmin 06\arcsec$ &1997 Dec &CO, HH
&48,67 & - & - &750 &67 &HHL75,HH 354\\

IRAS 22134
&$22^h 13^m 24\fs2$ & $+58\degr 34\arcmin 12\arcsec$ &1996 May &CO &68
& CS,\cdo &86,88 &2\,600 &87 &S134\\

L1221
& $22^h 26^m 37\fs2$ & $+68\degr 45\arcmin 52\arcsec$ &1996 May &CO, HH
&69,70 & CS,\hco, HCN,\cdo &69 &200 &69 &HH 363\\

L1251\tablefootmark{e}
& $22^h 37^m 40\fs8$ & $+74\degr 58\arcmin 38\arcsec$ &1996 May &CO, HH
&71,72 &\nh,CS,\cdo &54,55,73 &300 &74 &-\\

NGC 7538
& $23^h 11^m 35\fs2$ & $+61\degr 10\arcmin 37\arcsec$ &1993 Jan &CO &75
&HCN,CS,\cts,\hco &76,77,85 &2\,700 &75 &-\\

\noalign{\smallskip}
\hline
\noalign{\smallskip}
\end{tabular} 
}
\tablefoot{ 
\tablefoottext{a}{ Position where the observations were centered}. 
\tablefoottext{b}{CO = Molecular outflow; HH = Isolated Herbig-Haro object; jet = 
Optical outflow; \hh\ = Molecular hydrogen outflow.} 
\tablefoottext{c}{ See Paper I for \hho\ results on IRAS 00213+6530 in this 
region}
\tablefoottext{d}{See Paper I for \hho\ results on IRAS 00259+6510 in this 
region}
\tablefoottext{e}{Additional \nh\ data were obtained in 1990 February  (see 
Paper I).}
\tablebib{(1) Yang
et al.\ 1990; (2) Yang et al.\ 1991; (3) Yang et al.\ 1995; (4) Estalella
et al.\ 1993; (5) Carpenter, Snell \& Schloerb 1990; (6) Yang 1990; (7)
Snell et al.\ 1990; (8) Carpenter et al.\ 1993; (9) Henning et al.\ 1994;
(10) Cesaroni et al.\ 1991; (11) Strom et al.\ 1986;  (12) Mundt et al.\
1991; (13) Mitchell et al.\ 1994; (14) Mundt et al.\ 1987; (15) Hayashi et
al.\ 1994; (16) Ohashi et al.\ 1991; (17) Ohashi et al.\ 1996; (18) Onishi
et al.\ 1998; (19)  Garnavich et al.\ 1992; (20) Snell 1981; (21)
Moriarty-Schieven et al.\ 1995; (22) Devine, Reipurth \& Bally 1999; (23)  
Liljestr\"om et al.\ 1989; (24) Reipurth \& Heathcote 1990; (25)
Liljestr\"om 1991; (26) Davis et al.\ 1997; (27) Hoddap \& Ladd 1995; (28)
Reipurth et al.\ 1993; (29)  Reipurth \& Graham 1988; (30) Cesaroni, Felli
\& Walmsley 1999; (31) Zinchenko et al.\ 1997; (32) Wilking et al.\ 1990;
(33) Harju et al.\ 1993; (34) Tatematsu et al.\ 1993; (35) Yun \& Clemens
1994a; (36)  Moreira \& Yun 1995; (37) Wang et al.\ 1995; (38) Launhardt
et al.\ 1998; (39) Codella \& Scappini 1998; (40) Afonso, Yun \& Clemens
1998; (41) Launhardt \& Henning 1997; (42) Reipurth et al.\ 1996; (43)
Blitz et al.\ 1982; (44)  Reipurth \& Oldberg
1991; (45) Reipurth 1989; (46) Gredel \& Reipurth 1993; (47) Yang et al.\
1997; (48) Parker et al.\ 1991; (49) Fukui 1989; (50) Kelly \& Macdonald
1996; (51) Kelly \& Macdonald 1995; (52) Reipurth \& Eiroa 1992; (53)
Armstrong \& Winnewisser 1989; (54) see Paper 1; (55) Morata et al.\
1997; (56) Sandell, H\"oglund \& Kislyakov 1983; (57) Herbig \& Jones
1983; (58) Staude \& Neckel 1992; (59) Scappini et al.\ 1998; (60)
Bachiller, Fuente \& Tafalla 1995; (61) Gregersen et al.\ 1997; (62)
Wilking et al.\ 1989; (63) Levreault 1988; (64) Wilking et al.\
1993; (65) Serabyn et al.\ 1993; (66) Weikard et al.\ 1996; (67) Reipurth
et al.\ 1997; (68) Dobashi et al.\ 1994; (69) Umemoto et al.\ 1991; (70)
Alten et al.\ 1997; (71) Sato \& Fukui 1989; (72) Eiroa et al.\ 1994b; (73)
Sato et al.\ 1994; (74) Kun \& Prusti 1993; (75) Kameya et al.\ 1989; (76)
Cao et al.\ 1993; (77) Kameya et al.\ 1986; (78) Choi, Panis \& Evans II
1999; (79) Moriarty-Schieven et al.\ 1992; (80) Megeath \& Wilson 1997; 
(81) Swift, Welch \& Di Francesco 2005; (82) Leurini et al.\ 2007; 
(83) Choi \& Tang 2006; (84) Beltr\'an et al.\ 2008; (85) Sandell et al.\ 2005;
(86) Beuther et al.\ 2002; (87) Sridharan et al.\ 2002; (88) Dobashi \& Uehara 2001} 
}  
   \label{regions}
\end{table*}

\begin{figure}
\resizebox{\hsize}{!}{\includegraphics{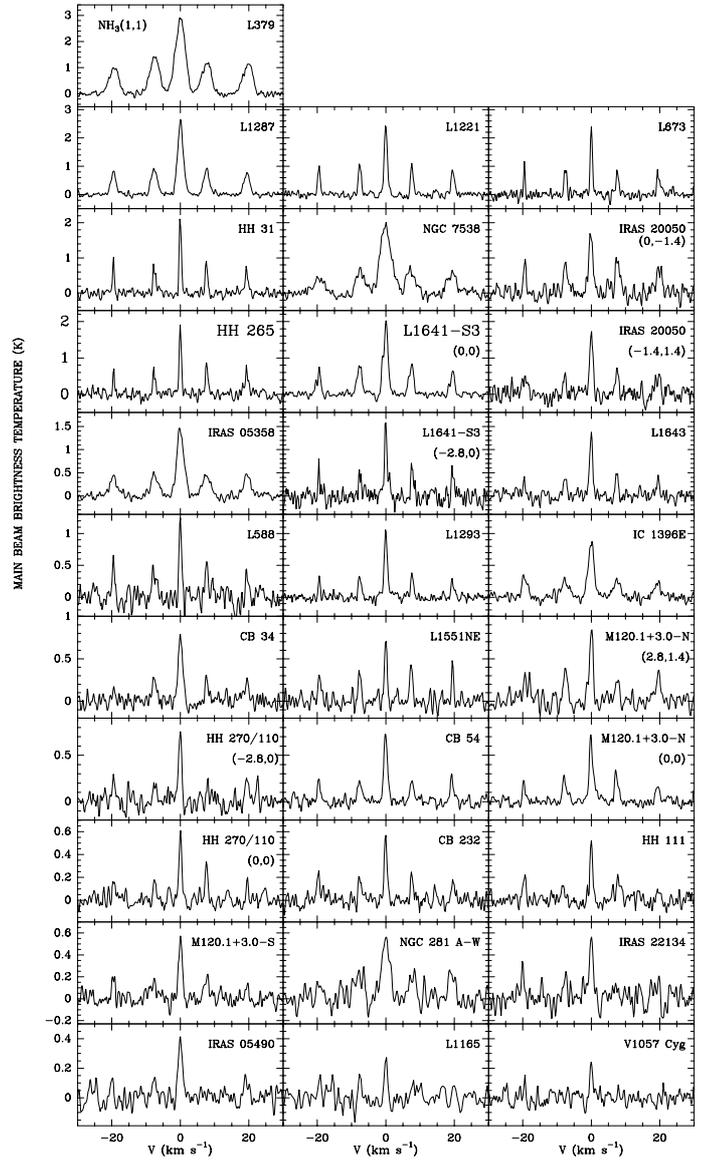}}
   \caption[ ]{Spectra of the $(J,K)=(1,1)$ inversion transition of the \nh\ 
    molecule toward the positions given in Table~\ref{line} for the detected
    sources. The vertical axis is the main beam brightness
    temperature and the horizontal axis is the velocity relative to the center
of the main line 
    (as given in Table~\ref{line}). For M120.1+3.0-N, L1641-S3, HH270/110 and
IRAS 20050, spectra at two different positions are
    shown. The spectrum of V1057 Cyg corresponds to the average of positions 
    with detected emission in a five-point grid.}
  \label{esp11}
\end{figure}

\begin{figure}
   \resizebox{\hsize}{!}{\includegraphics{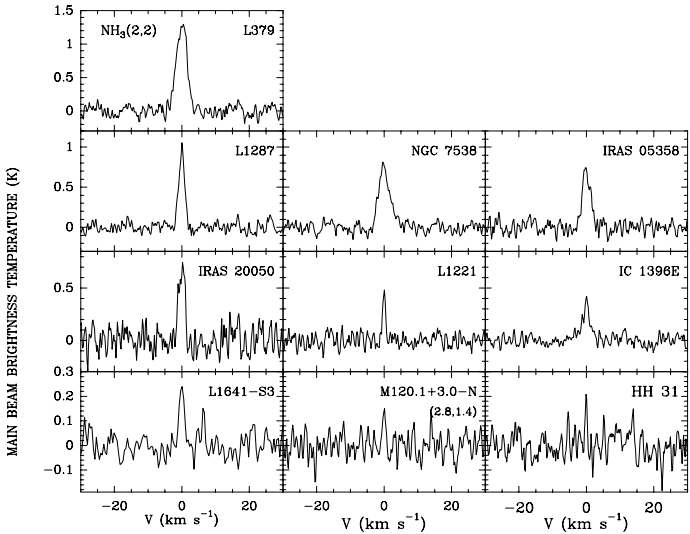}}
   \caption{Same as Fig.~\ref{esp11}, for the $(J,K)$=(2,2)
    inversion transition toward the positions given in Table~\ref{dosdos}}
     \label{esp22}
\end{figure}

We searched for ammonia emission in the 40 regions listed in
Table~\ref{regions}. In all cases, we first made measurements on a five-point
grid centered on the position given in Table~\ref{regions}, with a full beam
separation between points. The \nh(1,1) line was detected in 27 sources.  The
\nh(2,2) line was observed in 15 sources and was detected in 10 of them. The
observed spectra of the \nh(1,1) and \nh(2,2) lines at the position of the
emission peak are shown in Figs.~\ref{esp11} and \ref{esp22}.

In Tables~\ref{line} and \ref{dosdos} we give the \nh(1,1) and \nh(2,2)  line
parameters obtained from a multicomponent fit to the magnetic hyperfine
structure at the position of the emission peak. The intrinsic line widths
obtained range from $\sim0.3$ \kms\ to $\sim 3.6$ \kms\ for the \nh(1,1) line
and from $\sim 0.5$ \kms\ to $\sim 3.8$ for the \nh(2,2) line. The values of
optical depth obtained are in the range 0.1-5 for the \nh(1,1) and 0.1-0.5 for the \nh(2,2) line. 
We also used a single Gaussian fit to the main line to obtain the main beam brightness temperature at 
the position of the emission peak. The values obtained for the main beam brightness temperature for the 
detected sources range from $\sim0.3$ K to $\sim3$ K for the \nh(1,1) line and 
from $\sim0.2$ K to $\sim1$ K for the \nh(2,2) line.

\begin{figure}
\begin{center}
 \resizebox{7cm}{!}{\includegraphics[15,50][328,440]{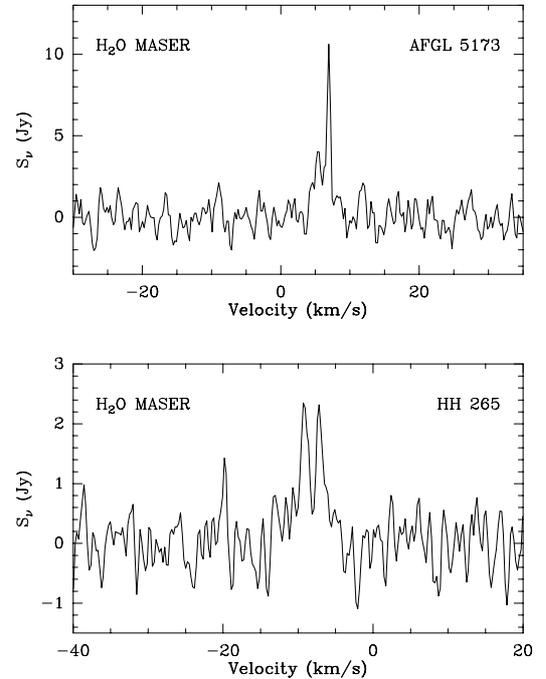}}
  \caption{Spectra of the \hho\ masers detected in 1996 May 16 in the 
   regions HH 265 (bottom) and AFGL 5173 (top) toward the positions given in 
   Table~\ref{maser}}
   \label{maser4}
\end{center}
\end{figure}

Additionally, we searched for the $6_{16}$-$5_{23}$ \hho\ maser line (at the
frequency of 22.235080 GHz) toward the reference position of the six sources
listed in Table~\ref{maser}. The \hho\ observations were carried out in 1996 May 16 and 17 with 
the same spectrometer and bandwidth used for the \nh\ observations. We reached a typical (1$\sigma$)
sensitivity of 1 Jy per spectral channel.  Of the six sources
observed in \hho, we only detected significant ($>3 \sigma$) \hho\
emission in two of them, HH 265 and AFGL 5173. The spectra of these \hho\ masers are shown
in Fig.~\ref{maser4}. In Table~\ref{maser} we give the maser line parameters obtained from a Gaussian fit.

\begin{table*}
\caption[ ]{\nh (1,1) line parameters}
\centering
\begin{tabular}{l c c c c c c c} 
\hline\noalign{\smallskip}
Region &
Position\tablefootmark{a} &
\multicolumn{1}{c}{${V_{\rm LSR}}$\tablefootmark{b}} &
${T_{\rm MB}({\rm m})}$\tablefootmark{c} &
${\Delta V}$\tablefootmark{d} &
${\tau_{\rm m}}$\tablefootmark{e} &
${A\tau_{\rm m}}$\tablefootmark{f} &
${N(1,1)}$\tablefootmark{g} \\
& (arcmin) & \multicolumn{1}{c}{($\rm km\,s^{-1})$} &(K) &
$(\rm km\,s^{-1})$ & & (K) &$(10^{13}\,{\rm cm^{-2}})$ \\
\noalign{\smallskip}
\hline
\noalign{\smallskip}
M120.1+3.0-N
& $(2.8,1.4)$ & $-18.78\pm0.02$ & $0.89\pm0.07$ & $0.90\pm0.05$
& $1.7\pm0.3$ &$1.9\pm0.1$ & 4.7 -- 16.4 \\
& $(0,0)$ & $-20.23\pm0.01$ & $0.65\pm0.04$ & $1.18\pm0.04$ &
$1.0\pm0.1$ & $1.06\pm0.05$ &3.5 -- 12.2 \\

M120.1+3.0-S
& $(0,0)$ & $-17.41\pm0.02$ & $0.57\pm0.05$ &$1.10\pm0.05$ &
$0.8\pm0.2$ & $0.83\pm0.07$ & 2.5 -- 9.0 \\

L1287
& $(0,0)$ & $-17.63\pm0.05$ & $2.51\pm0.05$ &$1.77\pm0.01$ &
$0.90\pm0.03$ &$3.80\pm0.04$ & 18.7 -- 30.9\\

L1293
& $(0,0)$  &$-17.67\pm0.01$ & $1.06\pm0.05$ & $0.79\pm0.03$ &
$0.7\pm0.2$ & $1.7\pm0.1$ & 3.6 -- 7.7\\

NGC 281 A-W
& $(0,0)$ & $-30.32\pm0.04$ & $0.54\pm0.07$ & $2.2\pm0.1$ &
$1.3\pm0.3$ & $0.93\pm0.07$ & 5.6 -- 27.0 \\

HH 156
& (0,0) & \multicolumn{1}{c}{-} &$\leq0.2$ & - & - & - & -\\

HH 159
& (0,0) & \multicolumn{1}{c}{-} &$\leq0.4$ & - & - & - & -\\

HH 158
& (0,0) & \multicolumn{1}{c}{-} &$\leq 0.3$ &- & - & - & -\\

HH 31
& $(-7,0)$ & $+6.86\pm0.01$ & $2.2\pm0.1$ & $0.42\pm0.01$ &
$2.6\pm0.2$ &$6.7\pm0.3$ & 7.7 -- 16.1 \\

HH265 
& $(-1.4,1.4)$ & $+6.68\pm0.01$ & $1.9\pm0.1$ & $0.40\pm0.02$ &
$2.5\pm0.3$ & $5.7\pm0.4$ & 6.3 -- 14.0 \\

L1551 NE
& $(0,0)$ &$+6.64\pm0.02$ &$0.79\pm0.09$ &$0.47\pm0.03$  &$5.0\pm0.8$
&$3.5\pm0.4$ & 4.6 -- 22.6 \\

L1642
& $(0,0)$ & \multicolumn{1}{c}{-} & $\leq0.1$ & - & - & - & -\\

L1634
& $(1.4,0)$ &$+8.00\pm0.01$ & $1.4\pm0.1$ & $0.81\pm0.04$ &
$0.6\pm0.2$ &$2.0\pm0.2$ &4.6 -- 8.5 \\

HH 59
& $(0,0)$ & \multicolumn{1}{c}{-} & $\leq0.3$ &- &- &- & -\\

IRAS 05358 
& $(0,0)$ & $-16.89\pm0.01$ &$1.44\pm0.05$ & $2.32\pm0.03$ &
$0.67\pm0.07$ & $1.95\pm0.05$ &12.6 -- 24.6\\

L1641-S3
& $(0,0)$ & $+4.96\pm0.01$ & $2.0\pm0.1$ & $0.68\pm0.04$ &
$1.3\pm0.3$ & $3.9\pm0.3$ &7.4 -- 14.4\\
& $(-2.8,0)$ & $+3.76\pm0.01$ &$1.8\pm0.2$ &$0.30\pm0.02$ & $3.3\pm0.5$ &
$7.0\pm0.3$ &5.8 -- 13.2\\ 

HH 68
& $(0,0)$ & \multicolumn{1}{c}{-} & $\leq0.3$ &- &- &- & -\\
 
CB 34
& $(0,0)$ & $+0.72\pm0.02$ & $0.77\pm0.06$ & $1.37\pm0.06$ &
$0.4\pm0.2$ & $0.99\pm0.07$ & 3.8 -- 8.3 \\

HH 270/110
& $(0,0)$ & $+8.86\pm0.02$ & $0.67\pm0.06$ &$0.65\pm0.04$ &$1.3\pm0.3$ &
$1.3\pm0.1$ & 2.4 -- 8.9 \\
&$(-2.8,0)$ & $+8.70\pm0.03$ &$0.8\pm0.1$ &$0.8\pm0.1$ &
$0.4\pm0.5$ &$1.1\pm0.2$ & 2.6 -- 4.8 \\

IRAS 05490
& $(0,0)$ & $+0.78\pm0.03$ & $0.41\pm0.05$ & $1.5\pm0.1$
&$0.1\pm0.4$ & $0.45\pm0.02$ &1.9 -- 3.0 \\

HH 111
&$(0,0)$ & $+8.72\pm0.02$ & $0.56\pm0.06$ & $0.78\pm0.08$
&$0.4\pm0.3$ & $0.7\pm0.1$ & 1.6 -- 3.9 \\

HH 113
&$(0,0)$ &\multicolumn{1}{c}{-} &$\leq0.4$ &- &- &- & - \\

AFGL 5173
& $(0,0)$  &\multicolumn{1}{c}{-} &$\leq0.2$ &- &- &- & -\\

CB 54 
& $(0,0)$ & $+19.55\pm0.01$ & $0.73\pm0.04$ & $1.14\pm0.04$ &
$0.8\pm0.2$ & $1.11\pm0.06$ & 3.5 -- 10.8 \\

L1709 
& $(0,0)$  &\multicolumn{1}{c}{-} & $\leq0.4$ &- &- &- & -\\

L379
& $(0,0)$ & $+18.89\pm0.01$ & $2.87\pm0.05$ & $2.84\pm0.02$ &
$1.86\pm0.03$ & $5.98\pm0.05$ & 47.2 -- 87.9 \\

L588
& $(0,0)$ &$+10.86\pm0.02$ & $1.3\pm0.1$ &$0.58\pm0.04$ &$2.4\pm0.4$
&$3.5\pm0.3$ & 5.6 -- 16.4 \\

CB 188
& $(0,0)$  &\multicolumn{1}{c}{-} &$\leq0.2$ &- & - & - & -\\

L673
& $(0,-1.4)$ & $+7.11\pm0.01$ &$2.5\pm0.1$ &$0.41\pm0.1$ &$2.5\pm0.3$
&$7.5\pm0.4$ & 8.4 -- 16.3 \\

HH221
& $(0,0)$  &\multicolumn{1}{c}{-} & $\leq0.2$ &- &- &- & -\\

L797
& $(0,0)$  &\multicolumn{1}{c}{-} &$\leq0.2$ & - & - &- & -\\

IRAS 20050
& $(0,-1.4)$ &$+6.86\pm0.02$ &$1.7\pm0.2$ & $0.93\pm0.04$ &$3.4\pm0.3$
&$5.6\pm0.4$ & 14.4 -- 38.7 \\
& $(-1.4,1.4)$ &$+5.06\pm0.02$ &$1.8\pm0.2$ & $0.96\pm0.04$ &$0.8\pm0.2$ 
& $2.6\pm0.2$ &7.1 -- 13.1 \\

V1057 Cyg\tablefootmark{h}
& $(0,0)$ &$+4.30\pm0.04$ & $0.3\pm0.1$ &$0.58\pm0.09$ 
&$\leq3$\tablefootmark{i}  &$0.3-0.7$\tablefootmark{j} &0.5 -- 14.5\\

CB 232
& $(0,0)$ & $+12.32\pm0.02$ & $0.58\pm0.05$ & $0.68\pm0.04$ &
$1.8\pm0.3$ & $1.3\pm0.1$ & 2.5 -- 11.8 \\

IC 1396E
& $(0,0)$ & $+0.53\pm0.02$ & $0.86\pm0.05$ & $1.89\pm0.04$ &
$0.8\pm0.1$ & $1.22\pm0.05$ & 6.1 -- 16.6\\

L1165
& $(0,0)$ &$-1.64\pm0.04$ & $0.35\pm0.08$ & $0.6\pm0.1$ &$3\pm1$ &
$0.9\pm0.2$ & 1.6 -- 13.4 \\

IRAS 22134
&$(0,0)$ &$-18.62\pm0.04$ &$0.55\pm0.09$ & $1.2\pm0.1$ &$0.4\pm0.4$ &
$0.7\pm0.1$ & 2.3 -- 6.0\\

L1221
& $(0,0)$ & $-4.36\pm0.01$ & $2.5\pm0.1$ & $0.71\pm0.01$ & $2.1\pm0.1$
&$6.1\pm0.2$ & 12.1 -- 23.4\\

L1251\tablefootmark{k}
& $(0,0)$ & - &$\leq0.2$ & - & - & - & - \\
 
NGC 7538
& $(0,-1.4)$ & $-56.22\pm0.02$ & $1.9\pm0.1$ & $3.57\pm0.05$ &
$0.35\pm0.06$ & $2.32\pm0.06$ & 23.1 -- 32.8 \\

\noalign{\smallskip}
\hline
\end{tabular}

\tablefoot{
\tablefoottext{a}{ Position of the emission peak, where line parameters
were obtained (in offsets from the position given
in Table~1).}
\tablefoottext{b}{Velocity of the line peak with respect to the local standard of 
rest.}
\tablefoottext{c}{Main beam brightness temperature of the main line of the transition,
obtained from a single Gaussian fit. For undetected sources a $3\sigma$ 
upper limit is given.}
\tablefoottext{d}{Intrinsic line width, obtained taking into account optical depth and
hyperfine broadening, but not the spectral resolution of the 
spectrometer.}
\tablefoottext{e}{Optical depth of the main line derived from the relative intensities
of the magnetic hyperfine components.}  
\tablefoottext{f}{Derived from the transfer equation, where
$A=f[J(T_{\rm ex})-J(T_{\rm bg})]$ is the ``amplitude'' (Pauls et al.\
1983), $f$ is the filling factor, $T_{\rm ex}$ is the excitation
temperature of the transition, $T_{\rm bg}$ is the background 
radiation temperature
and $J(T)$ is the intensity in units of temperature. Note that $A\simeq
fT_{\rm ex}$, for $T_{\rm ex}\gg T_{\rm bg}$}
\tablefoottext{g}{Beam-averaged column density for the rotational level $(1,1)$. Upper
limit is obtained from 
$$\left[\frac{N(1,1)}{\rm cm^{-2}}\right]=1.582~10^{13}
\frac{e^{(1.14/T_{\rm ex})}+1}{e^{(1.14/T_{\rm ex})}-1} \tau_{\rm m}
\left[\frac{\Delta V}{\rm km~s^{-1}}\right],$$ 
where $T_{\rm ex}$ is derived from the transfer equation assuming a 
filling factor $f=1$. If $T_{\rm
ex}\gg T_{\rm bg}$ the beam averaged column density is proportional to   
the ``amplitude" A, and the explicit dependence on $T_{\rm ex}$
disappears,  reducing to
$$\left[\frac{N(1,1)}{\rm cm^{-2}}\right]=
2.782~10^{13}\left[\frac{A\tau_{m}}{\rm K}\right] \left[\frac{\Delta
V}{\rm km~s^{-1}}\right],$$
providing  the lower limit for the beam-averaged column density (e.g., Ungerechts et al.\ 1986)}.
\tablefoottext{h}{Line parameters were obtained by averaging several positions of
a five-point map.}
\tablefoottext{i}{Obtained by adopting a $3\sigma$ upper limit for the intensity of the
satellite lines.}
\tablefoottext{j}{The highest value is obtained from the upper limit of $\tau_{\rm m}$
and the lowest value is obtained assuming optically thin emission.}
\tablefoottext{k}{This region was observed and mapped in Paper I. The undetection 
refers to the new observed positions.} }

\label{line}
\end{table*}

\begin{table*}
\caption[ ]{\nh (2,2) line parameters}
\begin{flushleft}  
\centering
\begin{tabular}{l c c c c c c c}
\hline\noalign{\smallskip}
Region &
Position\tablefootmark{a}&
${V_{\rm LSR}}$\tablefootmark{b} &   
${T_{\rm MB}({\rm m})}$\tablefootmark{c} &
${\Delta V}$\tablefootmark{d} &
${\tau_{\rm m}}$\tablefootmark{e} &
${A\tau_{\rm m}}$\tablefootmark{f} &
${N(2,2)}$\tablefootmark{g} \\
& (arcmin) & ($\rm km\,s^{-1})$ &(K) & $(\rm km\,s^{-1})$
& & (K) &$(10^{13}\, {\rm cm^{-2}})$ \\
\noalign{\smallskip}
\hline   
\noalign{\smallskip}

M120.1+3.0-N
& $(2.8,1.4)$ &$-18.4\pm0.1$ &$0.15\pm0.06$ &$0.9\pm0.2$&
$0.2\pm0.1$ &$0.17\pm0.03$ &0.2 -- 0.7\\
& $(0,0)$ &- &$\leq0.2$ &- &- &- &-\\

L1287
& $(0,0)$ & $-17.57\pm0.02$ & $0.98\pm0.05$ &$1.93\pm0.04$ &
$0.27\pm0.04$ & $1.13\pm0.02$ &2.9 -- 4.8\\

L1293
& $(0,0)$ & - &$\leq0.2$ &- & - & - & -\\

HH 31
& $(-7,0)$ & $6.94\pm0.04$ & $0.24\pm0.06$ & $0.51\pm0.08$ & 0.1
&$0.27\pm0.04$ &0.2 -- 0.4\\

HH 265
& $(-1.4,1.4)$ & - &$\leq0.2$ & - & - & - & -\\

IRAS 05358
& $(0,0)$ & $-16.79\pm0.04$ & $0.7\pm0.1$ & $2.6\pm0.1$ &
$0.28\pm0.05$ & $0.83\pm0.02$ &2.8 -- 5.5\\

L1641-S3
& $(0,0)$ & $+5.1\pm0.1$ & $0.3\pm0.1$ & $1.2\pm0.1$
&$0.1\pm0.1$ & $0.30\pm0.03$ &0.5 -- 1.0\\

CB 34
& $(0,0)$ &- &$\leq0.1$ &- &- &- &- \\

CB 54
& $(0,0)$ &- &$\leq0.2$ &- &- &- & -\\

L379
& $(0,0)$ & $+18.80\pm0.02$ & $1.3\pm0.1$ & $3.45\pm0.05$ &
$0.53\pm0.02$ & $1.7\pm0.02$ &7.7 -- 14.3\\ 

IRAS 20050
& $(0,0)$ & $+6.46\pm0.05$ & $0.7\pm0.1$ & $1.9\pm0.1$ & $0.5\pm0.1$ &
$0.93\pm0.04$ & 2.3 -- 5.5\\

CB 232
& $(0,0)$ &- &$\leq0.1$ &- &- &- & -\\

IC 1396E
& $(0,0)$ & $+0.42\pm0.06$ & $0.33\pm0.06$ & $2.4\pm0.2$ &
$0.23\pm0.06$ & $0.37\pm0.02$ &1.2 -- 3.2 \\

L1221
& $(0,0)$ & $-4.49\pm0.03$ & $0.5\pm0.1$ &$1.0\pm0.1$ &$0.19\pm0.03$
&$0.55\pm0.03$ & 0.7 -- 1.4\\

NGC 7538
& $(0,0)$ & $-56.91\pm0.03$ & $0.76\pm0.05$ & $3.8\pm0.1$ &
$0.13\pm0.07$ & $0.81\pm0.01$ &4.0 -- 5.9 \\

\noalign{\smallskip}
\hline
\end{tabular}
\end{flushleft}
\tablefoot{
\tablefoottext{a-d,f}{ See footnotes of Table~2.}
\tablefoottext{e}{ Optical depth of the (2,2) main line derived from the ratio of the 
(1,1) to (2,2) antenna temperatures and the optical depth of the (1,1) 
line, assuming the same excitation temperature for both transitions.}
\tablefoottext{g}{Beam-averaged column density for the rotational level (2,2). Upper
limit is derived from
$$\left[\frac{N(2,2)}{\rm cm^{-2}}\right]=7.469~10^{12}
\frac{e^{(1.14/T_{\rm ex})}+1}{e^{(1.14/T_{\rm ex})}-1} \tau_{\rm m}
\left[\frac{\Delta V}{\rm km~s^{-1}}\right],$$
assuming that both filling factor and excitation temperature
are the same for the (1,1) and (2,2) transitions. If $T_{\rm
ex}\gg T_{\rm bg}$, the beam-averaged column density is proportional to
the ``amplitude" A, and the explicit dependence on $T_{\rm ex}$
disappears, reducing to
$$\left[\frac{N(2,2)}{\rm cm^{-2}}\right]=
1.312~10^{13}\left[\frac{A\tau_{m}}{\rm K}\right] \left[\frac{\Delta
V}{\rm km~s^{-1}}\right],$$
providing the lower limit for the beam-averaged column density (e.g., Ungerechts et al.\ 1986)}.
}
\label{dosdos}
\end{table*}

\begin{table}
\caption{\hho\ maser line parameters}
\begin{flushleft}
\centering
\begin{tabular}{l c c c c}
\hline\noalign{\smallskip}
Region &
Position\tablefootmark{a} &  
\multicolumn{1}{c}{${V_{\rm LSR}}$\tablefootmark{b}} &
${S_{\nu}}$\tablefootmark{c} &
${\Delta V}$\tablefootmark{d}  \\  
& (arcmin) & \multicolumn{1}{c}{($\rm km\,s^{-1})$} &(Jy) &
$(\rm km\,s^{-1})$ \\
\noalign{\smallskip}
\hline
\noalign{\smallskip}
M120.1+3.0-N & (2.8,1.4) & -              & $\leq 1$    & - \\
HH 265       & (0,0)     & $-9.13\pm0.07$ & $2.2\pm0.6$ & $1.1\pm0.2$ \\
             & (0,0)     & $-7.14\pm0.07$ & $2.1\pm0.6$ & $0.9\pm0.2$ \\
L1634        & (1.4,0)   &  -             & $\leq1.5$   &    - \\
L1641-S3     &  (0,0)    &  -             & $\leq 1 $   &   - \\
AFGL 5137 & (0,0)     & $+6.92\pm0.02$ & $11\pm1$    & $0.61\pm0.06$ \\
L1221        & (0,0)     &    -           & $\leq1.8$   &  -   \\
\noalign{\smallskip}
\hline
\end{tabular}
\end{flushleft}
\tablefoot{Obtained from a Gaussian fit to the line profiles observed on May 1996. \\
  \tablefoottext{a}{Position observed where line parameters
have been obtained (in offsets from the position given
in Table~1).}
  \tablefoottext{b}{Velocity of the line peak with respect to the local standard of
rest.}
  \tablefoottext{c}{Flux density of the line peak. For undetected sources a $3\sigma$ upper
limit is given.}
  \tablefoottext{d}{Full width at half maximum.} 
}
  \label{maser}
\end{table}

\section{Results}\label{sources4}

In Table~\ref{parameters} we list the physical parameters of the molecular
condensations, derived from the \nh\ data given in Tables~\ref{line} and
\ref{dosdos}, following the procedures explained in the footnotes of
Table~\ref{parameters}. We mapped the \nh(1,1) emission in all the
detected regions, except in V1057Cyg and L1551NE.  Maps are shown in
Figs.~\ref{m120n}, \ref{m120s} to \ref{05358}, \ref{velog} to \ref{i20050} and
\ref{cb232} to \ref{ngc7538}.  In V1057 Cyg the emission is very weak in all
positions. 
The spectrum of this region shown in
Fig.~\ref{esp11} and the physical parameters listed in Table~\ref{parameters}
were obtained by averaging several points of a five-point map. In L1551NE, we
detected strong \nh\ emission in four positions, but we were not able
to map the region (see \S~\ref{l1551}). A summary of the relevant information,
taken from the literature, about the sources associated with these regions is
given in Table~\ref{information}.

We detected maser emission in the regions HH265 and AFGL 5173 (see Table~\ref{maser}). 
The position of the maser in AFGL 5173 coincides with that of IRAS 05553+1631, so that  
the maser could be excited by the IRAS source. We detected significant maser emission in 
the velocity range from 6.5 to 7.2 \kms. Brand et al.\ (1994) detected highly
variable \hho\ emission toward this source between 1989 March and 1991 January 
in the velocity  range from $-7.6$ to 13.1 \kms. However no \nh\ emission was detected toward the
source AFGL 5173 (see Table~\ref{line}). The results for individual sources are presented in Appendix~A.

\begin{table*}
 \caption[ ]{Physical parameters of the \nh\ condensations}
\begin{flushleft}
\centering
\begin{tabular}{l r r r c r@{\,$-$\,}l r r@{.}l}
\noalign{\smallskip}
\hline\noalign{\smallskip}
Region&
\multicolumn {2}{c}{Size\tablefootmark{a}} &
\multicolumn{1}{c}{${T_\mathrm{rot}}$\tablefootmark{b}} &
\multicolumn {1}{c}{$N({\rm H}_2)$\tablefootmark{c}} &
\multicolumn {2}{c}{$M$\tablefootmark{d}} &
\multicolumn {1}{c}{${M_\mathrm{vir}}$\tablefootmark{e}} &
\multicolumn {2}{c}{$n(\rm {H}_2)$\tablefootmark{f}} \\  \cline{2-3}
& \multicolumn{1}{c}{(arcmin)} &
\multicolumn{1}{c}{(pc)} &  
\multicolumn{1}{c}{(K)} & 
\multicolumn{1}{c}{($10^{22}\, $cm$^{-2}$)} & 
\multicolumn{2}{c}{(\mo)} & 
\multicolumn{1}{c}{(\mo)} &
\multicolumn{2}{c}{($10^{3}\, $cm$^{-3}$)} \\
\noalign{\smallskip}
\hline
\noalign{\smallskip}

M120.1+3.0-N$\;(2\farcm8,1\farcm4)$ & $3.9\times2.7$ &
$0.96\times0.67$ & 11.2 &$1.7-6.1$ &142&500 &68 &3&4 \\

M120.1+3.0-N$\;(0,0)$ & $3.0\times2.0$ &
$0.75\times0.50$ &$\leq15$ &$\geq0.9$ &\multicolumn{2}{c}{$\geq45$} &89
&$\sim2$&6 \\

M120.1+3.0-S
& $3.7\times2.4$ & $0.91\times0.59$ & $\sim13.5$ & $0.7-2.7$ & 51&181 
& $92$ & $\sim2$&8 \\

L1287
& $3.8\times1.9$ & $0.93\times0.48$ &17.4 &$4.6-7.6$ &259&429 &220
&11&3\\

L1293
& $2.7\times2.0$ & $1.26\times0.94$ &$\leq13$ &$\geq1.1$
&\multicolumn{2}{c}{$\geq168$} &71 &$\sim7$&7 \\

NGC 281 A-W
& $1.9\times1.6$ & $1.88\times1.62$ &$\sim20$
&$1.3-6.2$ &505&2\,413 &871&$\sim 1$&3 \\

HH31
& $2.8\times2.0$ & $0.11\times0.08$ &9.7 &$3.7-7.6$ &4&9 &2 &12&4\\

HH265
& $4.2\times2.3$ & $0.20\times0.11$ &$\leq9$
&$\geq3.3$ &\multicolumn{2}{c}{$\geq9$} &2 &$\sim 11$&3 \\

L1551 NE 
&$1.4\times1.4$ &$0.07\times0.07$ &$\sim 25$ &$1.0-5.1$ &0.6&3 &1.6
&$\sim1$&1\\

L1634  
&$2.6\times2.1$ &$0.34\times0.28$ &$\sim12$
&$1.6-2.9$ &19&35 &21 &$\sim13$&1\\

IRAS 05358
&$3.1\times1.9$ & $1.60\times0.98$ &20.6 &$2.9-5.6$ &575&1\,121 &707
&5&9 \\

L1641-S3\,($V\sim4.9$ km/s) &$3.8\times2.2$ &$0.53\times0.31$ &$12.6$
&$2.4-4.6$ & 49&96 &20 &10&1 \\
L1641-S3\,($V\sim3.8$ km/s) &$5.8\times2.8$ &$0.81\times0.39$ &13.5 
&$1.7-3.9$ &68&157 &5 &6&2 \\

CB 34
&$2.0\times1.7$ & $0.89\times0.76$ &$\leq12$ &$\geq 1.3$
&\multicolumn{2}{c}{$\geq111$} &162 &$\sim7$&8 \\

HH 270/110$\;(0,0)$ &$1.4\times1.6$ & $0.19\times0.21$ &$\sim 13.6$ 
&$0.7-2.6$ & 3.5&13.3 &9 &$\sim2$&5 \\

HH270/110$\;(-2\farcm8,0)$ &$2.8\times1.7$ & $0.37\times0.23$ &$\sim13.6$
  &$0.8-1.4$ & 8&15 &22 &$\sim10$&0 \\

IRAS 05490
&$2.6\times1.7$ &$1.59\times1.05$ &$\sim15.5$ &$0.5-0.8$ &104&168 &304 
&$\sim14$&3 \\
  
HH 111
&$1.6\times1.8$ &$0.22\times0.23$ &$\sim13$ &$0.5-1.2$ &3&8 &14 
&$\sim5$&6 \\

CB 54
&$1.7\times1.6$ & $0.74\times0.68$ &$\leq15$ &$\geq 1.0$
&\multicolumn{2}{c}{$\geq62$} &96 &$\sim3$&3 \\

L379
&$2.0\times2.0$ & $1.16\times1.16$ &17.8 &$11.4-21.2$ &1\,948&3\,628
&982 &7&6 \\

L588
&$4.1\times1.9$ & $0.37\times0.17$ &$\sim8$ &$4.1-11.8$ &32&94 &9 
&$\sim7$&5 \\

L673(SE)\tablefootmark{g}
&$2.2\times2.0$ & $0.19\times0.17$ &$\leq12$ &$\geq2.8$
&\multicolumn{2}{c}{$\geq12$} &3 &$\sim10$&9 \\

IRAS 20050$\;(0,-1\farcm4)$ &$3.1\times2.0$ & $0.63\times0.41$ &
$16.8$ & $3.6-9.7$ & 117&315 & 46 &3&6 \\

IRAS 20050$\;(-1\farcm4,1\farcm4)$ &$3.0\times2.0$
&$0.61\times0.41$ &$16.8$ & $1.8-3.3$ &56&103 &$49$ &8&2 \\

V1057 Cyg 
&$1.4\times1.4$ &$0.29\times0.29$ &$\sim10$ &$0.2-6.5$ &2&67 &10
&$0.3-0$&7 \\

CB 232
&$2.6\times2.5$ &$0.45\times0.43$ &$\leq11$ &$\geq 0.9$
&\multicolumn{2}{c}{$\geq23$} &21 &$\sim2$&2 \\

IC 1396E
&$3.2\times1.9$ & $0.70\times0.42$ &19.1 &$1.4-3.9$ &54&145 &191 &3&2 \\ 

L1165
&$3.5\times2.3$ & $0.76\times0.49$ &$\sim9$ &$0.9-7.3$ & 41&347 & 23 
&$\sim1$&3 \\

IRAS 22134
&$3.1\times2.3$ &$2.34\times1.74$ &$\sim15.8$ &$0.6-1.5$ &310&787 &283 
&$\sim4$&3 \\

L1221
&$2.1\times2.0$ & $0.12\times0.11$ &12.5 &$3.9-7.6$ &7&13 &6 &10&5\\

NGC 7538
&$2.5\times2.4$ &$1.96\times1.92$ &28.3 &$5.3-7.5$ &2\,514&3\,571
&2\,599 &12&2 \\ 
\noalign{\smallskip}
\hline
\noalign{\smallskip}
\end{tabular}
\end{flushleft}

\tablefoot{
\tablefoottext{a}{Major and minor axes of the half-power contour of the \nh\ emission. 
For sources L1551 NE and V1057 Cyg the size of the beam has been adopted.}
\tablefoottext{b}{Rotational temperature, derived from the ratio of column densities
in the (1,1) and (2,2) levels (given in Tables~1 and 2, respectively), for the sources where the (2,2) line was
detected. For the sources undetected in the (2,2) line, an upper limit was
obtained assuming optically thin emission. For sources not observed in the
(2,2) line, we assumed that $T_\mathrm{ex}(\mathrm{CO})
=T_\mathrm{rot}(22-11)=T_{k}$, where the CO data are from Yang et al.\ 1990 
(M120.1+3.0-S), Henning et al.\ 1994 (NGC 281 A-W), Moriarty-Schieven et
al.\ 1995 (L1551 NE), Reipurth \& Oldberg 1991 (HH 270/110 and HH 111),
Snell et al.\ 1990 (IRAS 05490), Parker et al.\ 1991 (L588 and L1165),
Levreault 1988 (V1057 Cyg) and Dobashi et al.\ 1994 (IRAS 22134). For
L1634, $T_{\rm rot}=12$ K has been adopted.} 
\tablefoottext{c}{Beam-averaged \hh\ column density, obtained from the \nh\ column
density adopting an \nh\ abundance of [\nh/\hh]= $10^{-8}$ (see Anglada et
al.\ 1995 for a discussion on \nh\ abundances). The \nh\ column density is
obtained assuming that only the rotational metastable levels of the \nh\
are significantly populated at their LTE ratios corresponding to
$T_k=T_R(22-11)$.}
\tablefoottext{d}{Mass of the condensation, derived from the beam-averaged \hh\ column   
density and the observed area.}
\tablefoottext{e}{Virial mass obtained from [$M_{\rm vir}$/\mo]=210[$R$/pc][$\Delta
V$/\kms]$^2$, where $R$ is the radius of the clump, taken as half
the geometrical mean of the major and minor axes, and $\Delta V$ is the
intrinsic line width given in Table~2.}
\tablefoottext{f}{Volume density, derived from the two-level model (Ho \& Townes
1983).}
\tablefoottext{g}{Parameters of the southeastern clump. Parameters of the northwestern 
clump are given in Paper I}. 
}
\label{parameters}
\end{table*}

\begin{table*}
\caption[]{Summary of properties of relevant sources in the regions detected in
\nh}
\resizebox{\hsize}{!}{
\centering
\begin{tabular}{l l r c c c c c c c c c}
\hline\noalign{\smallskip} 
Region &
IRAS &
\multicolumn{1}{c}{$L_{bol}$} &
Ref. &
Evolutionary &
Ref. &
Detection at &
Ref. & 
\hho &
Ref. &
Outflow &
Ref. \\
& &\multicolumn{1}{c}{(\lo)} & &status & &other wavelengths & &maser? &
&source? \\
\noalign{\smallskip}
\hline\noalign{\smallskip}

M120.1+3.0-N 
&00213+6530 &12.9 & 1 & - &- &mm,cm  &109 &Yes &2 & Yes &1\\
&00217+6533 &12.0 & 1 & - &- &-  &- &-   &-   & ? &3\\

M120.1+3.0-S
&00259+6510 &9.9  & 1 & - &- & - & - &No &4 &Yes &1\\
&00256+6511 &20.3 & 1 & - &- & - & - &- &- &? &3\\   

L1287
& 00338+6312 & 1\,800 &5 &Class I?\tablefootmark{a} &6 & NIR,MIR,smm,mm,cm &7,95,8,9,10  &Yes & 6 &? & 11\\

L1293
&00379+6248 &$<21$ & 12 &- &- & - & - &Yes &13 & Yes &12\\

NGC 281 A-W
&00494+5617 &8\,790 &14 &-\tablefootmark{a} & - &NIR,FIR,mm &7,15,14 &Yes &16 &Yes &11\\

HH 31
&04248+2612 &0.36 & 17 &Class I &17 &NIR,FIR,smm,mm &17,18,19,20 &No & 13 &Yes &21\\

L1551 NE 
& 04288$-$1802& 3.9 & 22 & Class 0 & 23 & NIR,smm,mm,cm & 24,19,20,25 &- & - &Yes &23\\

L1634
&05173$-$0555 & 17 & 26 &Class 0 &27 &NIR,FIR,smm,mm,cm &28,18,29,24,27  &No &30 &Yes &28\\
&IRS 7\tablefootmark{b} &0.03\tablefootmark{c} &27 &Class I/0 &27 &NIR,smm &28,27 &- &- &Yes &28\\

IRAS 05358
&05358+3543 &6\,300 & 31 &Herbig Ae/Be?\tablefootmark{a} &32 &NIR,MIR,smm,mm &33,32,104,96,31 &Yes &34 &Yes &11\\

L1641-S3
&05375$-$0731 &100 & 35 &Class I &36 &NIR,FIR,smm,mm,cm &36,37,29,35,38, &Yes &39 &Yes &40\\

CB 34
&05440+2059 &130 &41 & Class I &42 &NIR,smm,mm,cm &42,43,41,44 &Yes &58 &Yes &45\\
 
HH 270/110
&05487+0255 &7   &46 &Class I &46 &NIR,FIR,cm &47,46,48  &No & 49 &Yes &46\\
&05489+0256 &5.3 &50 &Class I &50 &NIR,mm,cm     &50,97,48    &- &-   &Yes &51\\

IRAS 05490
&05490+2658 &4\,200 &11 &Class I?\tablefootmark{a} &7 &NIR,FIR,cm &7,15 &No &16 &Yes &11\\

HH 111
&05491+0247 &25 &46 &Class 0 &52 &NIR,smm,mm,cm &53,26,54,55 &No &30 &Yes &46\\

CB 54
&07020$-$1618 &400 &43  &Class I & 41 &NIR,MIR,smm,mm,cm &56,105,42,43,57 &Yes & 58 &Yes & 45\\

L379 
&18265$-$1517 &$16\,000$ &59 &-\tablefootmark{a} &- &smm,mm,cm &59,8 &Yes &60 &Yes & 40 \\

L588
&18331$-$0035 &3.7 &26 &Class I &61 &NIR,mm &103,61 & - & - &? &61\\

IRAS 20050
&20050+2720 &206 &62 &Class 0 &63 &NIR,FIR,mm,cm &64,65,66,67 &Yes & 68 &Yes &63\\
&20049+2721 &236\tablefootmark{d} &- &- &- &NIR,cm &64,67 &- &- &? & - \\

V1057 Cyg 
&20571+4403 & 200 &69 &FU Or &70 &NIR,IR,FIR,cm,mm,smm &71,72,73,74 &Yes &75 &Yes & 76\\

CB 232
&21352+4307 &14 &41 &Class I &42 &NIR,mm,smm &42,41,77 &Yes &58 &Yes &45 \\

IC 1396E
&21391+5802 &440 &78 &Class 0 &78 &NIR,FIR,mm,smm,cm,X-ray  &79,80,78,98,106 &Yes &34 &Yes &40\\

L1165
&22051+5848 &120 &81 &Class I/FU Or &81 &NIR &81,82 &No &83 &Yes & 84,85\\

IRAS 22134
&22134+5834 &7\,943 &111 &-\tablefootmark{a} &- &NIR,FIR,mm  &107,86,108 &No &110 &Yes &86\\

L1221   
&22266+6845 &2.7 &87 &Class I &102 &NIR,mm &103,99 &No &88 &Yes &87\\

NGC 7538 
&IRS 1-3\tablefootmark{b} &$\sim 250\,000$ &89 &-$^a$ &-  &NIR,FIR,mm   &89,90 &Yes &91,92 &Yes &93\\
&IRS 9\tablefootmark{b}  & $\sim 60\,000$  &89 &-$^a$ &-  &NIR,FIR,mm,cm,smm   &89,100,101    &Yes &92    &Yes &93\\
&IRS 11\tablefootmark{b}        &10\,000   &67 &-$^a$ &-  &FIR,mm,smm   &89,90 &Yes &92    &Yes &93\\

\noalign{\smallskip}
\hline
\noalign{\smallskip}
\end{tabular}
}
\tablefoot{
\tablefoottext{a}{Probably young massive star/stars.}
\tablefoottext{b}{NIR source. No IRAS source at this position.}
\tablefoottext{c}{Submillimeter luminosity.} 
\tablefoottext{d}{IRAS luminosity.}
\tablebib{(1) Yang et al.\ 1990; (2) Han et al.\ 1998; (3) this work; (4) Anglada, Sep\'ulveda \& G\'omez 1997; (5) Moorkeja et al.\ 
1999; (6) Fiebig 1997; (7) Carpenter et al.\ 1993; (8) McCutcheon et al.\ 1995; (9) McMuldroch et al.\ 1995; (10) Anglada et al.\ 1994; (11) Snell et 
al.\ 1990; (12) Yang 1990; (13) Wouterloot et al.\ 1993; (14) Henning et al.\ 1994; (15) Carpenter, Snell \& Schloerb 1990; (16) Henning et al.\ 1992;  
(17) G\'omez et al.\ 1997; (18) Cohen et al.\ 1985; (19) Padgett et al.\ 1999; (20) Moriarty-Schieven et al.\ 1994; (21) Strom et al.\ 1986; 
(22) Chen et al.\ 1995; (23) Devine, Reipurth \& Bally 1999; (24) Hoddap 
\& Ladd 1995; (25) Rodr\'{\i}guez, Anglada \& Raga 1995; (26) Reipurth 
et al.\ 1993; (27) Beltr\'an et al.\ 2002a; (28) Davis et al.\ 1997; (29) Dent et al.\ 1998; (30) Felli, Palagi \& Tofani 1992; (31) Beuther et al.\ 2002; 
(32) Porras et al.\ 2000; (33) Yao et al.\ 2000; (34) Tofani et al.\ 1995; (35) Zavagno et al.\ 1997; (36) Chen \& Tokunaga 1994; 
(37) Price, Murdock \& Shivanandan 1983; (38) Morgan et al.\ 1990; (39) Wouterloot \& Walmsley 1986; (40) Wilking et al.\ 1990; (41) Launhardt \& 
Henning 1997; (42) Yun \& Clemens 1995; (43) Launhardt, Ward-Thompson \& Henning 1997; (44) Yun et al.\ 1996; (45) Yun \& Clemens 1994a; 
(46)Reipurth \& Olberg 1991; (47) Garnavich et al.\ 1997; (48) Rodr\'{\i}guez et al.\ 1998; (49) Palla et al.\ 1993; (50)  Reipurth, Raga \& Heathcote
1996; (51) Reipurth et al.\ 1996; (52) Cernicharo, Neri \& Reipurth 1997; (53) Gredel \& Reipurth 1993; (54) Stapelfeldt \& Scoville 1993;
(55) Rodr\'{\i}guez \& Reipurth 1994; (56) Yun \& Clemens 1994b; (57) Moreira et al.\ 1997; (58) G\'omez et al.\ 2006; (59)  Kelly \&
Macdonald 1996; (60) Codella, Felli \& Natale 1996; (61) Chini et al.\ 1997; (62) Gregersen et al.\ 1997; (63) Bachiller, Fuente \& Tafalla
1995; (64) Chen et al.\ 1997; (65) Di Francesco et al.\ 1998; (66) Choi, Panis \& Evans II 1999; (67) Anglada, Rodr\'{\i}guez \& Torrelles   
1998a; (68) Brand et al.\ 1994; (69) Kenyon 1999; (70) Herbig 1977; (71) Greene \& Lada 1977; (72) Kenyon \& Hartmann 1991; (73)  Rodr\'{\i}guez \& 
Hartmann 1992; (74) Weintraub, Sandell \& Duncan 1991; (75) Rodr\'{\i}guez et al.\ 1987; (76) Evans II et al.\ 1994; (77) Huard et al.\ 1999; 
(78) Sugitani et al.\ 2000; (79) Wilking et al.\ 1993; (80) Saraceno et al.\ 1996; (81) Reipurth \& Aspin 1997; (82) Tapia et al.\ 1997; (83) Persi,
Palagi \& Felli 1994; (84) Parker et al.\ 1991; (85) Reipurth et al.\ 1997; (86) Dobashi et al.\ 1994; (87) Umemoto et al.\ 1991; (88) Claussen et 
al.\ 1996; (89) Werner et al.\ 1979; (90) Akabane et al.\ 1992; (91) Genzel \& Downes 1977; (92) Kameya et al.\ 1990; (93) Kameya et al.\ 1989; 
(94) Minchin \& Murray 1994; (95) Quanz et al.\ 2007; (96) Leurini et al.\ 2007; (97) Choi \& Tang 2006; (98) Beltr\'an et al.\ 2002b; (99) Lee \& Ho 2005; 
(100) S\'anchez-Monge et al.\ 2008; (101)Sandell et al.\ 2005; (102) Lee et al.\ 2002;(103) Connelley, Reipurth \& Tokunaga 2007; (104) Longmore et al.\ 2006; 
(105) Ciardi \& G\'omez-Mart\'{i}n 2007; (106) Getman et al.\ 2007; (107) Kumar et al.\ 2006; (108) Wu et al.\ 2007; (109) Busquet et al.\ 2009; 
(110) Sridharan et al.\ 2002; (111) Williams et al.\ 2005}
}

\label{information}
\end{table*}

\section{General discussion}\label{results}

\subsection{Location of the exciting sources of the outflows}

Through the $(J,K)=(1,1)$ and $(2,2)$ inversion transitions of the ammonia
molecule we studied the dense gas in a sample of 40 regions with signs of
star formation, as indicated by the presence of outflow activity. We
detected ammonia emission in 27 regions and mapped 25 of them.
This high ratio of detections $(67,5\%)$ is a clear indication of the strong association
between outflow activity and \nh\ emission. This result also confirms the young
nature of the powering sources of the outflows included in our sample, 
because they appear to be still associated with (and most of them embedded in) 
the dense gas from which they have been formed.

In almost all the molecular outflow regions that we mapped in \nh, the
emission peaks close ($< 0.1$ pc) to the position of an object that was previously
proposed as an outflow driving source candidate. The association with the ammonia emission
peaks gives further support to the identification of these candidates as the outflow
driving sources, following the criterion
proposed by Anglada et al.\ (1989). The region IRAS 05490+2658 is the only
region associated with a molecular outflow where the \nh\ emission maximum is far
($\sim
0.7$ pc) from the position of the proposed exciting source; since the \nh\
emission 
peaks close to the center of symmetry of the outflow, for this region we suggest
that the
exciting source could be an undetected embedded object located
close to the \nh\ emission maximum.

For the sources of our sample that are only associated with optical signs of
outflow the \nh\ emission is generally weak. Among the nine proposed exciting
sources of optical outflow that we observed, only HH270 and HH290 IRS are
found close to an ammonia emission peak.  In all the other cases, ammonia
emission is not detected or there is no known object near the ammonia
maximum that could be a good candidate to drive the optical outflow.

\subsection{Physical parameters of the dense cores}

The  sizes of the  condensations mapped in  \nh\ generally range from $\sim  0.1$ pc to $\sim 1$ pc.   A somewhat higher value
of 2 pc is obtained for the regions NGC 281  A-W, IRAS 22134+5834 and NGC 7538,  the most
distant sources of our sample. We found evidence that several of the
condensations mapped are elongated (as was noted by
Myers et al.\ 1991). However, in many regions our angular resolution is
not good  enough to allow us  to further discuss the morphology of
the sources. A high  angular resolution interferometric study may be relevant for the sources
of our  sample that appear  compact in our present single-dish study.
Nevertheless, in several of the mapped regions (L673, M120.1+3.0-N, IRAS
20050+2720 and L1293), we can distinguish several clumps in the observed \nh\
structure. In particular, in the regions
M120.1+3.0-N   and  IRAS 20050+2720,  two   clumps   with  different
velocities, but gravitationally bound, can be identified.

For the values of the kinetic temperature obtained for these regions (see
Table~\ref{parameters}), the expected thermal line widths are $\leq 0.2$ \kms. 
This value is significantly lower than the intrinsic line widths we have
obtained, which range from 0.3 to 3.8 \kms (see Table \ref{line}). This result
suggests that the star formation process probably introduces a significant
perturbation in the molecular environment.  The lower value for the intrinsic
line width is found in the component at $V_{\rm LSR}=3.8$ \kms\ of the L1641-S3
region, whose line widths are almost thermal.  Although toward the position of
the source IRAS 05375-0731 both the 3.8 \kms\ and 4.9 \kms\ ammonia components
are observed, we argue in \S~\ref{l1641s3} that only the broad line emission at 4.9 \kms\ is likely associated
with this YSO, while no embedded sources are known to be associated with the
narrow line emission at 3.8 \kms.

\begin{figure}
\begin{center}
  \resizebox{\hsize}{!}{\rotatebox{-90}{\includegraphics{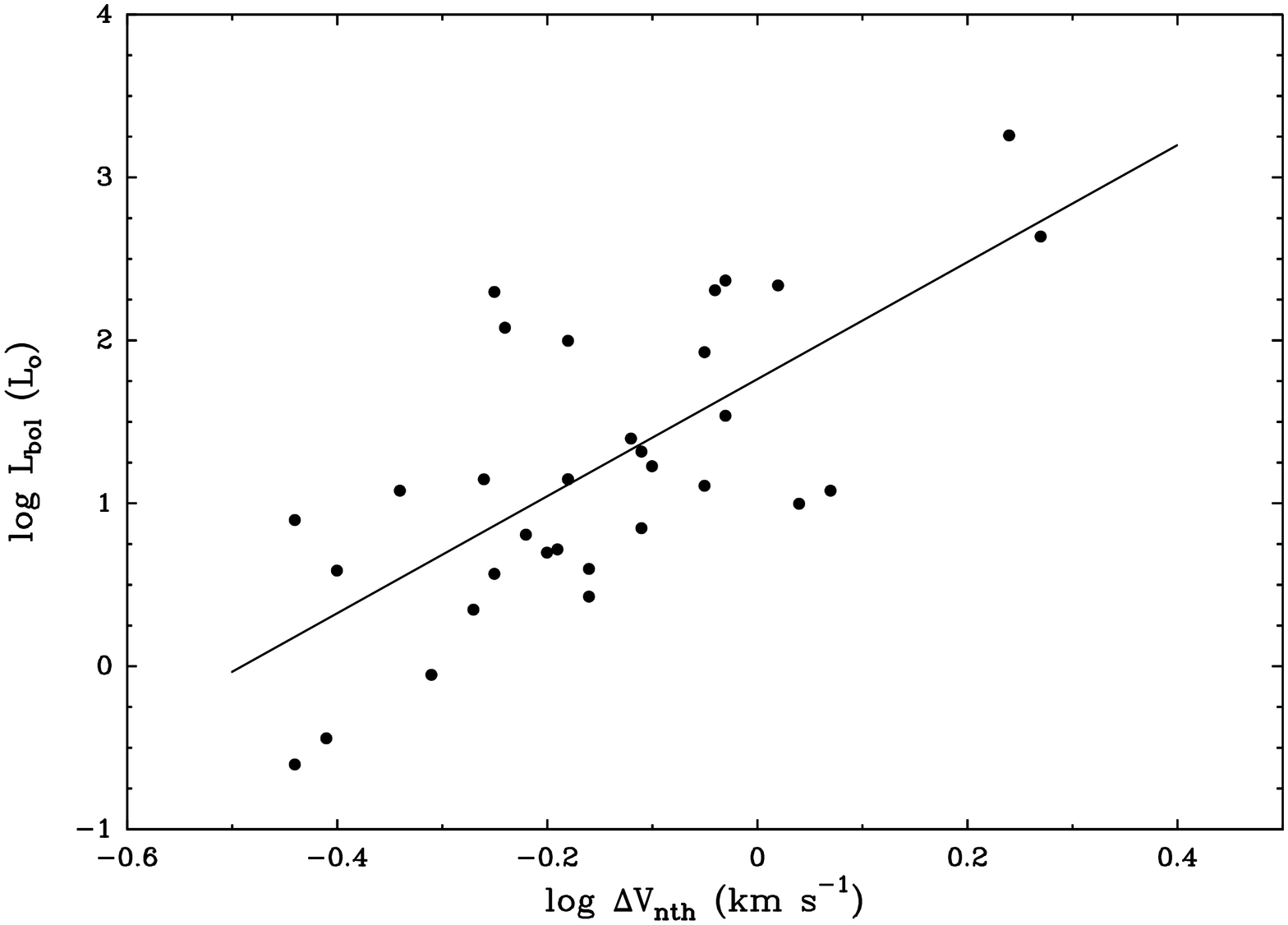}}}
  \caption[]{Bolometric luminosity vs. nonthermal line width for the observed
   regions with $D\leq1$ kpc (regions observed in Paper I are also included).}
 \label{corr}
 \end{center}
\end{figure}

The highest value of the line width is obtained for NGC 7538, which is
also  the  region with  the  highest value  of  the  mass and  kinetic
temperature of our sample.  In general, we found that the more massive
regions   present higher  values   of  the   line   width  (see
Tables~\ref{line}  and \ref{parameters}). Also,  higher values  of the
line   width  are   found   for  the   most   luminous  sources   (see
Table~\ref{information}). In Fig.~\ref{corr} we plot the luminosity  of the 
sources as a function of the nonthermal line width (subtracting the thermal
component using the derived rotational temperature  of  the region), for  the
sources  observed in  this paper and  in Paper  I. But because our sample of sources is limited 
by the sensitivity of the telescope, the most luminous sources are located mainly at larger distances
than the less luminous. In order to avoid this bias of luminosity with distance, we limited
our analysis to sources closer than 1 kpc. We found that the luminosity and the nonthermal line 
width are related by  
$\log(L_{\rm   bol}/L_{\sun})=(3.6\pm0.9)\,\log(\Delta V_{\rm nth}/\rm km~s^{-1})+(1.8\pm0.2)$ 
with a correlation coefficient of $0.7$. A similar correlation was found by Jijina et al.\ (1999).  
This correlation indicates that the most luminous sources produce a large perturbation in the 
surrounding material. 

We also note that regions associated with  CO molecular
outflow have mostly higher values of the line width than regions
with optical outflow only.

Partly because of our lack of  angular resolution,  we are not  able to
measure  the velocity gradient in our regions in detail. However, it is
remarkable  that in  L1287 our  results show  a strong
velocity gradient with sudden velocity  shifts of up to $\sim 1$ \kms\
between contiguous positions. A  high angular resolution VLA study of this
region (Sep\'ulveda et al.\ in preparation) shows that this region also exhibits 
a  complex  kinematics  at   small  scale.   Moreover,  in  two  cases
(M120.1+3.0-N   and   IRAS   20050+2720)  the   observed   velocity
distribution is compatible with  two clumps that are gravitationally bound. The
region L1641-S3 exhibits two velocity components separated by $\sim 1$
\kms\ that we interpreted as two distinct clumps of emission.

The  \hh\ column  densities we obtained  are generally
$\sim 10^{22}$  \cmd\ (assuming  [\nh/\hh]= $10^{-8}$),  implying mean  visual
extinctions of  $\sim 10$  mag. For L379  we obtained  the highest
\hh\ column  density ($\sim 10^{23}$  \cmd, corresponding to  a visual
extinction  of $\sim  100$ mag),  suggesting that  this object  is very
deeply embedded.

The  masses we obtained for  the observed  regions cover  a wide
range of values, from 1 to  3000 \mo\ (the highest value corresponds to
the NGC 7538  region). Most of the sources of  our sample are low-mass
objects, and the values of the mass obtained for their associated high-density
cores are smaller than 100 \mo.  In  general, the values derived for the mass
coincide with  the virial  mass within a  factor of 3.   This overall
trend suggests  that most  of the observed  condensations are  near the
virial   equilibrium  and   that   the  assumed   \nh\  abundance   is
adequate. However,  we found  four sources (HH 265, L588, L673, IRAS 20050+2720) for which  the derived
mass exceeds  the virial mass by  a large factor  ($>5$).  This result
could  imply   that  these  clouds   are  still  in  the   process  of
gravitational collapse.

\begin{figure}
\begin{center}
  \resizebox{\hsize}{!}{\rotatebox{-90}{\includegraphics{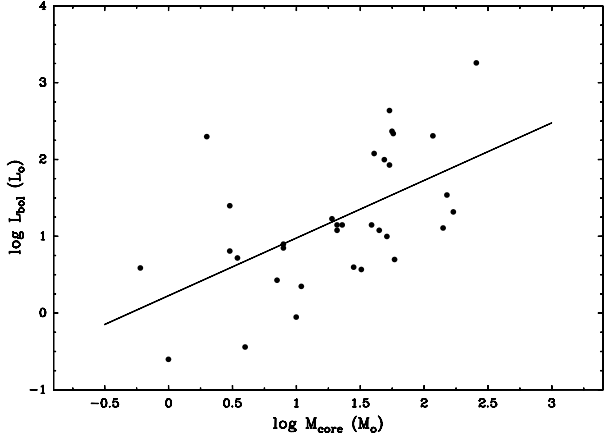}}}
\end{center} 
  \caption[]{Diagram of bolometric luminosity vs. mass of the core for the observed
   regions with $D\leq1$ kpc (regions observed in Paper I are also included)}
  \label{lbolmass}
\end{figure}

Recently, Wu et al. (2010) find a strong correlation between the luminosity of a sample of high-mass 
sources and the mass of the cores, traced by several high-density tracers. Although their sample spans 
a wide range of distances, they assume that there is no distance effect on the clump masses. 
In order to check this correlation, and to ensure that there is no distance bias in our sample (see above), we selected
sources closer than 1 kpc. In Fig.~\ref{lbolmass} we plot the luminosity of the source as a function 
of the mass of the core for the sources observed in this paper and in Paper I, using only sources 
with $D\leq1$ kpc. We found that both parameters are related by
 $\log(L_{\rm   bol}/L_{\sun})=(0.8\pm0.2)\,\log(M_{\rm core}/M_{\sun})+(0.2\pm0.3)$ with a correlation
 coefficient of $0.6$. The correlation we found for this sample is lower than the value
we would obtain for the complete sample, including the more distant sources. Although for the complete 
sample the correlation can be caused by the distance bias of the sample (more luminous sources are, 
in general, more distant), for the sources closer than $1$ kpc there is still a significant correlation 
between luminosity and core mass, and this correlation is not caused by any distance bias. 
This result indicates that the sources formed in massive clumps can accrete more material and 
form more massive stars that are also more luminous.

\section{Evolutive differences in the outflow sources}\label{evol}

We detected and mapped the  \nh\ emission in 24 out of 30 regions
associated with molecular  outflow in  our sample $(80\%)$. In  four of  the six
regions where  we failed in  detecting ammonia emission,  the evidence
for CO outflow is  weak. In the 24 regions  associated with molecular outflow,
the \nh\  emission is usually strong; the ammonia emission is  faint ($T_{\rm
MB}
\leq  0.5$  K;  see  Table~\ref{line})  only in  three  regions  (IRAS
05490+2658, V1057  Cyg and L1165). On  the other hand,  in the regions
without   molecular   outflow,  the   ammonia   emission  is   usually
undetectable or very  faint. These results agree well
with the results we obtained  in Paper I, where we studied 
the relationship between the type  of outflow and the intensity of the
\nh\ emission from a small sample of sources.

In order to substantiate the  relationship between the type of outflow
and the intensity of the ammonia emission,  here we continue the study we
began  in  Paper I,  with a more complete sample of regions. The sample
includes the regions 
observed in the present paper, the sources reported in Paper I, and also the
results 
of other Haystack \nh\ observations reported in
the literature.  We studied  the distribution of the intensity of
the \nh\ emission, as measured by the main beam brightness temperature
toward  the  outflow  exciting   source  in  this  large  sample  of
regions. As the main beam  brightness temperature is a good measure of
the intensity of the \nh\ emission  only for sources that fill the beam
of  the telescope, we included  in the sample only  the sources whose
angular size  of the  ammonia emission  is higher  than the
telescope beam. Thus, we used only sources with $D \leq1$ kpc. 

Our final sample is presented in Table~\ref{histograma}. It contains a total of
79 sources, with 30 sources associated with only molecular outflow, 40 sources associated
with both molecular and optical outflows and 9 sources with only optical outflow.

\begin{figure}
\begin{center}$
 \begin{array}{cc}
 \resizebox{!}{12cm}{\includegraphics{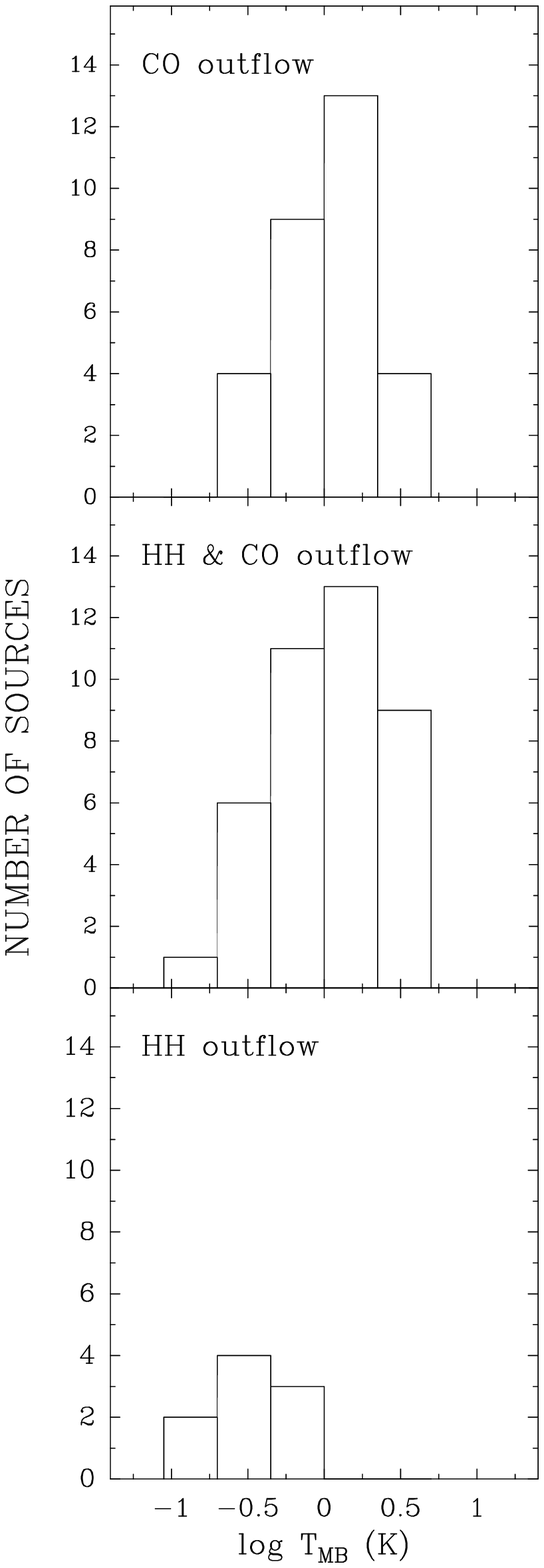}} &
   \resizebox{!}{12cm}{\includegraphics{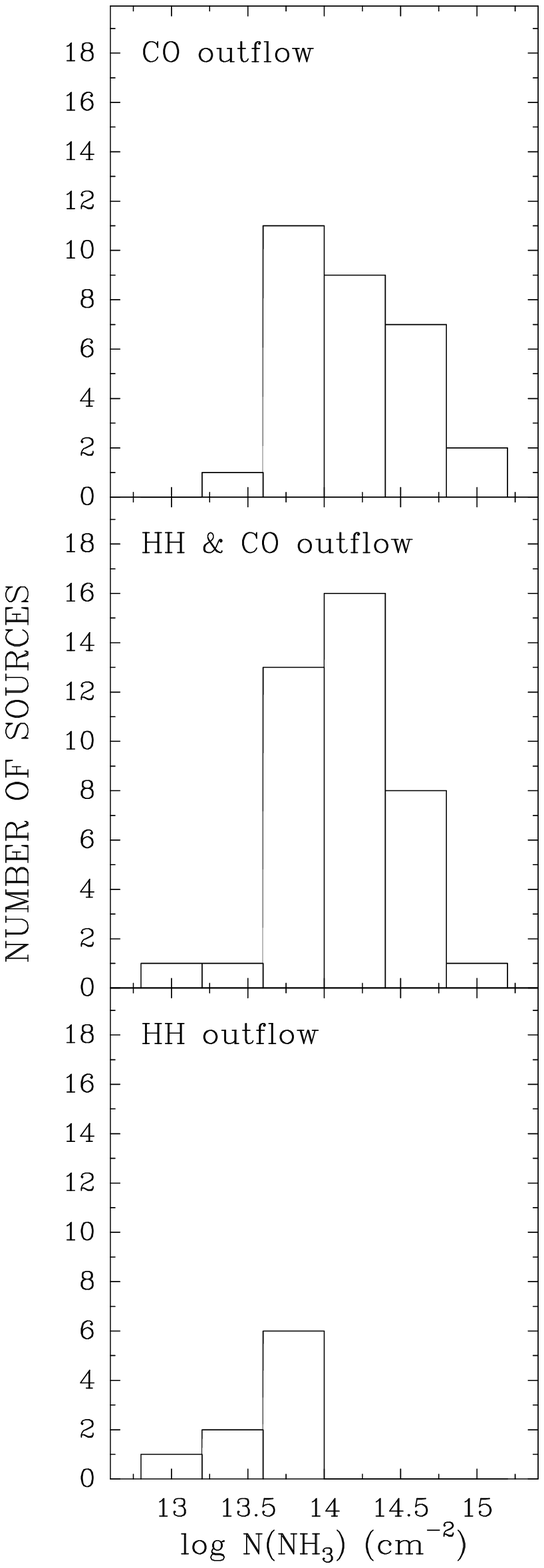}}
\end{array}$
\end{center}
   \caption[]{Distribution of the \nh\ main beam brightness temperature (left) 
     and the \nh\ column density (right)
     for sources with only molecular outflow (top), sources with both
     molecular and optical outflow (middle), and sources with only optical
     outflow (bottom).}
  \label{tmb}
\end{figure}

In Fig.~\ref{tmb} (left) we present the  distribution of the \nh\  main beam
brightness temperature  toward the  position of the  proposed outflow
exciting  source  (Table~\ref{histograma})  for  the three  groups  of
sources.  The mean values of  the \nh\ brightness temperature are 
$\langle T_{\rm MB} \rangle=0.42$~K for regions with only  optical outflow,
$\langle T_{\rm MB} \rangle=1.35$~K for regions with optical and molecular
outflow, and $\langle T_{\rm MB} \rangle=1.34$~K for sources with molecular outflow only.

Clearly the sources with only optical outflow tend to present lower values of the
\nh\ brightness  temperature, while the distribution  for sources with
molecular outflow is shifted to higher values of the brightness temperature. 
The displacement to higher values of $T_{\rm MB}$ is similar for sources with
only molecular outflow as for sources with both optical and molecular outflow. 
This was an expected result, because recent studies with high sensitivity detectors are revealing 
weak HH objects toward regions of high visual extinction, where previous
observations failed in the detection.
In  our study, we did not take into  account differences in the
brightness  of  the  Herbig-Haro  objects  or in the strength  of  the
molecular outflow. We  conclude, therefore, that the ammonia  emission is in
general more intense in  molecular outflow  sources than  in sources
without molecular outflow.

A similar result is obtained for the derived ammonia column densities.
In  Fig.~\ref{tmb} (right) we  show  the distribution of  the derived ammonia  column
density (Table~\ref{histograma}) for  the three  groups of sources. We note that
the  distribution for sources with CO outflow is
shifted to higher values of the \nh\ column density, while for sources with
only HH  outflow the  distribution tends to  lower values.   The mean
values  for the \nh\  column density are  $\langle N$({\rm \nh})$\rangle =
5.2\times10^{13}$~\cmd\ for
regions   with  optical  outflow only,   $\langle N$({\rm \nh})$\rangle =
1.94\times10^{14}$~\cmd\  for
sources  with molecular and  optical outflow  and $\langle N$({\rm \nh})$\rangle
= 2.49\times10^{14}$~\cmd\ for  sources with molecular outflow only.

Recently, Davis et al.\ (2010) find from a study of outflows and their exciting sources in the Taurus region 
that sources  driving CO outflows have redder near-IR colours than sources driving HH jets, 
and they conclude that CO outflow sources are more embedded in the high-density gas than the HH optical 
outflow sources. This result agrees well with ours.
 
All these  results can  be interpreted  as an indication that molecular
outflow sources are younger, since they are associated with a larger amount of
high-density gas.  As  the star  evolves,  the surrounding  material
becomes less dense, decreasing the ammonia column density, and at the  same
time making the Herbig-Haro objects detectable. At the time when the  molecular
outflow has disrupted and swept out the molecular material surrounding the YSO,
both the CO outflow and the \nh\ column density are expected to be weak, and
only  the Herbig-Haro objects would be observable.

The  ammonia emission and the observational  appearance of outflows trace an evolutive sequence  
of sources. Molecular  and optical outflow  would   be  phenomena  that   dominate  observationally  at
different stages of the YSO evolution.  In younger objects, molecular outflows
will be  prominent, while optical outflows will  progressively show up as the YSO evolves.

\section{Conclusions}\label{conclus4}

We detected the ammonia emission  in 27 sources of a sample of 40
sources associated with molecular  and/or optical outflows, and we were
able  to map 25  of them. We also searched for  \hho\ maser
emission toward 6 sources, and detected new \hho\ masers
in HH 265  and AFGL 5173.   Our main conclusions  can be
summarized as follows:

\begin{enumerate}

\item In  all molecular outflow  regions mapped, the \nh\
emission peak falls very close to the position of  a very good candidate
for the outflow excitation (except in the case of IRAS 05490+2658 where we
propose an alternative location for the exciting source). On the other hand, the
sources associated with optical outflow (except HH 270 IRS and HH 290 IRS) are not
associated with an ammonia emission peak.

\item Four regions (HH 265, L588, L673 and IRAS 20050+2720) could be in the process of 
gravitational collapse at scales $\geq0.2$ pc, as their derived masses exceed the virial mass by a 
factor $>5$.  The rest of ammonia condensations appear to be close to the virial equilibrium.

\item In several regions the ammonia structure presents more than one clump. In
the M120.1+3.0-N and IRAS 20050+2720 regions, two clumps with different
velocities, which are gravitationally bound though were identified.

\item We identified a high-density clump where the HH 270/110 jet can suffer the
collision responsible for the deflection of the jet.

\item We were able to separate the \nh\ emission from the L1641-S3 region
into two overlapping clouds, one with signs of strong perturbation, probably
associated with the driving source of the CO outflow, and a second, quiescent
clump, which probably is not associated with star formation.

\item In general,  the observed \nh\ condensations are very cold, with
line widths dominated by nonthermal (turbulent) motions. Among the observed
sources, the more massive regions appear to produce a larger perturbation in
their molecular high density environment.
 
\item We found that generally the more luminous objects are associated with broader ammonia 
lines. A correlation between the nonthermal
component of the line width and the luminosity of the associated object, 
$\log(L_{\rm   bol}/L_{\sun})=(3.6\pm0.9)\,\log(\Delta V_{\rm nth}/\rm km~s^{-1})+(1.8\pm0.2)$ 
was found for sources with $D\leq$ 1 kpc.

\item We found that there is a significant correlation between the luminosity of the source 
and the mass of the core and that this correlation is not caused by any distance bias in the sample. 
Both parameters are related by 
$\log(L_{\rm   bol}/L_{\sun})=(0.8\pm0.2)\,\log(M_{\rm core}/M_{\sun})+(0.2\pm0.3)$.

\item The ammonia brightness
temperature and column density of the sources decrease as the outflow activity
becomes prominent in the optical. These results give an evolutive scheme in
which young objects progressively lose their surrounding high-density gas. 
The  ammonia emission and the observational  appearance of outflows trace an evolutive sequence  
of sources.

\end{enumerate}

\begin{acknowledgements}
I.S. acknowledges the Instituto de Astrof\'{i}sica de Andaluc\'{i}a for the
hospitality during part of the preparation of this paper.  G.A., R.E., R.L. and
J.M.G. acknowledge support from MCYT grant AYA2008-06189-C03 (including FEDER
funds). G.A. acknowledges support from Junta de Andaluc\'{i}a, Spain. 

\end{acknowledgements}

\longtab{7}{
\begin{longtable}{l l l r c c c c}
\caption[]{Regions associated with molecular or optical outflow observed
in \nh}\label{histograma}\\
\hline\noalign{\smallskip}
Source & Outflow & Ref. & $T_{\rm MB}$\tablefootmark{a}  & $N$({\rm \nh})\tablefootmark{b}     &Ref. & $D$ & Ref. \\
       & associated &   &(K)               & ($10^{14}$ cm$^{-2}$) &     &(pc) & \\
\noalign{\smallskip}
\hline\noalign{\smallskip}
\endfirsthead
\caption[]{Continued}\\
\hline\noalign{\smallskip}
Source & Outflow & Ref. & $T_{\rm MB}$\tablefootmark{a}  & $N$({\rm \nh})\tablefootmark{b}     &Ref. & $D$ & Ref. \\
       & associated &   &(K)               & ($10^{14}$ cm$^{-2}$) &     &(pc) & \\
\noalign{\smallskip}
\hline\noalign{\smallskip}
\endhead 
\hline
\endfoot
\endlastfoot
 
M120.1+3.0-N (IRAS 00213+6530)
& CO      &1    &0.65   &0.9    &2   &850   &1  \\
M120.1+3.0-S (IRAS 00259+6510)
& CO      &1    &0.57   &0.7    &2   &850   &1  \\
L1287
& CO      &1    &2.51   &4.6    &2   &850   &1  \\ 
L1293
&CO       &1    &1.06   &1.1    &2   &850   &1  \\
L1448 IRS1
& CO, HH  &72,70  & 0.3   &0.3    & 4   & 350  & 3   \\
L1448 IRS2
& CO, HH  &70     & 1.9   &1.8    & 4   &350   & 3   \\
L1448 IRS3
& CO, HH  &23,5   & 3.1   &4.2    & 4   &350   & 3    \\
L1448 C 
& CO, HH  &23,70  & 2.5   &3.3    & 4   & 350  & 3    \\
GL490
& CO  &6        &$\leq0.5$ &$\leq0.6$ & 7   &900   & 8     \\
L1455 IRS1
& CO, HH  &9,70   &1.9    &1.0    & 4   &350   & 3    \\
L1455 IRS2
& CO, HH  &9,70   &2.5    &1.6    &4    &350   &3     \\
L1489
& CO, HH  &10,71   &0.8    &1.8    & 11  &140   & 8 \\
HH 156
& HH  & 1   &$\leq0.2$ &$\leq0.4$ & 2  &140   & 1 \\
HH 159
&CO, HH &1  &$\leq0.4$ &$\leq0.8$ &2  &160   & 1 \\
HH 158
&HH     &1  &$\leq0.3$ &$\leq0.6$ &2  &160   &1 \\
HH 31
&CO?,HH     &1  &0.56      &0.7       &2  &160   &1 \\
L1524 (Haro 6-10)
& CO, HH  & 73, 14 &$\leq0.6$  &$\leq1$ &4   &140   & 13  \\
L1551 IRS 5
& CO, HH &15, 16  &2.72   &2.3   &7    &140 & 13 \\
HL Tau
& CO, HH &17, 18  &$\leq1$  &$\leq1$  &7  &140 & 13 \\
L1551 NE
& CO, HH & 1 &0.79 &1.0 &2 &160 &1 \\
L1642
&CO, HH & 1 &$\leq0.1$ & $\leq0.1$ & 2 & 125 & 1 \\
L1527
& CO, HH &20,74,19 & 2.12   &5.0    &11   &140 &13 \\
L1634 (IRAS 05173--0555)
& CO, HH &1       &1.0     &1.1    &2   &460 &1 \\
L1634 (IRS 7)
& CO, HH &83,84    &0.59    &0.8    &2   &460 &1 \\ 
RNO 43 (IRAS 05295+1247)
& CO, HH &21,22  &0.40    &0.4   &2   &400 & 24 \\
HH 83
&CO, HH  & 26,25 &0.46   &0.4--0.5   &85   &470 &13  \\   
HH 84
& HH  & 25       &0.3   &0.5--1   &85   &470 &13   \\
HH 33/40
& HH    &27      &$\leq0.3$ &$\leq0.5$ &28   &470 &13  \\
HH 86/87/88
& HH  & 25       &$\leq0.2$ &$\leq0.1$ &85   &470 &13 \\
HH 34
& CO, HH  &30,29  &1.3  &1.1    &4   &480 &31 \\ 
L1641-N
& CO, HH  &32, 69      &2.2  &2.8       &85  &480 &33 \\
HH 38-43
& HH   & 27     &$\leq0.5$  &$\leq0.5$  &4   &480 &31 \\
Haro 4-255 FIR
& CO  & 72       &2.2   &2.1      &4   &480 &31 \\
L1641-S3(high velocity)
& CO  &1       &2.0   &2.4      &2  &480 &1 \\
HH 68
&HH   &1       &$\leq0.3$  &$\leq0.4$ &2 &460 &1 \\
B35
& CO  & 10      &1.2   &4.0      &11  &500 &34 \\
HH 26 IR
& CO, HH & 36,35 &2.8 &2.5      &7   &470 &13 \\   
HH 25 MMS
& CO, HH & 75,35 &2.4 &2.1      &7   &470 &13 \\
NGC 2071
& CO  & 37      &2.44  &2.5     &7   &500 &8 \\  
HH 270 IRS
& HH     &1     &0.67 &0.7     &2  &460 &1  \\
IRAS 05487+0255 
&CO      &1     &0.8  &0.7     &2  &460 &1    \\
HH 111
& CO, HH &1     &0.56 &0.5     &2  &460 &1 \\
Mon R2  
& CO   & 38     &1.2   &1.2     &7   &800 &8 \\
Mon R2-N
& CO   & 76     &1.0   &1.0       &7   &800 &8 \\
GGD 12-15
& CO  & 39      &1.52   &2.0      &7  &1000 &8 \\  
RMon
& CO, HH &40, 35 &$\leq0.5$  &$\leq0.6$ &7   &800  &13 \\
NGC 2264 (HH 14-4/5/6)
& HH  &41        &$\leq1$  &$\leq1$  &4   &800  &13 \\
HH 120  
& CO, HH &43,77,42  &1.8  &2.5    &44   &400  &45 \\
L1709
& CO  &1     &$\leq0.4$  &$\leq0.6$ &2 &160 &1 \\
L43     
& CO  &72         &2.7   &2.2    &4   &160   &46 \\
L100
& CO   &47       &0.5   &0.3--0.5    &85  &225   &48 \\
L483
& CO  & 47       &4.54   &14    &85   &200  &49  \\
L588
& CO, HH &1     &1.3    &4.1   &2   &310  &1  \\
R CrA (HH 100-IR)   
& CO, HH &72,50,79  &2.8    &2.0     &4  &130  &51 \\
R CrA (IRS 7)
&CO, HH &72,50,78   &1.6    &1.8     &4  &130  &51 \\
L673    
& CO  & 52      &2.1  &$\geq2.2$    &2,85  &300  &3  \\
CB 188
& CO  &1   &$\leq0.2$ &$\leq0.4$   &2  &300  &1 \\
HH 32a  
& CO, HH &17,35  &$\leq0.6$ &$\leq0.7$ &4   &300  &13 \\
L778
& CO & 10      &1.8   &6.3     & 11   &250  &8 \\
B335
&CO, HH  &53,54 & 1.2  &7.9     & 11   &250  & 13 \\
L797
&CO     &1 &$\leq0.2$ & $\leq0.4$ &2 &700 &1 \\
IRAS 20050
&CO     &1 &1.7       &3.6        &2 &700 &1 \\
V1057 Cyg
&CO     &1 &0.3       &0.2        &2 &700 &1 \\
L1228   
& CO, HH & 55,56  &2.71  &5.3   &85  &300  &56 \\
V1331 Cyg
& CO, HH  &72,80   &0.5  &2.3    &4   &700 &57 \\
L1172   
& CO  & 10   &1.8  &7.9    & 11 &440 &8 \\
CB 232
& CO  &1    &0.58 &0.9    &2  &600 &1 \\
IC 1396 E
&CO   &1    &0.86 &1.4    &2  &750 &1 \\
NGC 7129
&CO, HH & 58, 59   &0.52  &0.36    &7   &1000 &13 \\
HHL73 (IRAS 21429+4729)
& CO   &60     &1.66    &3.2  &85   &900 &61  \\
HHL73 (IRAS 21432+4719)
& CO,HHL,HH &60,61,81  &1.25 &1.1     &28  &900 & 61  \\
HHL73 (IRAS 21441+4722)
& CO   & 60       &0.7  &$\geq0.9$    &85  &900 & 61 \\
L1165
&CO,HH  &1     &0.35  &0.9   &2   &750 &1  \\
S140N (IRAS 22178+6317)
& CO    &32     &0.88  &1.4   &28   &900 & 63  \\
S140N (Star 2)
& CO, HH &82,62 &1.2   &1.9   &28   &900 &63 \\
L1221
&CO, HH   &1   &2.5   &3.9   &2   &200 &1 \\     
L1251 (IRAS 22343+7501)
& CO, HH  & 64,65  &0.29  &1.7        &85   &300 &66 \\  
L1251 (IRAS 22376+7455) 
& CO, HH  & 64,67  &1.71  &3.1        &85  &300 &66 \\
L1262
& CO      & 47     &1.57  &$\geq4.9$  &85   &200 &68  \\
\hline\noalign{\medskip}

\multicolumn{8}{l}{\parbox{\LTcapwidth}{\footnotesize{\textbf{Notes.} Regions with distance $\leq1$ kpc. \\ $^a$ Main brightness temperature at the position of the suspected exciting source. $^b$ Lower limit of the beam-averaged column density at the position of 
the suspected exciting source.}

\tablebib{(1) see Table~1; (2) this paper; (3) 
Herbig \& Jones 1983; (4) Anglada et al.\ 1989; (5) Eiroa et al.\ 1994a; 
(6) Snell et al.\ 1984; (7) Torrelles et al.\ 1983; (8) Fukui et al.\
1993; (9)  Goldsmith et al.\ 1984; (10) Myers et al.\ 1988; (11) Benson \&
Myers 1989; (12) Strom et al.\ 1986; (13) Reipurth 1994; (14) Elias 1978; 
(15) Snell et al.\ 1980;  (16)  Mundt \& Fried 1983; (17) Edwards \& Snell
1982; (18) Mundt et al.\ 1988;  (19)  Eiroa et al.\ 1994a; (20) Heyer et
al.\ 1987; (21) Edwards \& Snell 1984;  (22) Jones et al.\ 1984; 
(23) Bachiller et al.\ 1990; (24) Maddalena \& Morris 1987; (25)  
Reipurth 1989; (26) Bally et al.\ 1994; (27) Haro 1953; (28)  
Verdes-Montenegro et al.\ 1989; (29) Haro
1959; (30) Chernin \& Masson 1995; (31) Genzel et al.\ 1981; (32) Fukui et
al.\ 1986; (33) Chen et al.\ 1993; (34) Felli et al.\ 1992;  (35) Herbig
1974; (36) Snell \& Edwards 1982; (37) Bally 1982; (38) Loren 1981;  (39)
Rodr\'{\i}guez et al.\ 1982; (40) Cant\'o et al.\ 1981; (41) Adams et al.\
1979; (42) Cohen \& Schwartz 1987; (43) Olberg et al.\ 1989; (44) Persi et
al.\ 1994; (45) Petterson 1984; (46) Chini 1981; (47) 
Parker et al.\ 1988; (48)  Reipurth \& Gee 1986; (49) Ladd et al.\ 1991a;
(50) Strom et al.\ 1974;  (51)  Marraco \& Rydgren 1981; (52) Armstrong \&
Winnewisser 1989; (53)  Frerking \& Langer 1982; (54) Vrba et al.\ 1986;
(55) Haikala \& Laureijs 1989; (56)  Bally et al.\ 1995; (57)
Chavarr\'{\i}a-K 1981; (58) Loren 1977; (59) Ray et al.\ 1990; (60)
Dobashi et al.\ 1993; (61) Gyulbudaghian et al.\ 1987; (62)  Eiroa et al.\
1993; (63) Crampton \& Fisher 1974; (64) Sato \& Fukui 1989; (65) 
Bal\'azs et al.\ 1992; (66) Kun \& Prusti 1993; (67) Eiroa et al.\ 1994b;
(68)  Parker et al.\ 1991; (69) Reipurth et
al.\ 1998; (70) Bally et al.\ 1997;  (71) G\'omez et al.\ 1997; (72) 
Levreault 1985; (73) Hogerheijde et al.\ 1998; (74) Tamura
et al.\ (1996); (75) Gibb \& Davis 1998; (76) Tafalla et al.\ 1997; (77)
Nielsen et al.\ 1998; (78)  Anderson et al.\ 1997; (79) Whittet et al.\
1996; (80) Mundt \& Eisl\"offel 1998; (81) Devine, Reipurth \& Bally 1997;
(82) Davis et al.\ 1998; (83) Davis et al.\ 1997; (84) Hoddap \& Ladd
1995; (85) Anglada, Sep\'ulveda \& G\'omez 1997.}}}

\end{longtable}}

\appendix
\section{Results for individual sources}

\subsection{M120.1+3.0-North}

This region is associated with a bipolar molecular outflow (Yang et al.\ 1990)
and contains several low-luminosity objects. Two of these objects, IRAS
00213+6530 and IRAS 00217+6533, fall inside the outflow lobes.
On the basis of the geometrical position of IRAS 00213+6530, close to the
emission peak of the blue-shifted gas, and its cold IR colors, Yang et al.\
(1990) favor this source as the driving source of the outflow.
 We observed  the region around both IRAS sources in ammonia.

\begin{figure}
\begin{center}
\resizebox{8cm}{!}{\includegraphics[83,205][515,525]{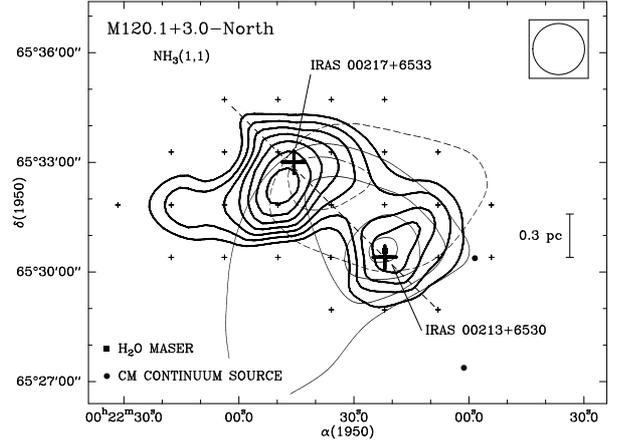}}
    \caption{Contour map of the main beam brightness temperature of the main
     line of the ammonia $(J,K)=(1,1)$ inversion transition (thick line) in
     the M120.1+3.0-North region. The lowest contour level is 0.3 K, and
     the increment is 0.1 K. The observed positions are indicated with
     small crosses. The half-power beam width of the telescope is shown as
     a circle. The positions of several relevant objects in the region are
     indicated. The CO bipolar outflow (thin line) is from Yang et al.\ (1990) 
     (solid contours indicate blueshifted gas, and dashed contours indicate 
     redshifted gas). Dashed straight line is the axis of the pos-vel diagram of Fig.~\ref{grad}}
  \label{m120n}
\end{center}
\end{figure}

\begin{figure}
\begin{center}
\resizebox{8cm}{!}{\rotatebox{-90}{\includegraphics[63,43][570,780]{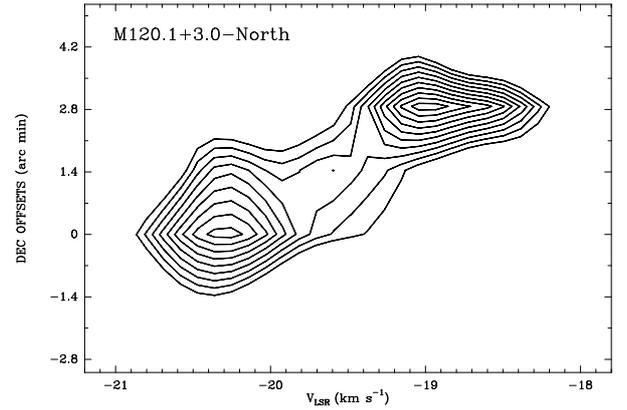}}}
    \caption{Position-velocity diagram of the \nh(1,1) main line along an axis
    passing toward the two maxima (P.A.$\sim45\degr$) of the 
     M120.1+3.0-North condensation. The 
     lowest contour level is 0.3 K and the increment is 0.05 K.}
  \label{grad}
\end{center}
\end{figure}

The \nh\ structure (Fig.~\ref{m120n}) consists of two sub-condensations, each
one peaking very close to the position of an IRAS source. This suggests
that both IRAS sources are embedded in the high density gas. Our results show
that the velocity is different for each clump (see Table~\ref{line} and 
Fig.~\ref{grad}). The observed difference in velocity is consistent with a
gravitationally bound rotational motion of the two clumps. 

The association of IRAS 00213+6530 with an ammonia emission maximum supports its
identification as
the outflow exciting source. However, we note that IRAS 00217+6530 falls very
close to the position of an ammonia emission peak, it lies close to the
emission peak of the outflow redshifted gas and its IRAS colors are
characteristic of
an embedded source (although the source appears confused in the 60 and 100 \mm\
IRAS bands). Therefore, based on these results, both IRAS sources are valid
candidates for the outflow excitation.

The radio continuum sources detected in the region (Anglada, G., private
communication) fall outside the ammonia condensation (see Fig.~\ref{m120n}), therefore they appear
to be unrelated to the star-forming region.

\subsection{M120.1+3.0-South}

This region is associated with several IRAS sources and with a CO bipolar
outflow (Yang et al.\ 1990). The outflow is asymmetric, with the red
lobe more intense than the blue one. Two sources, IRAS 00259+6510 and IRAS
00256+6511, lie inside the outflow lobes. Yang et al.\ proposed IRAS 00259+6510
as the driving source of the outflow. We observed the region
around both sources in ammonia. 

\begin{figure}
\begin{center}
\resizebox{8cm}{!}{\includegraphics[76,298][498,615]{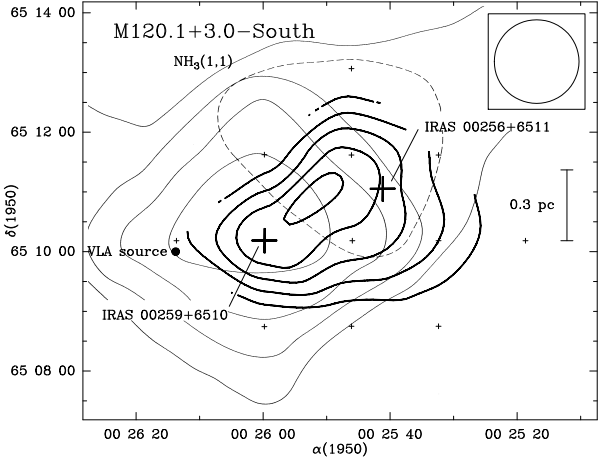}}
    \caption{Same as Fig.~\ref{m120n}, for the M120.1+3.0-South region.
     The \nh\ lowest contour is 0.2 K and the increment is 0.1 K.  CO bipolar
outflow map is
     also shown (Yang et al.\ 1990)(solid contours
     indicate redshifted gas, and dashed contours indicate blueshifted gas).}
  \label{m120s}
  \end{center}
\end{figure}

The \nh\ condensation (Fig.~\ref{m120s}) shows an elongated structure in the
NW-SE direction. Both IRAS 00259+6510 and IRAS 00256+6511 are located  close to
the ammonia emission maximum, suggesting that they are embedded sources. Both
sources have similar IRAS colors, but appear confused at 60 and 100 \mm.
Therefore, we cannot favor one of them as the driving source of the outflow.

A cm radio continuum source is detected at the edge of the ammonia condensation
(Anglada, G., private communication). Unfortunately, the information available
for this source is not enough to infer the nature of the emission.

\subsection{L1287}\label{1287}

The dark cloud L1287 is associated with an energetic bipolar molecular outflow
(Snell et al.\ 1990, Yang et al.\ 1991).  At the center of the outflow lies the
source IRAS 00338+6312, that has been proposed as the outflow exciting source
(Yang et al.\ 1990). The brightest visible object in the region, RNO 1 (Cohen
1980), lies $\sim40''$ NE of the nominal IRAS position. However, because of the
low angular resolution of the IRAS data, several young stellar objects 
(a FUOri binary system RNO 1B/1C and several radio continuum sources) fall
inside the IRAS error ellipsoid. Kenyon et al.\ (1993) proposed the FU Ori 
star RNO 1C as the outflow exciting source. However, additional studies by 
Anglada et al.\ (1994) favored a jet-like radio continuum source, VLA 3, located
very close to the IRAS nominal position and to the symmetry center of the polarization
pattern (Weintraub \& Kastner 1993), as the most likely candidate to be the exciting
source. 
The detection of \hho\ maser emission associated with VLA 3 (Fiebig 1995), and  
the interpretation of the \hho\ velocity pattern as a infalling disk (Fiebig
1997) further supports VLA 3 as the outflow exciting source.
 
The region was observed in HCN, \hco\ (Yang et al.\ 1991), CS (Yang et al.\
1995; McMuldroch et al.\ 1995), and in \nh\ (Estalella et al.\ 1993). Estalella et al.\
(1993) found a gradient in the NW-SE direction, which was interpreted as caused by the rotation of the core.

\begin{figure}
\begin{center}
 \resizebox{8cm}{!}{\includegraphics[80,227][570,635]{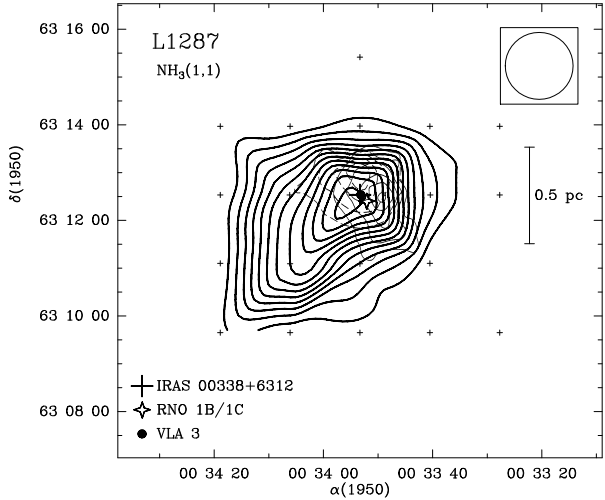}}
    \caption{Same as Fig.~\ref{m120n}, for the L1287 region. The \nh\
     lowest contour is 0.3 K and the increment is 0.2 K. The map of the CO
     bipolar outflow is from Snell et al.\ (1990).}
  \label{l1287}
  \end{center}
\end{figure}

The condensation we mapped (Fig.~\ref{l1287}) is clearly elongated in the
northwest-southeast direction, perpendicular to the CO outflow axis. We have
found a velocity gradient of $\sim 1.23$ \kmp\ in the NW-SE direction. Both results
agree well with the results of Estalella et al.\ (1993). The ammonia emission
peaks near the position of IRAS 00338+6312, RNO 1B/1C and VLA 3. However, because of
the small projected angular separation between all these objects ($\sim
5$\arcsec$-10$\arcsec), we cannot distinguish from our data which of these
sources is the best candidate for exciting the outflow in terms of its
proximity to the ammonia emission peak. 
This region was studied in \nh\ with high angular VLA resolution
(Sep\'ulveda 2001; Sep\'ulveda et al. in preparation). These observations revealed that the central
region toward the sources has a complex structure and exhibits a complex kinematics.

\subsection{L1293}

Yang (1990) discovered a bipolar molecular outflow in this region and proposed
IRAS 00379+6248 as its driving source.

\begin{figure}
\begin{center}
\resizebox{8cm}{!}{\rotatebox{-90}{\includegraphics[52,25][558,554]{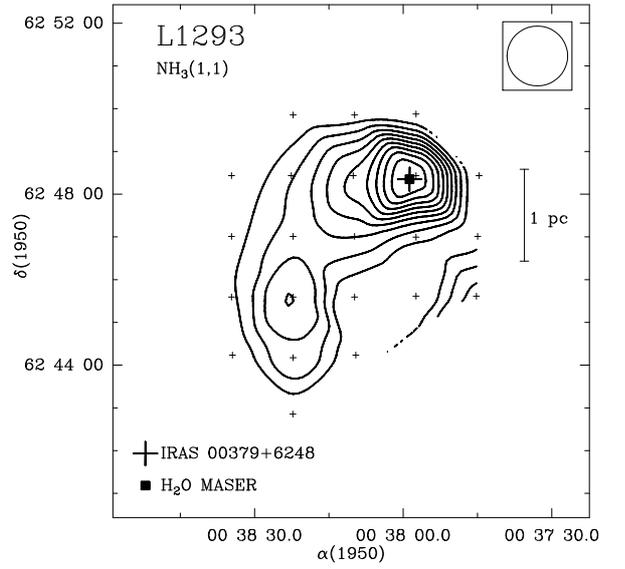}}}
     \caption{Same as Fig.~\ref{m120n}, for the L1293 region. The \nh\
      lowest contour level is 0.2 K, and the increment is 0.1 K.}
  \label{l1293}
  \end{center}
\end{figure}

The \nh\ structure presents two emission maxima separated by $\sim4'$
(Fig.~\ref{l1293}). The strongest ammonia peak coincides with the position of
IRAS 00379+6248. The HCN, \hco\ and \tco\ emission also peak toward the IRAS
position (Yang
1990). This IRAS source is not detected at 12 \mm\ and its infrared
flux increases steeply toward longer wavelengths. These IR results, along with
its association with strong \nh\ emission and with an \hho\ maser (Wouterloot
et al.\ 1993), suggest that IRAS 00379+6248 is a young stellar object, deeply
embedded in the high density gas, and the most plausible exciting source of the
outflow. The weaker emission peak is not associated with any known object.

\subsection{NGC 281 A-West}

This region is associated with a bipolar molecular outflow proposed to be
driven by the luminous source IRAS 00494+5617 (Snell et al.\ 1990; Henning et
al.\ 1994). A near-IR cluster (Carpenter et al.\ 1993) and several \hho\ maser
spots are found in association with the IRAS source (Henning et al.\ 1992).
Henning et al.\ (1994) modeled the observed spectral energy distribution of the
source 
from 1~\mm\ to 1 mm, concluding that it is a very good candidate for a
deeply embedded and very young protostellar object.

\begin{figure}
\begin{center}
\resizebox{8cm}{!}{\includegraphics[55,285][499,647]{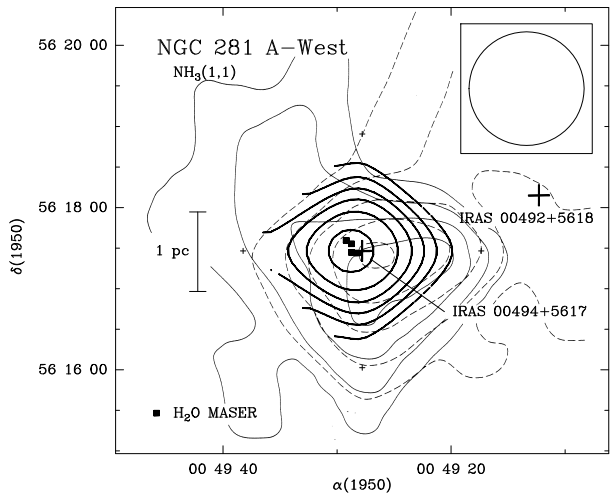}}
     \caption{Same as Fig.~\ref{m120n}, for the NGC 281 A--West region.
      The \nh\ lowest contour level is 0.15 K, and the increment is 0.075
      K. The CO bipolar outflow is from Snell et al.\ (1990) (solid
      contours indicate redshifted gas, and dashed contours indicate
      blueshifted gas)}
  \label{ngc281}
  \end{center}
\end{figure}

We detected an ammonia clump (Fig.~\ref{ngc281}), which appears unresolved
with our beam. The \nh\ emission peaks at the position of IRAS 00494+5617 . Our
results agree with the $40''$ angular resolution \nh\ map of Henning
et al.\ (1994),
which reveals that the ammonia clump is elongated along the east-west direction
with the emission peaking toward the position of the IRAS source. CS emission
mapped by Carpenter et al.\ (1993) with an angular resolution of $\sim 50''$,
also peaks toward the position of the IRAS source. These results, along with
the spectral energy distribution of the source, suggest that IRAS 00494+5617 is
a very
young object deeply embedded in the high density gas and that it is the most
likely candidate to excite the outflow.

Our ammonia results suggest that there is no significant amount of dense gas
in association with the source IRAS 00492+5618, located $\sim 2'$ to the west
of IRAS 00494+5617.

\subsection{HH 31}

The HH 31 jet is a sinusoidal chain of knots having a linear extent of $\sim
0.2$ pc (Herbig 1974; G\'omez et al.\ 1997). Cohen \& Schwartz (1983) found
four near-IR sources (IRS1, IRS2, IRS3 and IRS4) in the vicinity of the
jet, being IRS~2, that coincides with IRAS 04248+2612, the proposed exciting
source of the jet. This source has been detected at millimeter and
sub-millimeter wavelengths (Moriarty-Schieven et al.\ 1994) and apparently
drives a small molecular outflow (Moriarty-Schieven et al.\ 1992), although no
map has been published. In near-IR images, IRAS 04248+2612 (IRS~2) appears as
a bipolar reflection nebula (Padgett et al.\ 1999). We searched for
ammonia emission toward the four near-IR sources.

\begin{figure}
\begin{center}
\resizebox{8cm}{!}{\rotatebox{-90}{\includegraphics[110,39][572,764]{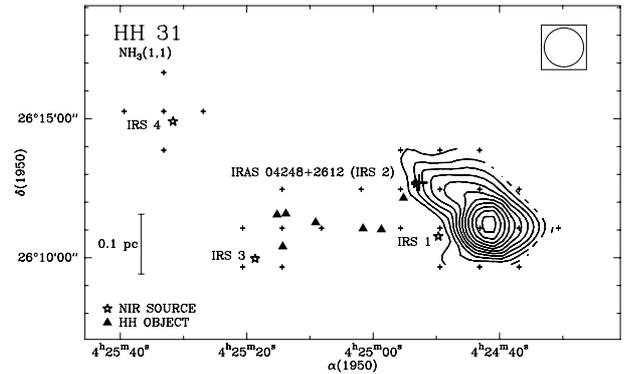}}}
    \caption{Same as Fig.~\ref{m120n}, for the HH 31 region. The \nh\
     lowest contour is 0.3 K and the increment is 0.2 K. The position of the HH
     31 knots is from G\'omez et al.\ (1997)}
  \label{hh31}
  \end{center}
\end{figure}

The ammonia condensation (Fig.~\ref{hh31}) is elongated in the NE-SW direction,
and it agrees with the \nh\ map shown by Benson \& Myers (1989).
 The \nh\ emission peak is displaced
$\sim 3'$ ($\sim 0.14$ pc) to the SW of the HH 31 IRS2 position. To our
knowledge, no source has been reported toward the position of the \nh\
emission peak. We suggest that high sensitivity observations could reveal a
deeply embedded object at this position. 
The sources HH 31 IRS2 and IRS1 lie at the edge of the condensation. We have
not detected significant emission toward IRS3 and IRS4, which lie far away
(more than $8'$) from the condensation.

\subsection{HH 265}\label{265}

HH 265, located in the L1551 cloud, is an isolated Herbig-Haro object whose
exciting source still remains unknown. Swift et al.\ (2005, 2006) mapped the
region in \nh\ and CS. From their results, these authors suggest that the cloud
is likely a prestellar core showing signs of undergoing the first phases of
gravitational collapse. 

We discovered an \hho\ maser (see
Fig.~\ref{maser4}) toward the position of the HH object. The maser shows two
velocity components, whose line parameters are given in Table~\ref{maser}. The
\hho\ maser emission suggests the presence of a nearby exciting
source, which could be also responsible for the excitation of HH 265.
However, it is also suggested that HH265 could be the end of a jet emanating
from the source LkH$\alpha$ 358 (Moriarty-Schieven et al.\ 2006; Movsessian et
al.\ 2007). 

\begin{figure}
\begin{center}
\resizebox{8cm}{!}{\rotatebox{-90}{\includegraphics[54,37][554,587]{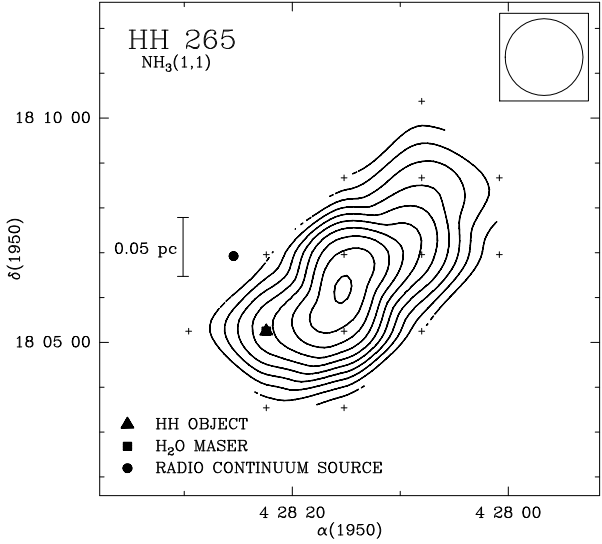}}}
    \caption{Same as Fig.~\ref{m120n}, for the HH 265 region.  The \nh\
     lowest contour is 0.3 K and the increment is 0.2 K}
  \label{hh265}
  \end{center}
\end{figure}

Our ammonia map (see Fig.~\ref{hh265}) shows that both the \hho\ maser and the
HH object fall inside the ammonia condensation, but they are displaced by
$\sim1\farcm5$ ($\sim 0.08$ pc) to the SE of the position of the ammonia
maximum. We suggest that a sensitive search in the submm, mm, or cm range in the
vicinity of the \nh\ emission peak could reveal an embedded object, 
responsible for the excitation of HH 265. The mass derived for this region (see
Table~\ref{parameters}) exceeds the virial mass by a factor of five, a result
that may indicate that a significant fraction of the cloud is still undergoing
the process of gravitational collapse toward a central, embedded protostar, in
good agreement with the results obtained by Swift et al.\ (2005, 2006).

A 20 cm source, located $\sim2'$ to the NE of HH 265, was detected by Snell \&
Bally (1986). However, this source lies outside the ammonia condensation and
was not detected at shorter wavelengths (6 cm and 2 mm; Snell \& Bally 1986),
suggestive of a negative spectral index, characteristic of background
extragalactic sources. Unfortunately, sensitive observations reaching the
position of the ammonia maximum are not available.

\subsection{L1551 NE}\label{l1551}

L1551 NE is a young stellar object in the L1551 molecular cloud. It is located
very close ($\sim 2\farcm5$) to the well-studied embedded source L1551 IRS 5.
The proximity to the red lobe of the large IRS 5 outflow has made difficult the
study of
the L1551 NE outflow itself. Moriarty-Schieven, Buttner \& Wannier (1995)
suggest the presence 
of a weak molecular outflow from this source and Devine, Reipurth \& Bally
(1999)
concluded that L1551 NE drives an HH flow (HH 454) and that probably drives the
objects HH 28 and 29, that were previously attributed to IRS 5.

We detected intense ammonia emission toward the position of L1551 NE (see
spectrum in Fig.~\ref{esp11}), but the proximity to IRS 5, which is associated
with a strong \nh\ condensation (Torrelles et al.\ 1983) makes it difficult to
separate both components. Higher angular resolution observations of high
density tracers are needed to detect the structure of dense gas around
L1551 NE.

\subsection{L1634}

\begin{figure}
\begin{center}
\resizebox{8cm}{!}{\rotatebox{-90}{\includegraphics[67,35][562,685]{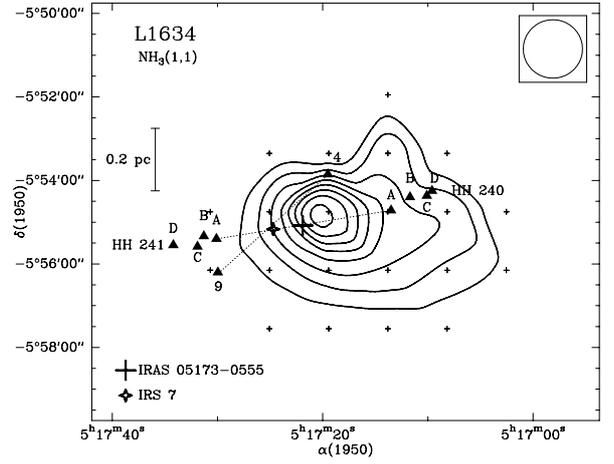}}}
    \caption{Same as Fig.~\ref{m120n}, for the L1634 region. The \nh\
     lowest contour is 0.3 K and the increment is 0.15 K. Dashed lines
     indicate outflow axes discussed in text. IRAS 05173--0555
     is the proposed exciting source for the HH240-241 outflow and the
     source IRS 7 is the proposed exciting source for the second
     outflow (knots 4 and 9)}
  \label{l1634}
  \end{center}
\end{figure}

L1634 contains two \hh\ bipolar jets (Hoddap \& Ladd 1995). One of them, HH
240-241, is constituted by several \hh\ knots symmetrically located from the
source IRAS 05173$-$0555, which has been proposed as the driving source (Hoddap
\& Ladd 1995; Davis et al.\ 1997). The jet extends from east (knots HH
241A-D) to west (knots HH 240A-D). Knots HH 240A and HH 241A were
previously known as RNO 40 and RNO 40E (Jones et al.\ 1984).
CO$(J=3\rightarrow2)$ observations (Davis et al.\ 1997) reveal the presence 
of a molecular outflow
associated with IRAS 05173$-$0555, a cm and a submm source, proposed as a Class
0 
(Beltr\'an et al.\ 2002). The second bipolar jet only has two knotty bow shocks 
(knots 9 and 4; Hoddap \& Ladd 1995;
see Fig.~\ref{l1634}). The near-IR source IRS 7 located near the center
of the jet, has been proposed as the powering source of this outflow (Hoddap \&
Ladd 1995; Davis et al.\ 1997). The CO outflow (Lee et al.\ 2000) shows a
distribution 
similar to that of the \hh\ jets. 

Our map (Fig.~\ref{l1634}) shows that both, IRAS 05173-0555 and
IRS 7 are associated with dense gas. The \nh\ emission peak is located close
($\sim 0\farcm7 \simeq 0.09 $ pc) to the position of the IRAS source.  This
source has a steeply rising spectral energy distribution through the IRAS bands
and it is detected at cm, mm, and submm wavelengths (see references in
Table~\ref{information}). These results together with its association with the
ammonia core suggest that IRAS 05173$-$0555 is a very young object deeply
embedded in the high density gas and that it is a very good candidate for
exciting the optical jet HH240-241 and the molecular outflow. The near-IR
source IRS 7 is displaced by $\sim 1\farcm4$ ($\sim 0.19$ pc) to the SE of the
position of the emission maximum. The association of IRS 7 with high density
gas suggests that it is a very young stellar object as suggested by Beltr\'an et
al.\ (2002).

We noted that the central velocity of the ammonia lines increases to the west
of the peak position. We found a velocity shift of $\sim 0.31$ \kms\ between
the ammonia peak position and the position of knot HH 240 A. This knot has a
large
proper motion away from the IRAS source (Jones et al.\ 1984). The velocity
shift detected could results from the interaction between the jet and the dense
gas. Indeed, Whyatt et al.\ (2009) find that the HH objects are illuminating the
molecular gas, enhancing the emission of the \hco\ associated with a dense
molecular condensation within the ammonia core.

\subsection{IRAS 05358+3543}

IRAS 05358+3543 was proposed as the exciting source of a bipolar molecular
outflow (Snell et al.\ 1990). Observations at higher angular resolution resolved
the outflow into at least three different outflows, two of them forming a
quadrupolar system (Beuther et al.\ 2002). Tofani et al.\ (1995) detected four
\hho\ maser
spots close to the IRAS position. Millimeter and submillimeter emission around
the IRAS source resolved at least four cores in the region within separations
between 4''-6''. At least two of these mm cores will be likely the exciting
source of the molecular outflows (Beuther et al.\ 2007, Leurini et al.\ 2007). 

\begin{figure}
\begin{center}
\resizebox{8cm}{!}{\includegraphics[69,291][497,655]{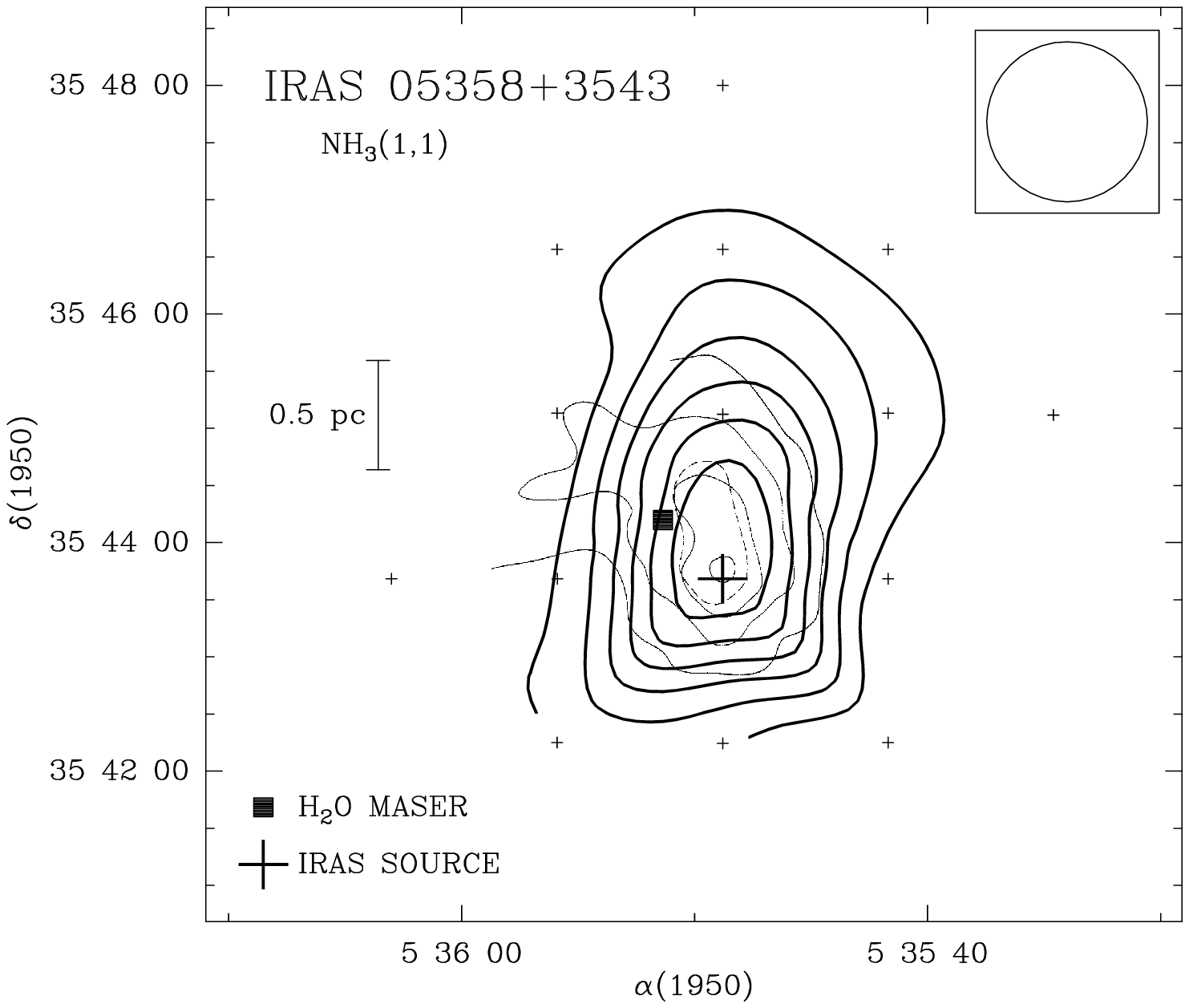}}
    \caption{Same as Fig.~\ref{m120n}, for the region around IRAS 05358+3543. 
     The \nh\ lowest contour is 0.3 K and the increment is 0.2 K. The CO 
     bipolar outflow is from Snell et al.\ (1990)}
  \label{05358}
  \end{center}
\end{figure}

We found an \nh\ condensation elongated in the north-south direction
(Fig.~\ref{05358}). The \nh\ emission peaks at the position of the IRAS source.
This positional coincidence, as well as its proximity to \hho\ maser
emission, along with the fact that its infrared emission increases steeply
toward longer wavelengths, suggest that IRAS 05358+3543 is a very young
stellar object, deeply embedded in the high density gas, favoring this object
as the driving source of the molecular outflow. Our angular resolution doesn't
allow us to infer about the subcores into the region.

\subsection{L1641-S3}\label{l1641s3}

L1641-S3 is a bipolar CO outflow located in the southern part of the L1641
cloud (Fukui et al.\ 1989, Wilking et al.\ 1990, Morgan et al.\ 1991). The
outflow is centered on the source IRAS 05375$-$0731 (= FIRSSE-101), which has
been proposed as its exciting source. The source has been detected in 
the near-IR, centimeter, millimeter and submillimeter wavelengths ranges  
with an spectral energy distribution of a Class I source (see references in
Table~\ref{information}). \hho\ maser emission (Wouterloot \& Walmsley 1986) has
been detected toward the IRAS source. An \hh\ giant flow is found probably
associated with the IRAS source (Stanke et al.\ 2000) 

\begin{figure}
\begin{center}
  \resizebox{8cm}{!}{\includegraphics[15,55][487,482]{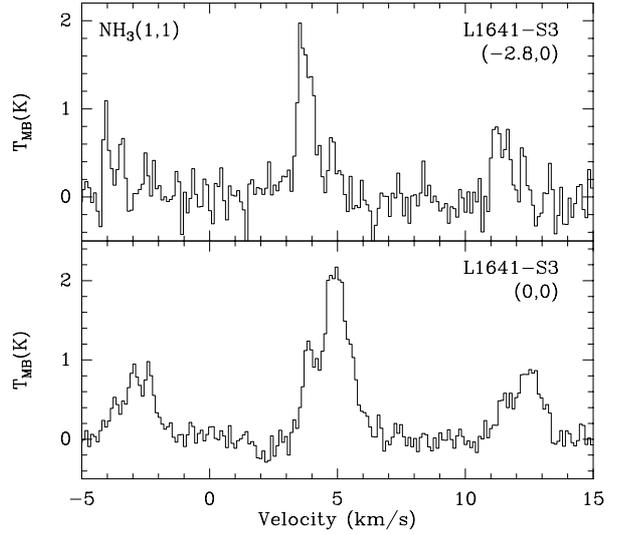}} 
  \caption[]{Spectra of the \nh(1,1) emission at $(0,0)$ and (-2.8,0), the
  positions of the emission maximum for each velocity component of L1641-S3.
  Offsets are with respect to the position given in Table~\ref{regions}.}
  \label{espectros}
  \end{center}
\end{figure}

In all the positions where emission is detected, the \nh\ spectra show two
velocity 
components at 3.8 and 4.9 \kms, in all the hyperfine lines. Each velocity
component
peaks at a different position. In Fig.~\ref{espectros} we show the observed
spectra at the position of the emission peak for each velocity component.

\begin{figure}
\begin{center}
\resizebox{8cm}{!}{\includegraphics{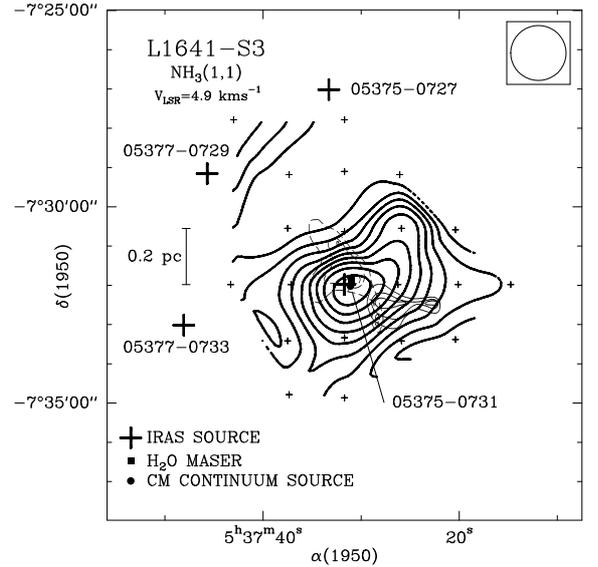}}
    \caption{Same as Fig.\ref{m120n}, but for the emission associated with the
    component at $\sim 4.9$ \kms\ for the L1641-S3 region. The
     \nh\ lowest contour is 0.3 K and the increment is 0.2 K. The map of
     the CO bipolar outflow is from Morgan et al.\ (1991)}
 \label{velog}
 \end{center}
\end{figure}

\begin{figure}
\begin{center}
\resizebox{8cm}{!}{\includegraphics[80,268][499,670]{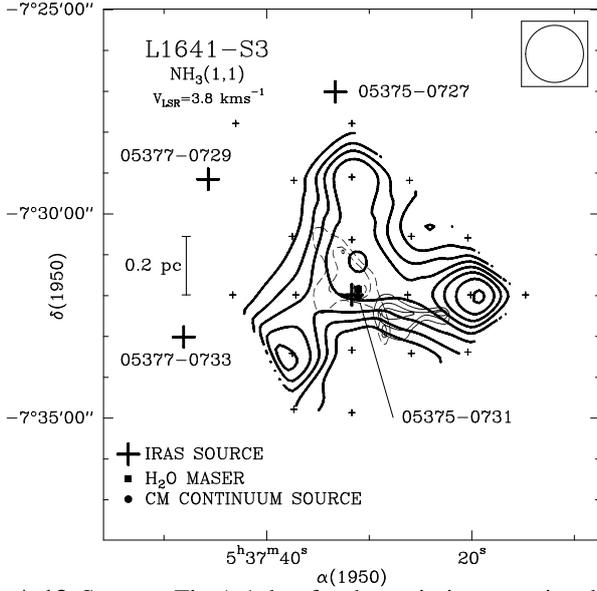}}
    \caption{Same as Fig.\ref{m120n}, but for the emission associated with
    component at $\sim 3.8$ \kms\ for the L1641-S3 region. The
     \nh\ lowest contour is 0.4 K and the increment is 0.2 K. The map of
     the CO bipolar outflow is from Morgan et al.\ (1991)}
 \label{velop}
 \end{center}
\end{figure}

 The region has also been mapped
in \nh\ with an angular resolution of $40''$ by Harju et al.\ (1993). 
These authors present a map of the overall emission, which is consistent with
our results, taken into account the difference in the beam 
sizes and the slight difference in the region covered by the maps. However 
these authors do not discuss the presence of  
two velocity components. Since the two velocity components are clearly defined
in 
our spectra, in Figs.~\ref{velog} and \ref{velop} 
we present separate maps of the two velocity components, and
in our analysis we will discuss separately each velocity component.

The map of the component at 4.9 \kms\ (Fig.~\ref{velog}) reveals a well defined
\nh\ condensation with the position of source IRAS 05375-0731 well centered in
the structure and coinciding with the maximum of emission. The main axis of this
condensation is elongated roughly in the NW-SE direction, perpendicular to the
outflow axis. We also detected \nh\ emission toward the NE of the region
mapped, near the positions of IRAS 05377-0729 and IRAS 05375-0727, suggesting
that
these sources may be also associated with high density gas. However, none of the
two sources is detected at submillimeter wavelengths (Dent et al.\ 1998) and
only IRAS 05375-0727 has a near-IR
counterpart (Strom et al.\ 1989). Unfortunately, our map is not completed around
the positions of these sources, so that we cannot establish their association
with high
density gas.

The map of the component at 3.8 \kms\ is shown in Fig.~\ref{velop}. The spatial
distribution presents an irregular morphology, with several local maxima. The
source IRAS 05375-0731 appears projected toward this structure, but it is not
as clearly associated with any particular feature.  

In summary, we observe that for the 4.9 \kms\ component, the position
of the source IRAS 05375-0731 is better centered on the \nh\ structure and
closer to the emission maximum  than for the 3.8 \kms\ component. In
addition, the main axis of the 4.9 \kms\ structure is aligned roughly
perpendicular to the outflow axis. Finally, the line widths of the 4.9 \kms\ 
component are broad (see Table \ref{line}), suggesting that the dense gas is
suffering a 
perturbation by an embedded object, while the line widths of the 3.8 \kms\
component are narrow, suggestive of a starless core. 
From these results, we conclude that the source IRAS 05375-0731 is likely
associated with the dense gas component at 4.9 \kms. A study of the local
heating through a high angular resolution mapping of the \nh(1,1) and \nh(2,2)
could confirm this association.

\subsection{CB 34}

The small Bok globule CB 34 (Clemens \& Barvainis 1988) is associated with the
source 
IRAS 05440+2059, which is the proposed driving source of a bipolar
molecular outflow (Yun \& Clemens 1994a). IRAS 05440+2059 has near-IR,
submillimeter, 
millimeter and centimeter counterparts (see references in
Table~\ref{information}). 
Near-IR images revealed a small aggregate of YSOs embedded in the cloud 
(Alves \& Yun 1995). Alves (1995) discovered, from optical and near-IR images, a
variable
object CB34V, which is identified as an embedded PMS object (Alves et
al.\ 1997). 

Moreira \& Yun (1995) discovered in this region four Herbig-Haro objects (HH
290S,
HH 290 N1, HH 290 N2 and HH 291) and several \hh\ structures (labeled Q1, Q2,
Q3, Q4, {\it hh291X} and {\it hh291Y}) . These authors 
suggested that the objects HH 290S/N1/N2 constitute an optical jet driven by an
embedded near-IR source HH 290 IRS, that the structure Q1-Q4 is a well
collimated \hh\ 
jet driven by an embedded object (labeled Q), and that
{\it hh291X}, {\it hh291Y} and HH 291 could be bright knots of an embedded
jet, whose driving source remains undetected.

\begin{figure}
\begin{center}
\resizebox{8cm}{!}{\includegraphics[50,257][498,655]{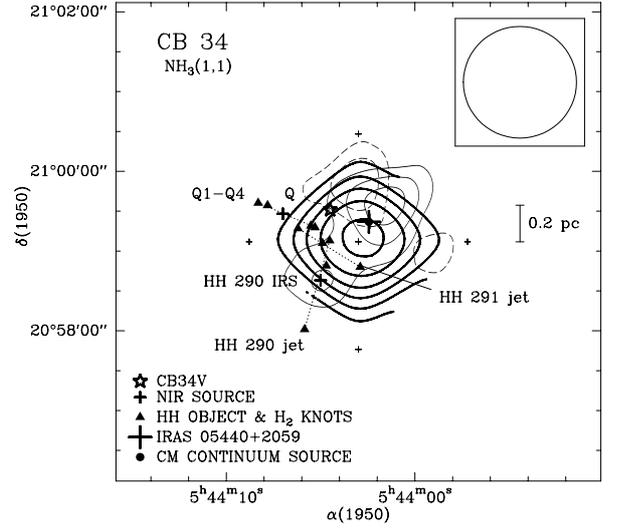}}
     \caption{Same as Fig.\ref{m120n}, but for the CB 34 region. The \nh\
      lowest contour is 0.15 K and the increment is 0.1 K. The three
      jets discussed in text are indicated by dotted lines. The map 
      of the CO bipolar outflow is from Yun \& Clemens (1994a).}
   \label{cb34}
   \end{center}
\end{figure}

The ammonia structure, unresolved with our beam, peaks close to the position of
IRAS 05440+2059 (see Fig.~\ref{cb34}), in good agreement with the results
obtained from
other high-density tracers (CS, Launhardt et al.\ 1998; HCN, Afonso et al.\
1998; \nh, Codella \& Scappini 1998). The sources HH 290 IRS and CB34V appear
in projection toward the ammonia core, suggesting that they are young stellar
objects embedded in the dense molecular gas. These results support the
identification of IRAS 05440+2059 as the exciting source of the molecular
outflow, and HH290 IRS as the exciting source of an optical jet.  The source Q
lies at the edge of the condensation.
The HH 291 jet, two of whose knots are only
detected in the near-IR, also appears projected toward the high density gas,
so 
it can be tracing an embedded jet as proposed by Moreira \& Yun (1995).
We suggest that the exciting source of this jet could be
located in the line connecting the knots and close to the position of the \nh\ 
emission maximum. High-resolution observations could reveal the position of
this embedded object.

\subsection{L1617}

\begin{figure}
\begin{center}
\resizebox{8cm}{!}{\includegraphics[60,273][504,634]{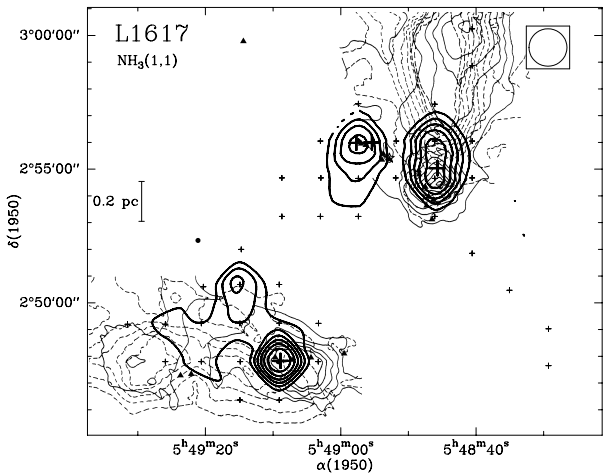}}
    \caption{Ammonia cores in L1617 (thick contours) overlapped on the CO
    outflow maps of Reipurth \& Olberg (1991) (thin contours). 
    Symbols used in this map are same as Fig.~\ref{m120n}.
    Close-ups of the clumps of the NW region (associated with HH 270/110)
    and SE region (associated with HH 111) are shown in Fig.~\ref{hh110} and 
    Fig.~\ref{hh111} respectively. Ammonia contour levels are the same as in
    these figures. Additional positions were observed near
    HH 113 ($\sim 12'$ east from HH111).}
  \label{l1617}
  \end{center}
\end{figure}

A map of the overall L1617 region where \nh\ emission is detected is shown in
Fig.~\ref{l1617}. The map encloses the ammonia condensations associated with HH
270/110 and with HH 111, as well as their molecular outflows.

\subsubsection{HH 270/110}

HH 110 is a well collimated jet located in the L1617 molecular cloud (Reipurth
\& Olberg 1991). Reipurth et al.\ (1996) discovered a second jet, HH
270, $\sim 3'$ to the NE of HH 110 and proposed the near-IR source HH 270 IRS as
its exciting source. The position of HH 270 IRS lies very close to the error
ellipsoid of IRAS 05489+0256 and both sources could be associated. VLA
observations of HH 270 IRS at cm wavelengths 
(Rodr\'{\i}guez et al.\ 1998) revealed that this source (named VLA1) is
elongated
along the axis of the HH 270 jet, suggesting that it traces the base of the
flow. Reipurth et al.\ (1996) suggest that the HH 270 jet suffers a grazing
collision 
with a nearby molecular cloud core, thus producing a deflected flow, 
which is manifested as the HH 110 jet. The kinematical studies (Riera et al.\
2003; L\'opez et al.\ 2005) provide some additional evidence of the interaction
between the outflow and the surrounding material. 

About $3'$ to the SW of HH 270 IRS, lies IRAS 05487+0255, a source with a
spectral energy distribution steeply rising toward longer wavelengths. IRAS
05487+0255 is associated with a near-IR  and a VLA source (Davis et al.\ 1994;
Garnavich et al.\ 1997; Rodr\'{\i}guez et al.\ 1998), driving a bipolar
molecular outflow  (Reipurth
\& Olberg 1991; see Fig.~\ref{l1617}) and a \hh\ jet (Davis et al.\ 1994;
Garnavich et al.\ 1997) running almost north-south. A second near-IR source,
powering another \hh\ jet extending in the north-south direction, is found a few
arcseconds to the south (Davis et al.\ 1994; Garnavich et al.\ 1997).

\begin{figure}
\begin{center}
\resizebox{8cm}{!}{\rotatebox{-90}{\includegraphics[51,35][556,620]{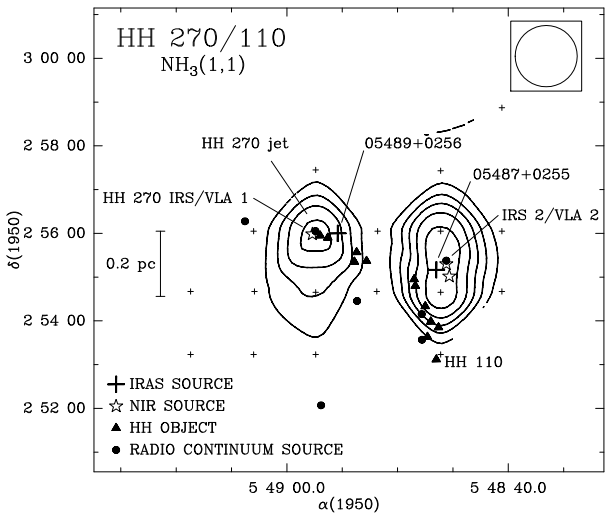}}}
    \caption{Same as Fig.~\ref{m120n}, for the HH 270/110 region. The \nh\
     lowest contour level is 0.2 K and the increment is 0.15 K. The sources
     discussed in text are indicated.}
  \label{hh110}
  \end{center}
\end{figure}

Our ammonia map (Fig.~\ref{hh110}) shows two high density clumps, separated by
$\sim 0.4$ pc, apparently corresponding with two local
maxima observed in the \tco\ extended structure mapped by Reipurth, Raga \&
Heathcote
(1996). The emission of the eastern \nh\ clump peaks at the position of  HH 270
IRS/VLA~1, suggesting that this object is embedded in the high density gas,  and
giving
support to its identification as the powering source of the HH 270 jet.   

The HH 110 flow is observed toward the SE edge of the western ammonia clump
(see Fig.~\ref{hh110}). The coincidence of the \nh\ clump at the point where the
HH 270/110 flow changes abruptly its direction, gives strong support to the
scenario proposed by Reipurth et al.\ (1996), where the HH 110 jet arises as a
result of the deflection of the HH 270 jet after a collision with a high-density
clump. Our \nh\ observations provide evidence for the presence of such a
high-density clump.  

The western \nh\ clump peaks near the positions the proposed exciting sources of
the molecular outflow and the \hh\ jets (see Fig.~\ref{hh110}). This result
suggest that these sources are  embedded objects, giving
support to their identification as the driving sources of the molecular
outflow and the \hh\ jets. 

The remaining five centimeter continuum sources detected in the region by
Rodr\'{\i}guez
et al.\ (1998) (see Fig.~\ref{hh110}) have negative spectral index,
characteristic of non-thermal emission. One of them (VLA 4) could be associated
with the knot HH 110 H and the others are probably background objects unrelated
with the star-forming region.

\subsubsection{HH 111}

\begin{figure}
\begin{center}
\resizebox{8cm}{!}{\rotatebox{-90}{\includegraphics[51,35][556,620]{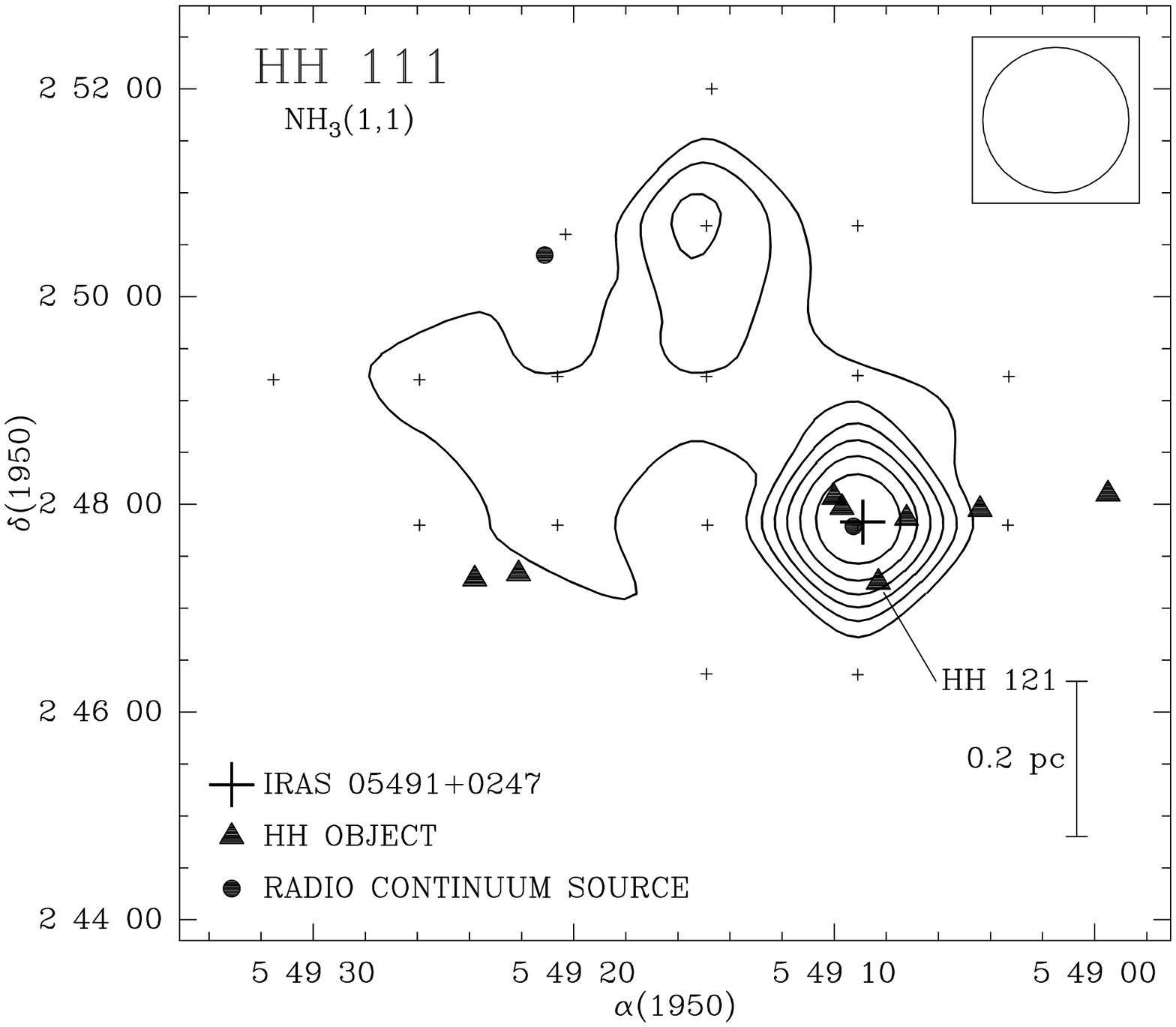}}}
    \caption{Same as Fig.~\ref{m120n}, for the HH 111 region. The \nh\
     lowest contour level is 0.15 K and the increment is 0.07 K.}
  \label{hh111}
  \end{center}
\end{figure}

HH 111, in the L1617 cloud, is a well collimated jet associated with a highly
collimated molecular
outflow apparently driven by IRAS 05491+0247 (Reipurth \& Oldberg 1991; see
Fig.~\ref{l1617}), which also has a centimeter counterpart, VLA~1 (Rodr\'{\i}guez
\& Reipurth 1994, Anglada et al.\ 1998b). Reipurth, Bally \& Devine (1997)
proposed that the HH 111 jet (together with HH 113 and HH 311) constitutes a
giant flow with a total extent of $7.7$ pc. Gredel \& Reipurth (1993) detected
an \hh\ bipolar jet, HH 121, which is almost perpendicular 
to HH 111 and appears to emanate from the IRAS/VLA 1 source. Cernicharo \&
Reipurth (1996) resolved the CO outflow into a quadrupolar structure along the
axes of the HH111 and HH 121 jets. At smaller scales, the source VLA 1 also
shows evidence of a similar quadrupolar structure (Reipurth et al.\ 1999). These
authors detected an additional centimeter source, VLA 2 ($\sim 3''$ NW of VLA
1), which exhibits some evidence of driving its own outflow. 

The \nh\ map (Fig.~\ref{hh111}) shows a condensation with the emission peaking
near the positions of the proposed triple system. The spectral energy
distribution of
the IRAS source is steeply rising toward longer wavelengths. Altogether this
suggests that the sources are deeply embedded in the high density gas.

The remaining radio continuum sources detected in the region (Anglada et al.\
1998b) are not associated with dense gas and have negative spectral index, 
indicating that probably almost all are non-thermal background sources
unrelated to the star-forming region.

\subsubsection{HH 113}

We observed a five-point grid around HH 113 (not shown in Fig.\ref{l1617}),
which is located $\sim 12'$ to the east of the HH 111 complex. We did not
detect significant emission in any of these positions. Reipurth, Bally \&
Reipurth (1997) suggest that HH 113 is the eastern boundary of the HH 111
complex. The lack of dense gas around this object, and that
there are no sources in its vicinity suggests a non local origin for this
object, giving support to its identification as part of the HH 111 complex.

\subsection{IRAS 05490+2658}

\begin{figure}
\begin{center}
\resizebox{8cm}{!}{\includegraphics[91,314][478,684]{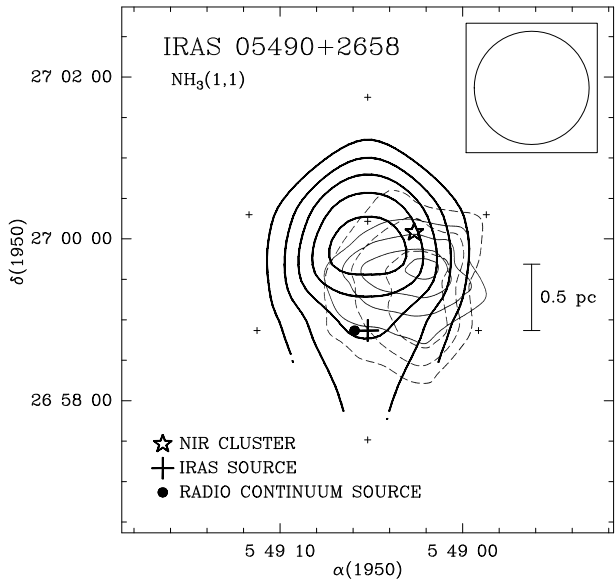}}
     \caption{Same as Fig.~\ref{m120n}, for the region around IRAS
      05490+2658. The \nh\ lowest contour is 0.2 K and the increment is
      0.1 K. The center of the near-IR cluster (Carpenter et al.\ 1993), 
      which extends over a region of 1 pc in size, is indicated. The CO bipolar 
      outflow is from Snell et al.\ (1990)}
  \label{i05490}
  \end{center}
\end{figure}

IRAS 05490+2658 lies $\sim 5'$ east of the \hii\ region S242.  This IRAS source
has been proposed as the exciting source of a poorly collimated molecular
outflow (Snell et al.\ 1990), although it is displaced $\sim 1'$ to the
SE of the geometrical center of the outflow. A 6 cm radio continuum source
has been detected close to the position of the IRAS source (Carpenter et al.\
1990). A near-IR cluster, extending over a region of $\sim 1$ pc in size, has
been detected in the region by Carpenter et al.\ (1993). 

The condensation we mapped in \nh\ (Fig.~\ref{i05490}) has the emission
peak displaced $\sim 1'$ ($\sim 0.7$ pc) to the north of the IRAS source
position, but it is very close to the center of the outflow 
and to the center of the near-IR cluster. This result suggests that some cluster
members 
could be embedded stellar objects, in agreement with the Carpenter et al.\
(1993) suggestion, and 
that the outflow exciting source could be located close to the \nh\ maximum and 
to the north of the IRAS position. Sensitive cm continuum observations toward
this position 
could reveal this object. 
  
\subsection{CB 54}

CB 54 is a Bok globule associated with the source IRAS 07020-1618, which has
been proposed as the exciting source of a highly collimated bipolar molecular
outflow (Yun \& Clemens 1994a). The IRAS source is double in the near-IR (two
components CB54YC1-I and CB54YC1-II separated by $12''$:Yun \& Clemens 1994b)
and it is detected also 
at cm (Yun et al.\ 1996; Moreira et al.\ 1997) and mm wavelengths (Launhardt \&
Henning 1997). Only CB54YC1-II is detected in the mid-infrared images, but three
new mid-infrared sources with no near-infrared counterpart were detected
spatially coincident with both the IRAS source and the center of the dense core
(Ciardi \& G\'omez Mart\'{i}n 2007). 

\begin{figure}
\begin{center}
\resizebox{8cm}{!}{\includegraphics[52,274][496,645]{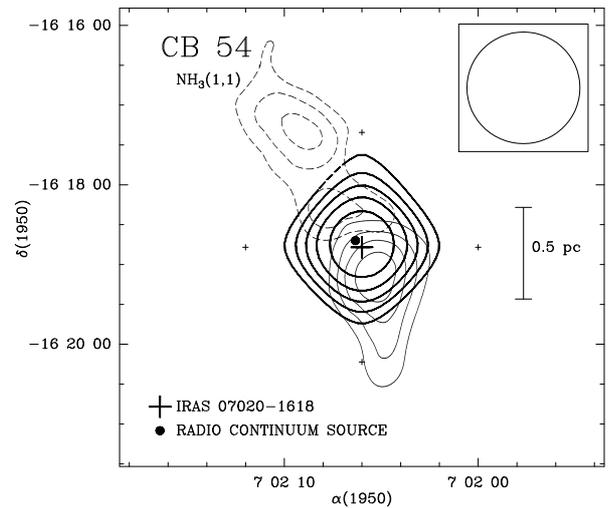}}
     \caption{Same as Fig.~\ref{m120n}, for the CB 54 region. The \nh\
      lowest contour is 0.25 K and the increment is 0.1 K. The CO bipolar
      outflow is from Yun \& Clemens (1994a)}
  \label{cb54}
  \end{center}
\end{figure}

We found a compact \nh\ condensation (Fig.~\ref{cb54}) with the emission
peaking at the position  of the IRAS source, which has a cm radio continuum counterpart. The spectral energy distribution of the IRAS souce is steadily rising at longer wavelengths. Altogether this suggests
that the IRAS source is a very young object deeply embedded in the high density
gas and favors it as the exciting source.

\subsection{L379}

The dark cloud L379 contains the bright source IRAS~18265-1517, which was 
proposed as the exciting source of a bipolar molecular outflow (Hilton et al.\
1986; Wilking et al.\ 1990).  The red- and blue-wing emission overlap for most
of the outflow extension, but the emission maxima are not coincident. This
structure has been interpreted as two outflows centered north and south of the
IRAS source (Kelly \& McDonald 1996). Observations at mm and submm wavelengths
have revealed two distinct clumps of dust continuum emission located several
arcsecs northwest and southwest, respectively from the IRAS nominal
position. This interpretation is supported by the two velocity components found
in \cdo\ spectra
(McCutcheon et al.\ 1995; Kelly \& Mcdonald 1996). Kelly \& McDonald
(1996) suggest that the dust clumps probably contain the driving sources of
the molecular outflows.

\begin{figure}
\begin{center}
\resizebox{8cm}{!}{\includegraphics[59,292][495,643]{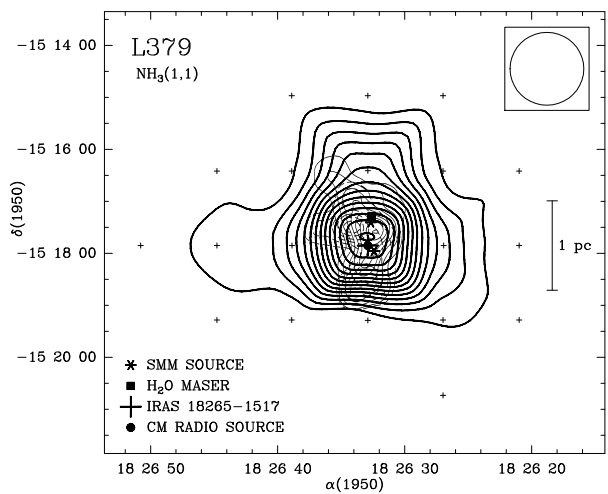}}
    \caption{Same as Fig.~\ref{m120n}, for the L379 region. The \nh\
     lowest contour level is 0.3 K and the increment is 0.2 K. The position of
     the two dust clumps is indicated by an asterisk. The map of
     the CO bipolar outflow is from Kelly \& Macdonald (1996).}
  \label{l379}
  \end{center}
\end{figure}

We found an \nh\ condensation (Fig.~\ref{l379}) with the emission peaking at
the position of IRAS~18265-1517. 
The spectral energy distribution of this IRAS
source rises steeply at longer wavelengths. Altogether this suggests that the IRAS
source is a deeply embedded object. Although our ammonia lines are broad, 
a hint of two velocity components can be appreciated in the satellite lines 
(see Fig.~\ref{esp11}), in agreement with the \cdo\ results (Kelly \& McDonald
1996).
Both submm sources appear to be associated
with \nh\ emission, but our angular resolution does not allow us to favor one of
them in terms of the proximity to the \nh\ maximum.

 The physical parameters we obtained for this region
(Table~\ref{parameters}) indicate that L379 is a massive region ($M\sim
2\,000-3\,700$ \mo). The estimated luminosity of the IRAS source is $L_{\rm
bol}\sim 1.6\times10^{4}$ \lo\ (Kelly \& McDonald 1996). The high mass
obtained from \nh, together with the high luminosity of the source could
indicate that
this source is a massive protostellar object, and thus that L379 is a high-mass
star-forming region.

\subsection{L588}

Reipurth \& Eiroa (1992) discovered two isolated Herbig-Haro objects, HH 108
and HH 109, in this region and proposed IRAS 18331-0035 as their driving source.
Ziener \& Eisl\"offel (1999) found that both HH objects
consist of several bright knots, some of them with \hh\ counterpart. Chini et
al.\ (1997) found two 1.3 mm sources, the stronger is coincident with the IRAS
source and the fainter, HH 108 MMS, has no counterpart.  At present it is
unclear which one of these two sources is the driving source of the HH objects.
Parker et al.\ (1991) detected broad line wings in CO spectra taken toward the
IRAS position. 

\begin{figure}
\begin{center}
\resizebox{8cm}{!}{\rotatebox{-90}{\includegraphics[50,33][547,547]{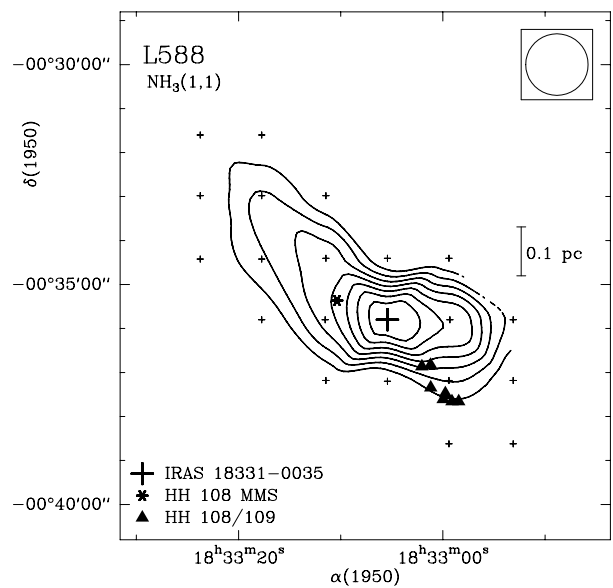}}} 
    \caption[]{Same as Fig.~\ref{m120n}, for the L588 region. The \nh\
     lowest contour is 0.3 K and the increment is 0.15 K. The position of
     the HH 108 and HH 109 knots are from Ziener \& Eisl\"offel (1999).}
  \label{l588}
  \end{center}
\end{figure}

The \nh\ map (Fig.~\ref{l588}) shows a condensation elongated in the NE-SW
direction, similarly to the molecular cloud mapped in
CO by Parker et al.\ (1991). The ammonia emission maximum is located at the
position of IRAS 18331-0035. This result, along with the spectral energy
distribution,
suggests that the IRAS source is a very young object deeply
embedded in the high-density gas. Although the source HH 108 MMS is located
inside the \nh\ condensation, it is displaced $\sim1\farcm2$ ($\sim 0.1$ pc) to
the NE of the emission peak.  Owing to its association with the \nh\ emission
peak, it appears that the source IRAS 18331-0035 is the deepest embedded
object and constitutes a very good candidate for the energy source of the HH
complex.

From our data we found that the mass of this region exceeds the virial mass by
more than a factor of five (see Table~\ref{parameters}). This result could
indicate that the cloud is in process of gravitational collapse.

\subsection{L673}

Armstrong \& Winnewisser (1989) detected an extended bipolar molecular outflow
in this region and proposed the source IRAS 19180+1116, which coincides with
the object RNO 109 (Cohen 1980), as its driving source. In a
previous work (see Paper I) we have observed in \nh\ a region of
$\sim10'\times7'$ around the object RNO 109.  Those observations revealed  an
ammonia structure of $\sim5'\times2'$ elongated from northwest to southeast. It 
consists of three subcondensations peaking at the positions of sources IRAS
19180+1116 (RNO 109), IRAS 19180+1114 and IRAS 19181+1114. The source IRAS
19180+1114 is located close to the strongest emission maximum. 

The region was mapped in CS by Morata et al.\ (1997).  The CS emitting region
is elongated in the NW-SE direction and is more extended
($16\farcm6\times7\farcm6$) than the region mapped in ammonia in Paper I. 
This extended CS emission encompasses the \nh\ condensation, and has the
emission peak 
displaced $\sim8'$ to the south-east of IRAS 19180+1114. 

\begin{figure}
\begin{center}
 \resizebox{8cm}{!}{\includegraphics[68,282][478,666]{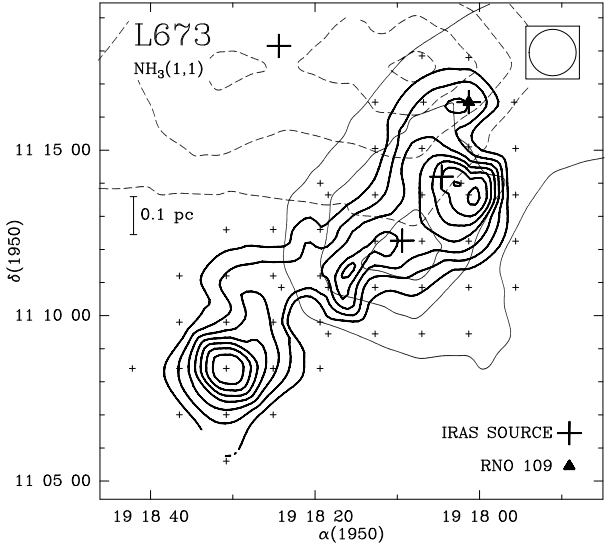}}
  \caption{Same as Fig.\ref{m120n}, but for the L673 region. The \nh\ lowest
   contour level is 0.45 K and the increment is 0.3 K. The \nh\ map obtained in
   previous observations is also included.  The IRAS sources
   associated with the ammonia structure are (from north to south) IRAS
   19180+1116(=RNO 109), IRAS 19180+1114 and  IRAS 19184+1118. The map of the CO
   molecular outflow is from Armstrong \& Winnewisser (1989)} 
\label{l673tot}
\end{center}
\end{figure}

To complete the study in \nh\ of this region, we carried out new observations,
covering the  region around the CS emission maximum.  In Fig.~\ref{l673tot} we
show the complete \nh\ map of
the region (including the data from Paper I). The new observations reveal that
the \nh\ emission further extends to the SE, where we found the strongest \nh\
maximum of the whole region. Up to now, no source has been found toward this 
position. Morata, Girart \& Estalella (2003, 2005) have found that this core
splits in multiple condensations, with no signs of star formation. Most of the
condensations are transient, in the sense that they are not gravitationally bound.

\subsection{IRAS 20050+2720}

Bachiller et al.\ (1995) mapped a molecular outflow consisting of three pairs
of lobes emanating from the vicinity of IRAS 20050+2720, suggesting that two or
three independent outflows are driven by young sources embedded in the
core. Chen et al.\ (1997) found a cluster of near-IR sources with three
subclusters, 
two of them are associated with IRAS 20050+2720 and IRAS 20049+2721,
respectively.
IRAS 20050+2720 was resolved at centimeter wavelengths in two components at
subarsec scale  
(Anglada et al.\ 1998a) and in four sources at millimeter wavelengths. Two of
them, 
which are separated $\sim 20''$, are suggested to be protostellar collapse
candidates
(Choi et al.\ 1999, Beltr\'an et al.\ 2008). An analysis of the velocity fields
of \hho\ masers (Furuya et al.\ 2005) 
indicates that one of these millimeter sources (MM1) is driving a powerful jet. 
IRAS 20049+2721 was barely detected in the IRAS  12 and 25 \mm\ bands (2.3 Jy in
both bands), but
is very bright at longer wavelengths (flux densities are 171.2 and 397.7 Jy in
the 60 and 100 \mm\ IRAS bands, respectively). A cm continuum source (Anglada
et al.\ 1998a) is detected in association with this source.
CS observations show the emission maximum at the position of IRAS 20050+2720, 
while only weak CS emission was detected toward IRAS 20049+2721 (Bachiller et
al.\ 1995) 

\begin{figure}
\begin{center}
\resizebox{8cm}{!}{\includegraphics[66,280][475,653]{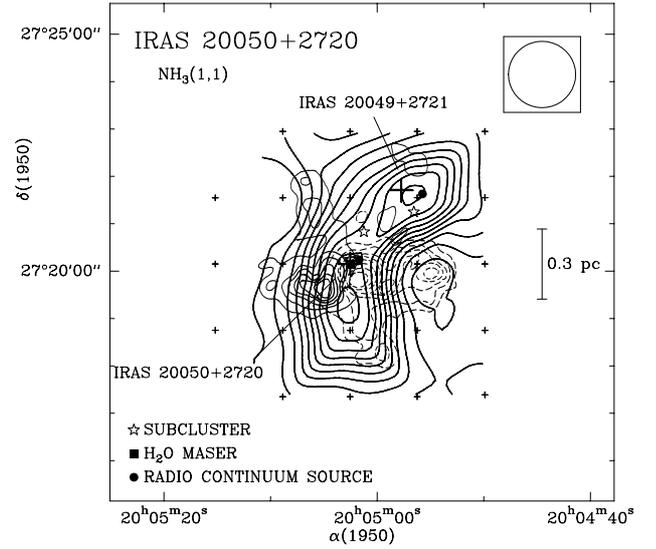}}
    \caption{Same as Fig.\ref{m120n}, but for the region around IRAS
     20050+2720. The \nh\ lowest contour level is 0.4 K and the increment
     is 0.15 K. The centers of the three near-IR subclusters found by Chen et 
     al.\ (1997) are indicated. The map of the CO multipolar outflow is from
     Bachiller et al.\ (1995).}
   \label{i20050}
   \end{center}
\end{figure}

\begin{figure}
\begin{center}
\resizebox{8cm}{!}{\rotatebox{-90}{\includegraphics{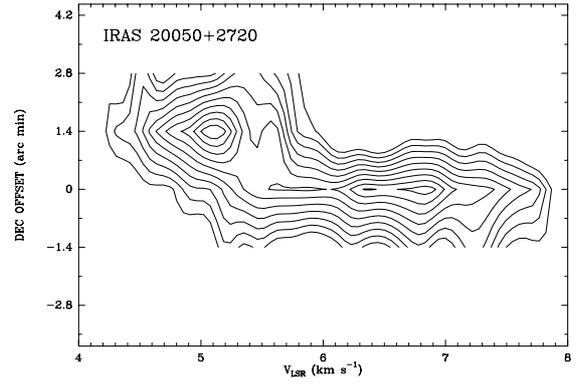}}}
  \caption{Position-velocity diagram of the \nh\ main line along the NW-SE
  direction (p.a.= $-45\degr$) centered on IRAS
  20050+2720. The $1\farcm4$ 
  offset corresponds to the position of IRAS 20049+2721. The 
  lowest contour level is 0.4 K and the increment is 0.15 K}
 \label{nw_se}
 \end{center}
\end{figure}

The \nh\ condensation (Fig.~\ref{i20050}) shows two strong emission peaks, very
close to the positions of the IRAS sources of the region, suggesting that both
sources are associated with high density gas. The velocities of the two \nh\
maxima 
differ by $\sim 2$ \kms\ (see Table~\ref{line}). In Fig.~\ref{nw_se} we show a
position-velocity diagram along 
the northwest-southeast direction. The structure of the
\nh\ emission is consistent with gravitationally bound rotational motion of two
clumps.

\subsection{V1057 Cyg}

V1057 Cyg belongs to the small group of the FU Orionis type stars. Before its
flare-up in 1970, it was a T Tauri star. A marginally resolved outflow was
reported by Levreault (1989) and Evans et al.\ (1994). They 
detected only a blue wing extending to the
north, but no contour map is shown. 

We detected very weak ammonia emission toward this source (see
Fig.~\ref{esp11} 
and Table~\ref{line})
and the line analysis was carried out by averaging several positions, so no
contour map could
be made. This weak emission indicates either a low column density gas or that
the \nh\ 
emission is very compact. 
The lack of a large amount of high-density gas agrees with
the fact that the source is optically visible.

\subsection{CB 232}

This Bok globule is associated with IRAS 21352+4307, which is proposed to be
the exciting source of a poorly collimated bipolar molecular outflow (Yun \&
Clemens 1994a).  The IRAS source has near-IR (Yun \& Clemens 1995) and 
millimeter counterparts (Launhardt \& Henning 1997), and is associated with two
compact submillimeter sources. One of them, SMM1, is
proposed as a Class 0 candidate (Huard et al.\ 1999).   

\begin{figure}
\begin{center}
\resizebox{8cm}{!}{\includegraphics[86,324][450,697]{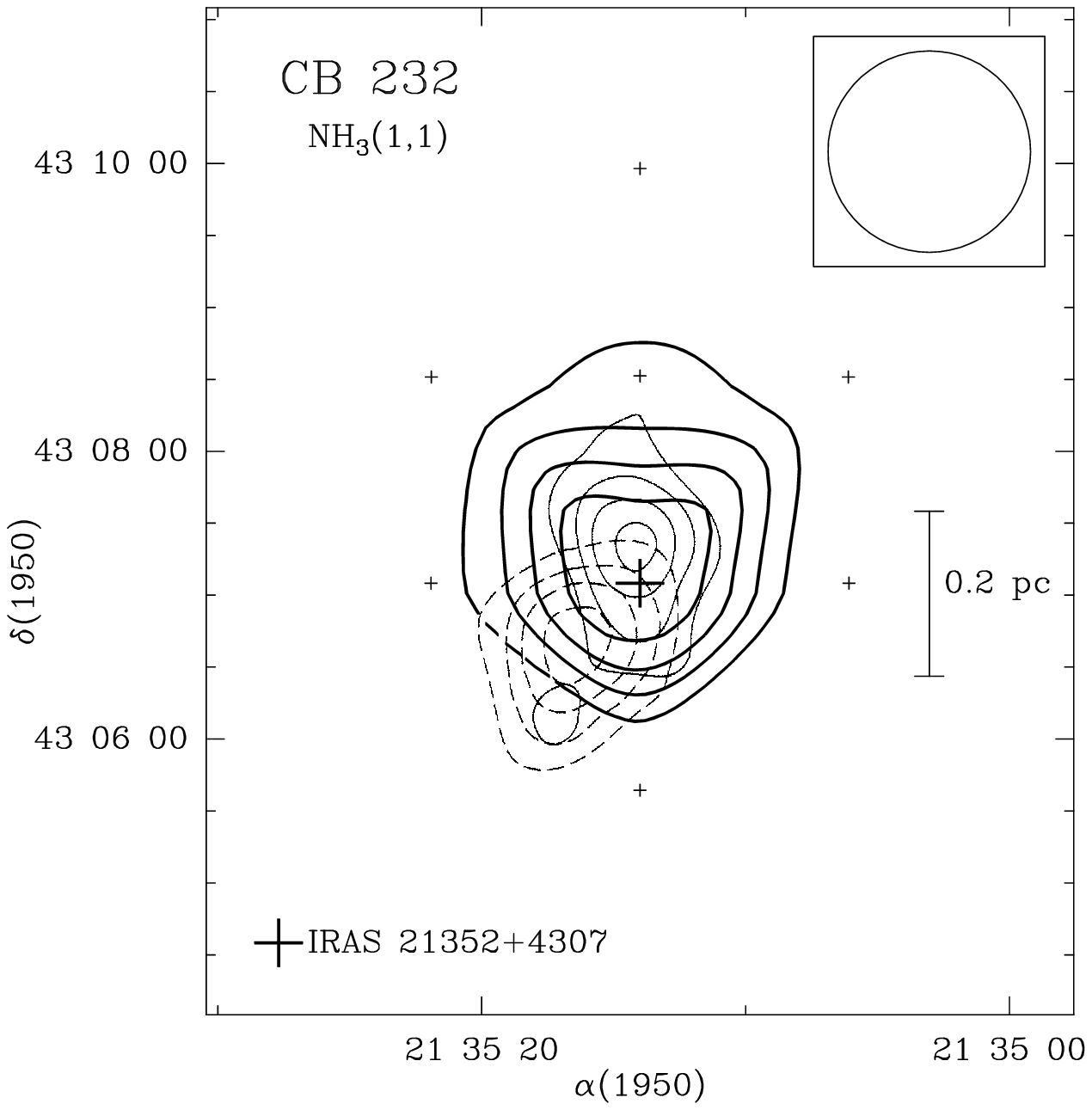}}
    \caption{Same as Fig.~\ref{m120n}, for the CB 232 region.
     The \nh\ lowest contour is 0.2 K and the increment is 0.1 K. The CO
     bipolar outflow is from Yun \& Clemens (1994a)}
  \label{cb232}
  \end{center}
\end{figure}

We have detected an ammonia condensation (see Fig.~\ref{cb232})  unresolved by
our beam, whose emission peak coincides with the position of the IRAS
source. Given that the spectral energy distribution of the IRAS source is rising
toward
longer wavelengths, our results suggest that the IRAS
source traces the location of one or several YSOs, deeply embedded in the high
density gas, and that the globule is a site of very recent star formation.

\subsection{IC 1396E}

IC 1396E is a bright-rimmed cometary globule located in the northern periphery
of the  
\hii\ region IC 1396.  Wilking et al.\ (1990) mapped a bipolar molecular
outflow and proposed IRAS 21391+5802, an intermediate-mass YSO which is found 
roughly at the center of the globule,
as the exciting source. This IRAS source is detected at near-IR, submm, mm, and
cm wavelengths
(Wilking et al.\ 1993; Beltr\'an et al.\ 2002).
The molecular outflow axis is oriented at a position angle of $75\degr$, 
which is similar to the position angle of the \hho\ maser bipolar outflow
observed at scales from 
1 to 500 AU (Patel et al.\ 2000).  

The region was mapped in \nh(1,1) and \nh(2,2) (with an angular resolution of
$40''$), 
and several other molecular lines by Serabyn et al.\ (1993). The \nh\ clump
mapped by 
these authors is elongated in the north-south direction and shows a temperature
gradient increasing 
outward from the center and reaching a maximum on the surface most directly
facing the stars
ionizing IC1396.   

\begin{figure}
\begin{center}
\resizebox{8cm}{!}{\includegraphics[55,213][501,585]{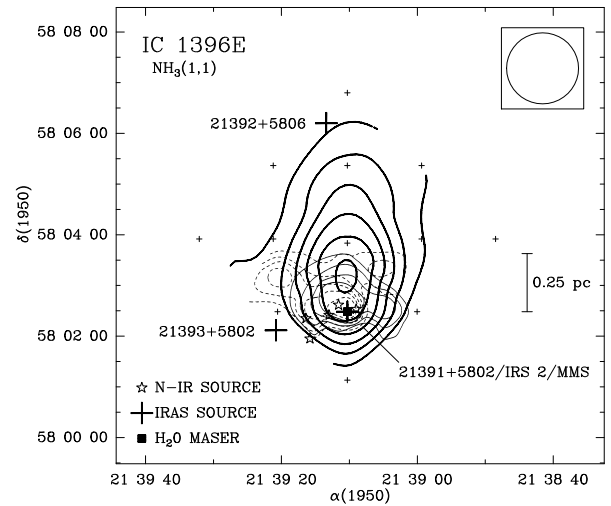}}      
     \caption{Same as Fig.\ref{m120n}, but for the IC 1396E region. The
      \nh\ lowest contour level is 0.2 K and the increment is 0.15 K. The
      map of the CO bipolar outflow is from Wilking et al.\ (1990) }
  \label{ic1396}
  \end{center}
\end{figure}

Our \nh\ map of the condensation (Fig.~\ref{ic1396}), obtained with a poorer
angular
resolution, is elongated in the north-south direction, in good agreement with
the 
one obtained by Serabyn et al.\ (1993). We found for this source a kinetic
temperature 
of $\sim19$ K, which is above the average for the sources studied in this paper.
Although 
since we only have observed a single position in the \nh(2,2) line we cannot
establish the presence
of the temperature gradient reported by Serabyn et al.\ (1993). The position of
the source 
IRAS 21391+5802 falls very close to the \nh\ emission peak. 
This positional coincidence, as well as the spectral energy distribution of the
source, suggest 
that it is a very young object, deeply embedded in the high density gas and the
best candidate
to drive the molecular outflow. There are two
other IRAS sources in the region, but they lie outside, near the edge of the
\nh\ condensation
(see Fig.~\ref{ic1396}). At present, little is known about these sources and
further studies are
required to investigate their nature and relationship with the molecular
condensation.

\subsection{L1165}

L1165 is a small cloud whose distance is not well established. Estimates by
different authors range from 200 pc to 750 pc. We will adopt a distance of 750
pc, based on the assumption that the cloud is part of the IC 1396 region
(Schwartz et al.\ 1991). Parker et al.\ (1991) discovered a bipolar molecular
outflow centered on IRAS 22051+5848, which was proposed as the exciting source.
This IRAS source, that has a near-IR counterpart (Tapia et al.\ 1997), is
located $\sim 15''$ to the NE of the reflection nebulosity GY 22 (Gyulbudaghian
1982; Reipurth et al.\ 1997). Reipurth et al.\ (1997) reported a HH object, HH
354, located $11'$ NE of the IRAS position and at the end of a cavity in the
molecular cloud, possibly excavated by the molecular outflow. These
authors proposed that HH 354, the cavity, the molecular outflow and the GY 22
nebulosity are all parts of a single giant outflow excited by the IRAS source.
From near-IR spectroscopy, Reipurth \& Aspin (1997) concluded that IRAS
22051+5848 (= HH 354 IRS) is a FUor candidate. The source IRAS 22051+5849, which
has also a near-IR counterpart (Tapia et al.\ 1997), lies $\sim1\farcm5$ north
of IRAS
22051+5848, well off the axis of the proposed giant outflow.  

\begin{figure}
\begin{center}
\resizebox{8cm}{!}{\includegraphics[84,205][501,579]{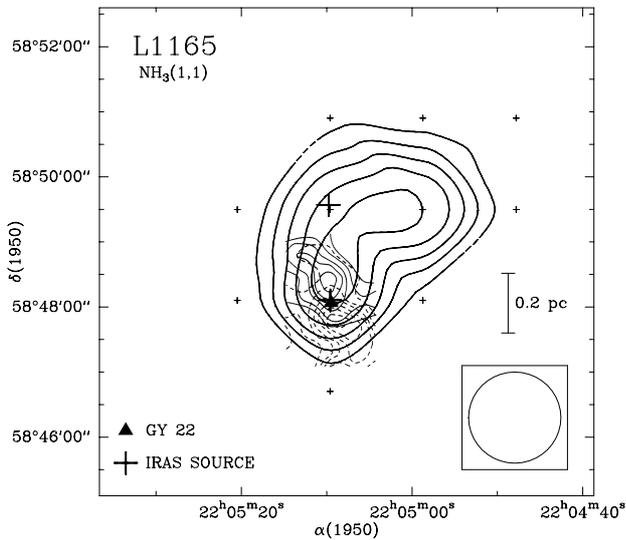}}
    \caption{Same as Fig.~\ref{m120n}, for the L1165 region. 
     The \nh\ lowest contour is 0.15 K and the increment is 0.05 K. The IRAS
     sources are, from north to south, IRAS 22051+5849 and IRAS 22051+4858
     (=HH 354 IRS). The CO bipolar outflow is from Parker et al.\ (1991)}
  \label{l1165}
  \end{center}
\end{figure}

The \nh\ map (Fig.~\ref{l1165}) shows a condensation with the emission peaking
very close to the position of IRAS 22051+5848. The IRAS colors of this source
are typical of embedded sources (Parker 1991) and it is surrounded by the
reflection nebulosity GY 22. Altogether suggest that IRAS 22051+5848 is a young
object embedded in the high density gas. IRAS 22051+5849 is located close 
to the emission maximum and also appears associated
with the dense gas, but its IRAS
colors corresponding to a blackbody at $T>1000$ K are suggestive of a background
source (Tapia et al.\ 1997). This source needs more accurate observations in
order to establish its relationship with the core and the outflow.

\subsection{IRAS 22134+5834}

The distance of this source is established to be 2.6 kpc (Sridharan et al.\ 2002), 
although previously a distance of 900 pc was assumed. Dobashi et al.\ (1994) discovered a 
molecular outflow associated with the source
IRAS 22134+5834, one of the most luminous sources in the Cepheus region. \hh\
images
(Kumar, Bachiller \& Davis, 2002) revealed a dense stellar cluster around the
IRAS source, that 
was interpreted as ring-shaped cluster (Kumar, Ojha \& Davis 2003).

\begin{figure}
\begin{center}
\resizebox{8cm}{!}{\includegraphics[77,257][491,631]{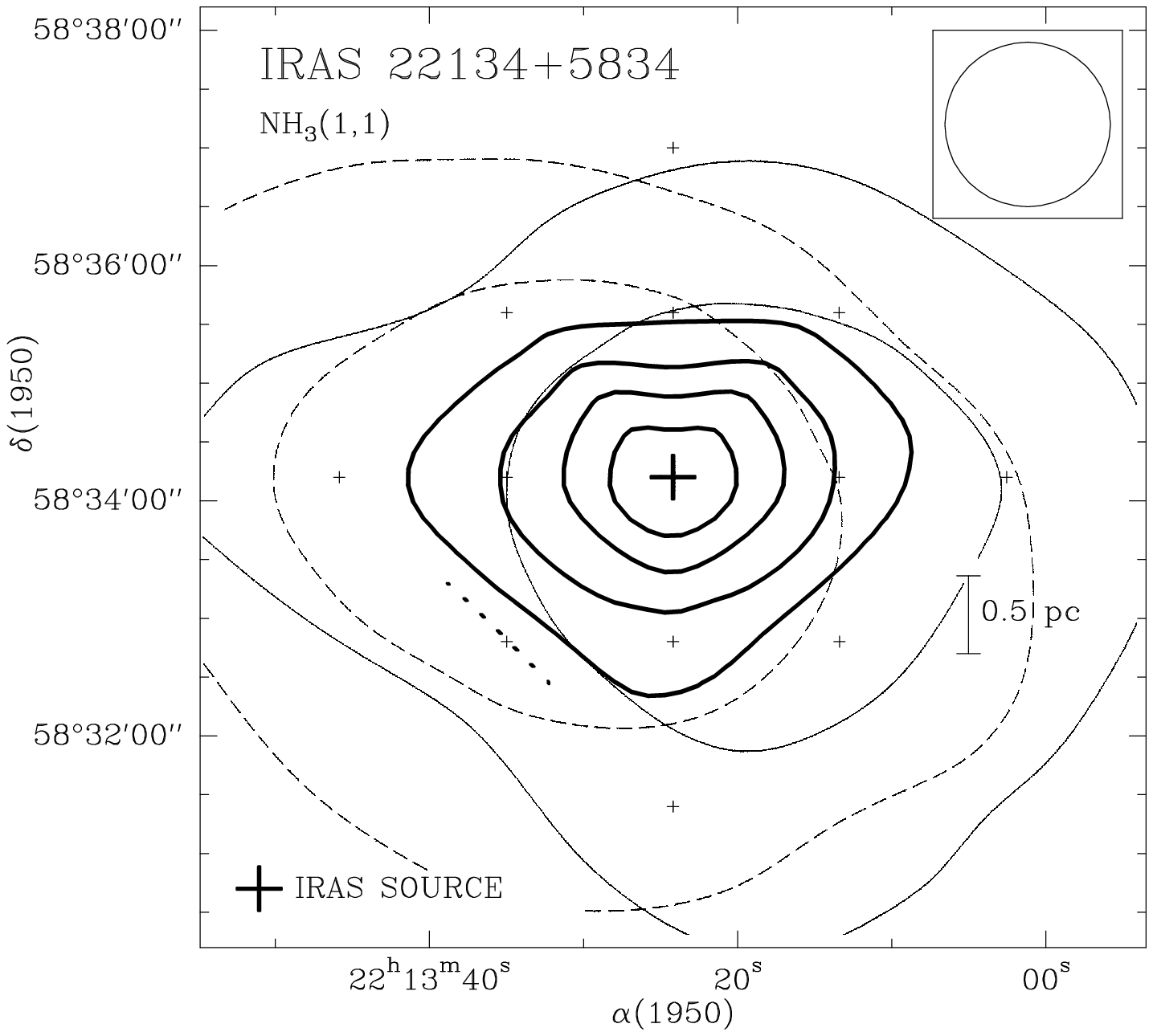}}
    \caption{Same as Fig.~\ref{m120n}, for the region around IRAS 22134+5834.
     The \nh\ lowest contour is 0.2 K and the increment is 0.1 K. The CO bipolar
     outflow is from Dobashi et al.\ (1994)}
  \label{s134}
  \end{center}
\end{figure}

The \nh\ distribution (Fig.~\ref{s134}) shows a compact condensation with the
emission maximum located at the IRAS position, in agreement with the \dco\ and
\tco\ maps obtained by Dobashi et al.\ (1994). The IRAS source is bright at FIR
wavelengths
and not at NIR wavelengths. This and its association with high-density gas, 
suggest that IRAS 22134+5834 is a very young stellar object, a possible massive
protostar as suggested by Dobashi et al.\ (1994) and Kumar et al.\ (2003).

\subsection{L1221}

L1221 is a small isolated  dark cloud associated with IRAS 22266+6845,
which has an energy distribution rising to longer wavelengths. Umemoto
et al.\  (1991) discovered a  bipolar molecular outflow  centered near
the position of the  IRAS source.
 The   outflow  shows  a   U-shaped  structure  open to the
northwest. However, CO observations with higher angular resolution showed that
the outflow may consist of two bipolar outflows, an east-west outflow associated
with the IRAS source and a north-south outflow originating about $25''$ to the
east of the IRAS source (Lee et al. 2002). 

The region was mapped with different high-density
tracer  molecules (CS,  \hco,  HCN, \cdo; Umemoto et al.\ 1991, Lee \& Ho 2005).
Alten  et al.\ (1997) 
discovered  an HH object,  HH 363, in the vicinity of the IRAS source. Anglada
et al.\ (2005) detected three cm continuum sources in this region, 
but none seems to be associated with the IRAS source. Two mm continuum sources
are detected, one of them (MM1) peaks around the IRAS source and toward one of
the three infrared sources detected by the {\it Spitzer Space Telescope} (Lee \&
Ho 2005)

\begin{figure}
\begin{center}
\resizebox{8cm}{!}{\includegraphics[64,217][490,566]{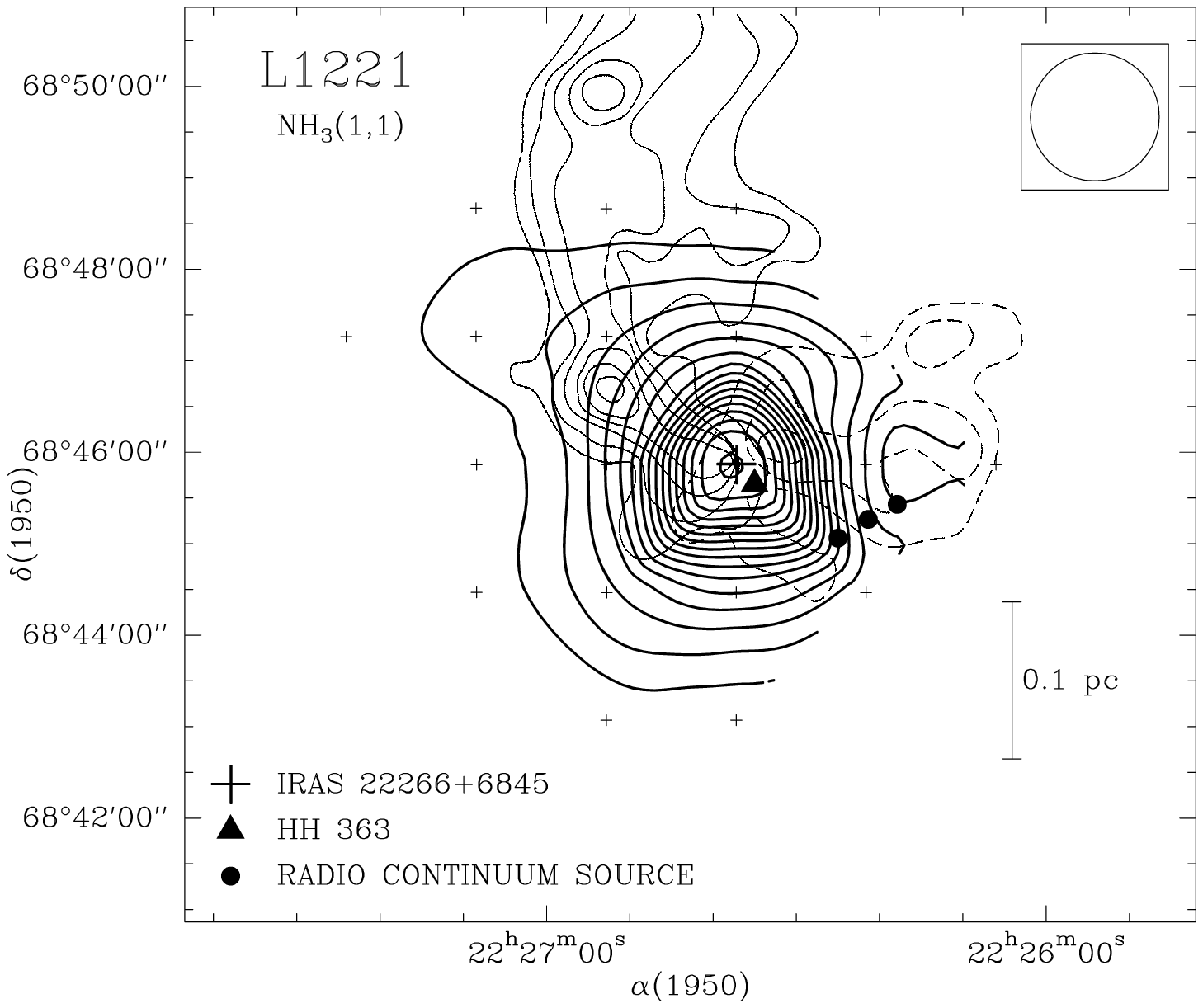}}
    \caption{Same as Fig.~\ref{m120n}, for the L1221 region. The \nh\
     lowest contour level is 0.25 K and the increment is 0.2 K. The map of
     the CO bipolar outflow is from Umemoto et al.\ (1991)}
  \label{l1221}
  \end{center}
\end{figure}

The  \nh\  emission (Fig.~\ref{l1221})  is  distributed  in a  compact
condensation centered on the IRAS source, with weak emission extending
to the NE. The size  of the  \nh\
condensation (see Table~\ref{parameters})  is similar to that obtained
in the CS and \hco\ maps (Umemoto et al.\ 1991), but  the emission peaks of the
CS, \hco, and
HCN cores are  located $\sim 54''$ ($\sim 0.06$  pc) to the south
of the \nh\ emission peak. A displacement between CS and \nh\ emission
peaks  has  been  found  in   several  other  regions,  and  has  been
interpreted by Morata et al.\ (1997) in terms of chemical evolution.

The presence of the IRAS source toward the ammonia peak, which is located 
at the center of symmetry of the outflow, favors IRAS 22266+6845 as its 
exciting source. This contrast with the proposal of 
Umemoto et al.\ (1991), who postulate the existence of an object located to the
south 
of the position of the CS emission peak as the outflow driving source.

\subsection{NGC 7538}

The NGC 7538 molecular cloud is an active site of high-mass star formation
containing five infrared sources  (IRS 1, 2, 3, 9 and  11) within an area 
of $3\farcm5  \times 3\farcm5$ (Werner et al.\ 1979).  Estimates  of the 
distance range from 2.2  to 4.7 kpc. We adopt  a  distance of  2.7  kpc 
(Kameya  et  al.\  1986). Campbell  \& Thompson (1984) found a high-velocity 
outflow near IRS 1. Kameya
et al.\ (1989) discovered three additional outflows in the region. Two
of them are associated with sources IRS 9 and IRS  11, but the third
one  was not associated  with any  known source.  Davis et  al.\ (1998)
detected a  collimated \hh\ jet associated  with the IRS 9 outflow, two
possible bow shocks related with the IRS 1 and IRS 9 outflows, and a number
of \hh\ compact knots which coincide with the IRS 11 outflow, and that could
be related  with it.  Davis et  al.\ (1989) also detected  a cavity to
the northwest of IRS 1.

\begin{figure}
\begin{center}
\resizebox{8cm}{!}{\includegraphics[36,276][495,619]{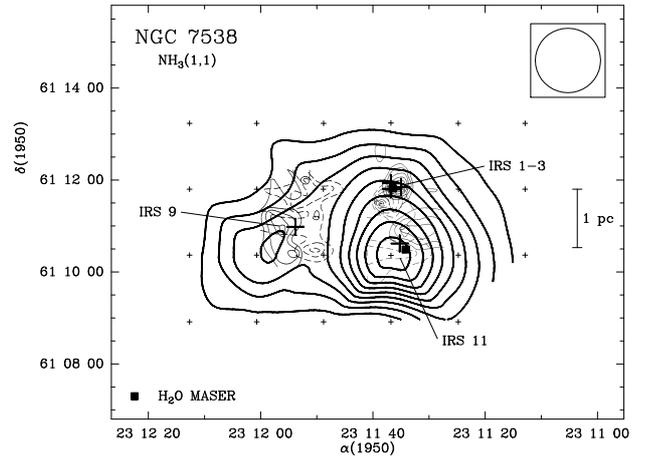}}
    \caption{Same as Fig.~\ref{m120n}, for the NGC 7538 region. The \nh\
    lowest contour level is 0.25 K and the increment is 0.2 K. The map of
    the CO  bipolar outflows are from Kameya et al.\ (1989)}
  \label{ngc7538}
  \end{center}
\end{figure}

The \nh\ map (Fig.~\ref{ngc7538}) shows a condensation elongated in
the east-west direction. The ammonia emission peak is located toward 
the position of IRS 11. A secondary emission peak is located near IRS 9. 
IRS 1-3 also appear projected toward the ammonia
condensation. The association of IRS 11 with the ammonia emission peak suggests
that this source is the most embedded object, in agreement with the CS
observations (Kameya et al.\ 1986).

\begin{figure}
\begin{center}
\resizebox{8cm}{!}{\rotatebox{-90}{\includegraphics[58,40][553,673]{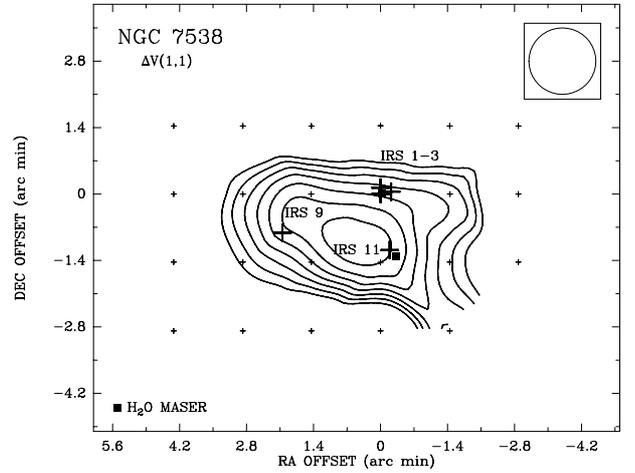}}
}
   \caption[]{A contour map of the \nh\ intrinsic line width for the NGC 7538
   region. The lowest contour level is 1.6 \kms\ and the increment is 0.3 \kms}
  \label{amplada}
  \end{center}
\end{figure}

The \nh\ line profiles show broad intrinsic line widths, with values ranging
from $\Delta V = 1.68$ \kms\ to $3.41$ \kms. The broadest intrinsic line width
is found very close to the IRS 11 position. Fig.~\ref{amplada} shows a contour
map of the \nh\ intrinsic line width. The broadening due to the \nh\ hyperfine
structure 
is about 0.5 \kms\ and the thermal broadening for the derived kinetic
temperature ($\sim
28.3$ K, see Table \ref{parameters}) is $\sim 0.27$ \kms. Both values are
significantly lower than the observed line width, which could indicate
turbulent motions of the gas, or that the dense gas is suffering an interaction
with the 
outflows.

\end{document}